\documentclass[12pt,a4paper]{article}
\usepackage{cite}
\usepackage{axodraw}
\usepackage{epsfig}

\let\log\ln

\advance\textheight by \topskip
\if@twoside \oddsidemargin -17pt \evensidemargin 00pt
\else \oddsidemargin 00pt \evensidemargin 00pt
\fi
\expandafter\ifx\csname mathrm\endcsname\relax\def\mathrm#1{{\rm #1}}\fi

\renewcommand{\theequation}{\thesection.\arabic{equation}}
\newcounter{saveeqn}

\makeatletter 
\@addtoreset{equation}{section}
\makeatother

\def\beq{\begin{equation}}
\def\eeq{\end{equation}}
\def\beqar{\begin{eqnarray}}
\def\eeqar{\end{eqnarray}}
\def\barr#1{\begin{array}{#1}}
\def\earr{\end{array}}
\def\bfi{\begin{figure}}
\def\efi{\end{figure}}
\def\btab{\begin{table}}
\def\etab{\end{table}}
\def\bce{\begin{center}}
\def\ece{\end{center}}

\def\nl{\nonumber\\}
\def\nln{\nonumber\\*[-1ex]\phantom{\fbox{\rule{0em}{2ex}}}}

\def\al{\alpha}

\def\ga{\gamma}
\def\de{\delta}

\def\la{\lambda}
\def\si{\sigma}

\def\refeq#1{\mbox{(\ref{#1})}}

\def\reffi#1{\mbox{Fig.~\ref{#1}}}
\def\reffis#1{\mbox{Figs.~\ref{#1}}}
\def\refta#1{\mbox{Table~\ref{#1}}}

\def\refse#1{\mbox{Sect.~\ref{#1}}}

\def\refapp#1{\mbox{Appendix~\ref{#1}}}
\def\citere#1{\mbox{Ref.~\cite{#1}}}
\def\citeres#1{\mbox{Refs.~\cite{#1}}}

\def\solid{\raise.9mm\hbox{\protect\rule{1.1cm}{.2mm}}}
\def\dash{\raise.9mm\hbox{\protect\rule{2mm}{.2mm}}\hspace*{1mm}}
\def\dot{\rlap{$\cdot$}\hspace*{2mm}}

\def\solid{\raise.9mm\hbox{\protect\rule{12mm}{.2mm}}}
\def\dash{\raise.9mm\hbox{\protect\rule{1.6mm}{.2mm}}\hspace*{1mm}}
\def\dot{\raise.9mm\hbox{\protect\rule{0.8mm}{.2mm}}\hspace*{0.8mm}}
\def\dashdot{\raise.9mm\hbox{\protect\rule{.3mm}{.2mm}}\hspace*{.8mm}\raise.9mm\hbox{\protect\rule{1.3mm}{.2mm}}\hspace*{.8mm}}

\newcommand{\GeV}{\unskip\,\mathrm{GeV}}

\newcommand{\TeV}{\unskip\,\mathrm{TeV}}
\newcommand{\aba}{\unskip\,\mathrm{ab}}

\def\mathswitchr#1{\relax\ifmmode{\mathrm{#1}}\else$\mathrm{#1}$\fi}

\newcommand{\PW}{\mathswitchr W}

\newcommand{\PZ}{\mathswitchr Z}

\newcommand{\PH}{\mathswitchr H}
\newcommand{\Pe}{\mathswitchr e}

\newcommand{\Pne}{\mathswitch \nu_{\mathrm{e}}}
\newcommand{\Pane}{\mathswitch \bar\nu_{\mathrm{e}}}

\newcommand{\Pd}{\mathswitchr d}

\newcommand{\Pt}{\mathswitchr t}

\newcommand{\Pep}{\mathswitchr {e^+}}
\newcommand{\Pem}{\mathswitchr {e^-}}

\newcommand{\PWp}{\mathswitchr {W^+}}
\newcommand{\PWm}{\mathswitchr {W^-}}
\newcommand{\PWpm}{\mathswitchr {W^\pm}}

\def\mathswitch#1{\relax\ifmmode#1\else$#1$\fi}

\newcommand{\MW}{\mathswitch {M_\PW}}

\newcommand{\MZ}{\mathswitch {M_\PZ}}
\newcommand{\MH}{\mathswitch {M_\PH}}

\newcommand{\Mt}{\mathswitch {m_\Pt}}

\newcommand{\PL}{\mathswitch {P_\PL}}

\newcommand{\NCt}{\mathswitch {N_{\mathrm{C}}^t}}
\newcommand{\scrs}{\scriptscriptstyle}
\newcommand{\sw}{\mathswitch {s_{\scrs\PW}}}
\newcommand{\cw}{\mathswitch {c_{\scrs\PW}}}

\newcommand{\GF}{\mathswitch {G_\mu}}

\def\ie{i.e.\ }

\newcommand{\Oa}{\mathswitch{{\cal{O}}(\alpha)}}

\newcommand{\SUtwo}{\mathrm{SU(2)}}
\newcommand{\Uone}{\mathrm{U}(1)}

\newcommand{\rR}{{\mathrm{R}}}
\newcommand{\rT}{{\mathrm{T}}}
\newcommand{\rL}{{\mathrm{L}}}

\newcommand{\ri}{{\mathrm{i}}}

\newcommand{\M}{{\cal{M}}}

\newcommand{\EM}{\mathrm{EM}}
\newcommand{\EW}{\mathrm{EW}}
\newcommand{\CM}{\mathrm{CM}}

\newcommand{\brems}{\mathrm{brems}}

\newcommand{\Born}{\mathrm{Born}}

\newcommand{\fact}{{\mathrm{fact}}}

\newcommand{\EVBA}{\mathrm{EVBA}}
\newcommand{\exact}{\mathrm{exact}}

\newcommand{\virt}{\mathrm{virt}}

\newcommand{\elm}{{\mathrm{em}}}
\newcommand{\ew}{\mathrm{ew}}

\newcommand{\SC}{{\mathrm{LSC}}}
\renewcommand{\SS}{{\mathrm{SSC}}}
\newcommand{\cc}{{\mathrm{C}}}

\newcommand{\pre}{{\mathrm{PR}}}

\newcommand{\bew}{b^{\ew}}

\newcommand{\cew}{C^{\ew}}

\newcommand{\loZW}{\log{\left(\frac{\MZ^2}{\MW^2}\right)}}

\newcommand{\lots}{\log{\left(\frac{|\mat|}{|\mas|}\right)}}
\newcommand{\lous}{\log{\left(\frac{|\mau|}{|\mas|}\right)}}
\newcommand{\lotu}{\log{\left(\frac{|\mat|}{|\mau|}\right)}}

\newcommand{\lWla}{l(\MW^2,\la^2)}

\newcommand{\PT}{P_{\mathrm{T}}}

\newcommand{\PPT}{\mathrm{T}}
\newcommand{\PPL}{\mathrm{L}}
\newcommand{\tot}{\mathrm{tot}}

\newcommand{\PCPU}{\mathrm{CPU}}

\newcommand{\mas}{r_{12}}
\newcommand{\mat}{r_{13}}
\newcommand{\mau}{r_{23}}

\newcommand{\As}{\frac{\mat-\mau}{\mas}}
\newcommand{\At}{\frac{\mas-\mau}{\mat}}
\newcommand{\Au}{\frac{\mat-\mas}{\mau}}

\newcommand{\proc}[8]{
\PW^{#1}_{#2}
\PW^{#3}_{#4}
\to 
\PW^{#5}_{#6}
\PW^{#7}_{#8}
}
\newcommand{\csproc}[8]{
W^{#1}_{#2}
W^{#3}_{#4}
W^{#5}_{#6}
W^{#7}_{#8}}

\newcommand{\csmel}[8]{
\M^{\csproc{#1}{#2}{#3}{#4}{#5}{#6}{#7}{#8}}}

\newcommand{\lah}{\la_\PH}

\newcommand{\genproc}{
\PW^-_{\la_1}
\PW^+_{\la_2}\to
\PW^-_{-\la_3}
\PW^+_{-\la_4}
}
\newcommand{\genproccrossed}{
W^-_{\la_1}
W^+_{\la_2}
W^+_{\la_3}
W^-_{\la_4}
}
\newcommand{\procLL}{
\swbos{-}{0}{1}
\swbos{+}{0}{2}\to
\swbos{-}{0}{3}
\swbos{+}{0}{4}
}

\newcommand{\procTT}{
\swbos{-}{\tau_1}{1}
\swbos{+}{\tau_2}{2}\to
\swbos{-}{-\tau_3}{3}
\swbos{+}{-\tau_4}{4}
}

\newcommand{\procmixTT}{
\swbos{-}{\tau_1}{1}
\swbos{+}{\tau_2}{2}\to
\swbos{-}{0}{3}
\swbos{+}{0}{4}
}
\newcommand{\procmixLL}{
\swbos{-}{0}{1}
\swbos{+}{0}{2}
\to
\swbos{-}{-\tau_3}{3}
\swbos{+}{-\tau_4}{4}
}

\newcommand{\procmixTL}{
\swbos{-}{\tau_1}{1}
\swbos{+}{0}{2}\to
\swbos{-}{-\tau_3}{3}
\swbos{+}{0}{4}
}

\newcommand{\procmixLT}{
\swbos{-}{0}{1}
\swbos{+}{\tau_2}{2}\to
\swbos{-}{0}{3}
\swbos{+}{-\tau_4}{4}
}

\newcommand{\cprocLL}{
\iwbos{-}{0}{1}
\iwbos{+}{0}{2}
\iwbos{+}{0}{3}
\iwbos{-}{0}{4}
}

\newcommand{\cprocTT}{
\iwbos{-}{\tau_1}{1}
\iwbos{+}{\tau_2}{2}
\iwbos{+}{\tau_3}{3}
\iwbos{-}{\tau_4}{4}
}

\newcommand{\cprocmixTT}{
\iwbos{-}{\tau_1}{1}
\iwbos{+}{\tau_2}{2}
\iwbos{+}{0}{3}
\iwbos{-}{0}{4}
}
\newcommand{\cprocmixLL}{
\iwbos{-}{0}{1}
\iwbos{+}{0}{2}
\iwbos{+}{\tau_3}{3}
\iwbos{-}{\tau_4}{4}
}

\newcommand{\cprocmixTL}{
\iwbos{-}{\tau_1}{1}
\iwbos{+}{0}{2}
\iwbos{+}{\tau_3}{3}
\iwbos{-}{0}{4}
}

\newcommand{\cprocmixLT}{
\iwbos{-}{0}{1}
\iwbos{+}{\tau_2}{2}
\iwbos{+}{0}{3}
\iwbos{-}{\tau_4}{4}
}

\newcommand{\losw}[1]{\log^{#1}{\left(\frac{|\mas|}{\MW^2}\right)}}
\newcommand{\lost}[1]{\log^{#1}{\left(\frac{|\mas|}{\Mt^2}\right)}}
\newcommand{\loHw}[1]{\log^{#1}{\left(\frac{\MH^2}{\MW^2}\right)}}
\newcommand{\lotw}[1]{\log^{#1}{\left(\frac{\Mt^2}{\MW^2}\right)}}

\newcommand{\swbos}[3]{\PW^{#1}_{#2}}
\newcommand{\iwbos}[3]{W^{#1}_{#2}}

\def\draftdate{\relax}
\def\mpar#1{\relax}
\def\mda{\relax}
\def\mua{\relax}
\def\mla{\relax}
\def\Mda{\relax}
\def\Mua{\relax}
\def\Mla{\relax}
\def\draft{
\def\thtystars{******************************}
\def\sixtystars{\thtystars\thtystars}
\typeout{}
\typeout{\sixtystars**}
\typeout{* Draft mode!
         For final version remove \protect\draft\space in source file *}
\typeout{\sixtystars**}
\typeout{}
\def\draftdate{\today}
\def\mua{\marginpar[\boldmath\hfil$\uparrow$]%
                   {\boldmath$\uparrow$\hfil}%
                    \typeout{marginpar: $\uparrow$}\ignorespaces}
\def\mda{\marginpar[\boldmath\hfil$\downarrow$]%
                   {\boldmath$\downarrow$\hfil}%
                    \typeout{marginpar: $\downarrow$}\ignorespaces}
\def\mla{\marginpar[\boldmath\hfil$\rightarrow$]%
                   {\boldmath$\leftarrow $\hfil}%
                    \typeout{marginpar: $\leftrightarrow$}\ignorespaces}
\def\Mua{\marginpar[\boldmath\hfil$\Uparrow$]%
                   {\boldmath$\Uparrow$\hfil}%
                    \typeout{marginpar: $\Uparrow$}\ignorespaces}
\def\Mda{\marginpar[\boldmath\hfil$\Downarrow$]%
                   {\boldmath$\Downarrow$\hfil}%
                    \typeout{marginpar: $\Downarrow$}\ignorespaces}
\def\Mla{\marginpar[\boldmath\hfil$\Rightarrow$]%
                   {\boldmath$\Leftarrow $\hfil}%
                    \typeout{marginpar: $\Leftrightarrow$}\ignorespaces}
\def\mpar##1{\marginpar{\hbadness10000%
                      \sloppy\hfuzz10pt\boldmath\bf##1}%
                      \typeout{marginpar: ##1}\ignorespaces}
\overfullrule 5pt
\oddsidemargin -15mm
\marginparwidth 29mm
}

\newcommand{\thismonth}{\ifcase\month\or January\or February\or March \or April
\or May \or June \or July \or August \or September \or \November \or 
\December\fi}

\makeatletter

\def\eqnarray{\stepcounter{equation}\let\@currentlabel=\theequation
\global\@eqnswtrue
\global\@eqcnt\z@\tabskip\@centering\let\\=\@eqncr
$$\halign to \displaywidth\bgroup\hskip\@centering
  $\displaystyle\tabskip\z@{##}$\@eqnsel&\global\@eqcnt\@ne
  \hskip 2\arraycolsep \hfil${##}$\hfil
  &\global\@eqcnt\tw@ \hskip 2\arraycolsep $\displaystyle\tabskip\z@{##}$\hfil
   \tabskip\@centering&\llap{##}\tabskip\z@\cr}
\def\appendix{\par
 \setcounter{section}{0} \setcounter{subsection}{0}
 \def\thesection{\Alph{section}}}

\makeatother

\def\slash#1{\setbox0\hbox{$#1$}\hbox to\wd0{\hss$/$\hss}\nobreak\hskip-\wd0\box0}

\begin{document}

\tolerance=100000
\thispagestyle{empty}
\setcounter{page}{0}

\thispagestyle{empty}
\def\thefootnote{\fnsymbol{footnote}}
\setcounter{footnote}{1}
\null
\draftdate\hfill  DFTT 19/2006\\
\strut\hfill MPP-2006-147\\
\strut\hfill PSI-PR-06-12\\
\strut\hfill hep-ph/0611289
\vskip 0cm
\vfill
\begin{center}
{\Large \bf Logarithmic electroweak corrections to \boldmath 
$\Pe^+\Pe^-\rightarrow\Pne\Pane\PW^+\PW^-$
\par} \vskip 2.5em
{\large
{\sc E.\ Accomando$^1$, A.\ Denner$^2$, S.\ Pozzorini$^3$}}%
\\[.5cm]
$^1$ {\it Dipartimento di Fisica Teorica, Universit\`a di Torino,\\
and INFN, Sezione di Torino,\\
Via P. Giuria 1, 10125 Torino, Italy} \\[0.5cm]
$^2$ {\it Paul Scherrer Institut, W\"urenlingen und Villigen\\
CH-5232 Villigen PSI, Switzerland} \\[0.5cm]
$^3$ {\it Max-Planck-Institut f\"ur Physik (Werner-Heisenberg-Institut)\\ 
F\"ohringer Ring 6, D-80805 M\"unchen, Germany}
\\[0.3cm]
\par
\end{center}\par
\vskip 2.0cm \vfill {\bf Abstract:} \par 
We consider $\PW$-boson scattering at high-energy $\Pep\Pem$ colliders
and study one-loop logarithmic electroweak corrections within the
Standard Model assuming a light Higgs boson.  We present explicit
analytical results for $\PWp\PWm\to\PWp\PWm$.  Using the equivalent
vector-boson approximation, we have implemented these corrections into
a Monte Carlo program for the process
$\Pe^+\Pe^-\rightarrow\Pne\Pane\PW^+\PW^-$.  The quality of the
equivalent vector-boson approximation and of the logarithmic
high-energy approximation for the electroweak corrections is discussed
in detail.  The impact of the radiative effects is quantitatively
analysed.  The corrections are negative and their size, typically of
the order of 10\%, increases with energy reaching up to $-20\%$ and
$-50\%$ at the ILC and CLIC, respectively.  
\vskip1cm
\noindent
November 2006
\par
\null
\setcounter{page}{0}
\clearpage
\def\thefootnote{\arabic{footnote}}
\setcounter{footnote}{0}

\def\mla{\marginpar[\boldmath\hfil$\rightarrow$]%
                   {\boldmath$\leftarrow $\hfil}%
                    \typeout{marginpar: $\leftrightarrow$}\ignorespaces}
\def\mua{\marginpar[\boldmath\hfil$\uparrow$]%
                   {\boldmath$\uparrow$\hfil}%
                    \typeout{marginpar: $\uparrow$}\ignorespaces}
\def\mda{\marginpar[\boldmath\hfil$\downarrow$]%
                   {\boldmath$\downarrow$\hfil}%
                    \typeout{marginpar: $\downarrow$}\ignorespaces}

\section{Introduction}
\label{Introduction}

One of the foremost open questions in particle physics concerns the
mechanism of electroweak symmetry breaking. Depending on its
realization in nature, the understanding of this subtle mechanism must
be approached in different ways. If the Standard Model (SM) with a
light Higgs boson is realized, the Higgs boson can be directly
produced and its properties investigated. If the Higgs boson is heavy
or absent, a complementary approach must be pursuit. In this case,
information can be extracted from the scattering of longitudinally
polarized gauge bosons. In fact, at high energies, longitudinal vector
bosons unveil their origin as Goldstone bosons and, by virtue of the
equivalence theorem, reflect the dynamics of electroweak symmetry
breaking.

Accordingly, vector-boson scattering (VBS) looks very much different
in these different scenarios of electroweak symmetry breaking.  In the
SM with a light Higgs boson the gauge sector remains weakly
interacting and the cross section can be reliably predicted within
perturbation theory.  In the alternative scenario, where a light Higgs
boson is absent, perturbative unitarity is violated, and the
longitudinal gauge bosons must become strongly interacting at high
energies thus allowing for non-perturbative restoration of unitarity
\cite{SEWSB}.

In the past years, a variety of models have been proposed to
parametrize the strongly-interacting electroweak gauge sector, and
to recover unitarity (see for instance \citere{Bagger:1993zf} and
references therein). The common prediction is an enhanced production
of longitudinal gauge bosons.  However, the phenomenological
consequences as well as the new particles the various models provide
can be sensibly different.  They can be classified into two main
groups \cite{Butterworth:2002tt}. In the most optimistic case, one
could expect many new resonances at future colliders. In the less
favorable scenario, the mass of any new particle could be much bigger
than the energy scale probed at the planned accelerators. In this
case, the indirect effect of such particles would only consist in a
slight increase of the VBS event rate at high energy compared to the
predictions of the SM with a light Higgs boson.

Several studies have been performed in order to estimate the possible
reach of the future lepton and hadron colliders
\cite{Bagger:1993zf,Butterworth:2002tt,sewsb-analyses1,Barger:1995cn,sewsb-analyses2,boos-1997,sewsb-analyses4}.
The answers strongly depend on energy and luminosity parameters.  Also
a good control of the SM background can prove essential, particularly
for the less favorable case.  In this respect, the scattering of
\PW~bosons within the SM with a light Higgs boson constitutes an
irreducible background to any new-physics signal pointing towards a
strongly-interacting VBS regime.

In this paper, we consider VBS within the SM with a light Higgs boson
at the planned $\Pep\Pem$ colliders. For the International Linear
Collider (ILC) \cite{ilc} we consider a centre-of-mass (CM) energy
$\sqrt{s}=1\TeV$ and for the Compact LInear Collider (CLIC)
\cite{Accomando:2004sz} $\sqrt{s}=3\TeV$.  More precisely, we focus on the
production of $\PW$-boson pairs plus neutrinos in the reaction
$\Pep\Pem\to\nu_\Pe\bar\nu_\Pe\PW^+\PW^-$.  For the projected
luminosity $L=1\aba^{-1}$, the experimental collaborations will
collect thousands of events coming from VBS in the high-energy domain,
where a possible strongly-interacting regime of the weak gauge sector
could manifest itself.

To match the envisaged statistical precision, the SM predictions have
to be computed beyond lowest order. Indeed, in the very same
high-energy region of interest, the electroweak radiative effects are
enhanced by electroweak Sudakov logarithms
\cite{Kuroda:1991wn,Denner:2000jv,Denner:2001gw,Pozzorini:2001rs,Fadin:2000bq},
\ie double and single logarithms of the ratio of the scattering energy
over the vector-boson mass (recent progress in the evaluation of
electroweak Sudakov logarithms is discussed in
\citere{Denner:2006jr}).  At $\Oa$, these corrections can reach
several tens of per cent, as confirmed by various analyses performed
for different processes at lepton and hadron colliders
\cite{Beccaria:2000fk,Layssac:2001ur,Dittmaier:2001ay,Accomando:2001fn,Accomando:2004de,Maina:2003is,Baur:2004ig,Kuhn:2004em,Kuhn:2006vh,Bernreuther:2006vg}.
Hence, in the case at hand, they have to be taken into account 
in order to search for small deviations between data and SM
predictions that might appear as a signal of strongly interacting
electroweak symmetry breaking.

So far, the process $\Pep\Pem\to\nu_\Pe\bar\nu_\Pe\PW^+\PW^-$ has been
computed at Born level, and found promising for investigating
electroweak symmetry breaking at high invariant masses of the produced
$\PW$-boson pairs \cite{boos-1997,moenig-2005,zerwas-2006}. The aim of
our work is to study the contributions of the electroweak Sudakov
logarithms, which represent the dominant electroweak corrections at
high energies.

For the calculation of the electroweak corrections we use the
equivalent vector-boson approximation following the approach of
\citere{kuss1997}. Within this approximation we only consider virtual
$\Oa$ corrections to the WW-scattering subprocess. These corrections
are calculated using the method of
\citeres{Denner:2000jv,Denner:2001gw} in the high-energy logarithmic
approximation.  As in \citeres{Denner:2000jv,Denner:2001gw}, the
virtual photonic corrections are split into a symmetric-electroweak
and a purely electromagnetic part, which originate from above and
below the electroweak scale, respectively.  The former part is in
practice obtained by setting the photon mass equal to $\MW$ in the
photonic virtual corrections, and is infrared finite.  The infrared
singularities are contained in the purely-electromagnetic part and are
cancelled when including soft-photon bremsstrahlung.  Both the
symmetric-electroweak and purely-electromagnetic parts are included in
our analytical results. However the latter is omitted in our numerical
studies since it strictly depends on the experimental setup.

The paper is organized as follows. In \refse{sec:setup} we define the
process and give the setup for the numerical evaluation.  In
\refse{sec:strategy} we describe the strategy of the calculation and
discuss the quality of the used approximations.  Numerical results are
presented in \refse{sec:results}, and \refse{sec:conclusions} contains
the summary. Explicit analytical results are listed in the appendices.

\section{Process definition and numerical setup}
\label{sec:setup}

\begin{figure}
  \unitlength 1cm
  \begin{center}
  \begin{picture}(16.,15.)
  \put(-0.7,-4){\epsfig{file=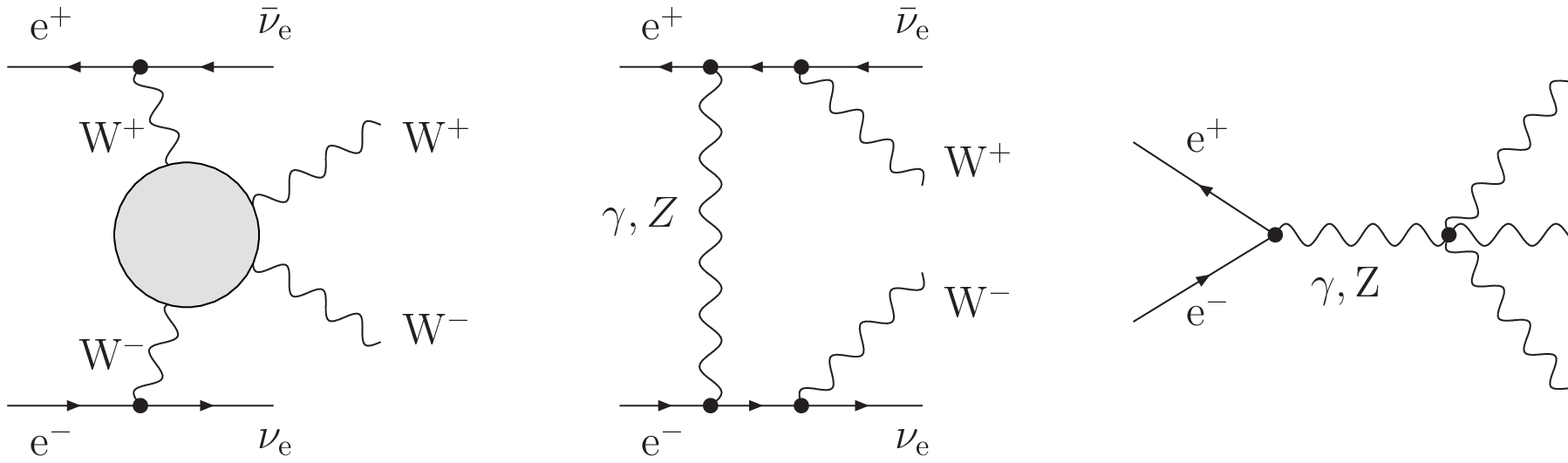,width=14cm}}
  \end{picture}
  \end{center}
\vspace{-10.5cm}
\caption{Feynman diagrams contributing to the process 
  $\Pep\Pem\to\Pne\Pane\PW^+\PW^-$. The first graph on the left is a
  generic representation of the $\PW\PW$-scattering signal.  The
  remaining two diagrams are examples of irreducible background.  }
\label{fi:graphs}
\end{figure}

We consider the production of a $\PW$-boson pair plus two neutrinos in 
electron--positron collisions:
\beq
\Pep\Pem\to\Pne\Pane\PW^+\PW^-.
\label{eq:process}
\eeq
This process contains the VBS subprocess $\PW^+\PW^-\to\PW^+\PW^-$
generically described by the first Feynman diagram in
\reffi{fi:graphs}.
The $\PW\PW$-scattering signal is not the only contribution to the
final state in \refeq{eq:process}. The irreducible background,
exemplified by the second and third Feynman graphs in
\reffi{fi:graphs}, is indeed sizeable and must be properly suppressed
in order to enhance the VBS signal-to-background ratio.  A set of
appropriate kinematical cuts to be imposed is summarized below.

In our analyses, we use the input values \cite{Yao:2006px}
\beqar
\MW &=& 80.403\GeV,\qquad \MZ = 91.1876\GeV, \nl
\MH &=& 120\GeV, \qquad \Mt = 174.2\GeV, 
\label{eq:SMpar}
\eeqar
for vector-boson, Higgs-boson, and top-quark  masses. 
All other
fermions are taken to be massless.  The sine $\sw$ and cosine $\cw$ of
the weak mixing angle are fixed by
\beq\label{eq:defcw}
\cw^2 = 1-\sw^2 = \frac{\MW^2}{\MZ^2}.
\eeq
Moreover, we adopt the so called $G_{\mu}$-scheme, which effectively
includes higher-order contributions associated with the running of the
electromagnetic coupling and the leading universal two-loop
$\Mt$-dependent corrections in the definition of $\al$.  Using
\beq
\GF=1.16637\times 10^{-5}\GeV^{-2},
\eeq
we have
\beq
\alpha=\sqrt{2}G_{\mu}\MW^2\sw^2/\pi=
1/132.38\ldots
.
\eeq

In the following sections, we present results for a $\CM$ energy
$\sqrt s=1\TeV$, which can be reached at the ILC, and for $\sqrt
s=3\TeV$ as is planned for CLIC. In both cases we assume an integrated
luminosity $L=1\aba^{-1}$.

Based on the study of \citere{boos-1997}, we have implemented a
general set of cuts, proper for ILC and CLIC analyses, defined as
follows.  In the listed cuts, the numbers outside parentheses are used
for $\sqrt{s}=1\TeV$, those within parentheses for $\sqrt{s}=3\TeV$.
\begin{itemize}
\item We require a $\PW$-boson transverse momentum $\PT(\PW^\pm )\ge
    100 (200)\GeV$, since the production of longitudinal vector
  bosons, \ie the VBS signal, is enhanced for large scattering angles
  and high energies. Moreover, this cut removes events dominated by
  $t$-channel photon exchange in subprocesses.
\item We require $|{\cos\theta (\PW^\pm)}|\le 0.8$, where
$\theta(\PW^\pm)$ is the angle of the produced $\PW^\pm$ boson with
respect to the incoming positron in the laboratory frame, since the
VBS signal is characterized by central $\PW$-boson production. This
cut also removes events dominated by $t$-channel photon exchange in
subprocesses.
\item We require $|y(\PWp)-y(\PWm)|\le2$, where $y(\PW^\pm)$ is the
  rapidity of the produced $\PWpm$ boson defined as
  $y=0.5\ln[(E+P_{\rL})/(E-P_{\rL})]$ and $E$ and $P_{\rL}$ are the
  W-boson energy and component of the momentum along the beam axis,
  respectively. This additional angular cut ensures the production of
  central W bosons also in the  
CM
frame of vector-boson scattering.
\item We require a transverse momentum of the W-boson pair
    $\PT(\PW\PW)\ge 40 (50)\GeV$.  This cut suppresses the
  reducible background coming from the process
  $\Pe^+\Pe^-\to\Pe^+\Pe^-\PW^+\PW^-$, \ie $\ga\ga$ fusion,
when the two produced $\Pe^\pm$ are emitted forward/backward.
\item We require a 
neutrino-pair invariant mass $M
    (\nu_\Pe\bar\nu_\Pe )\ge 150 (200)\GeV$. This cut removes 
  $\PW^+\PW^-\PZ$ production events, in which the neutrinos come from
  the $\PZ$-boson decay (see the last graph in \reffi{fi:graphs}).
\item We require a diboson invariant mass $M(\PW\PW)\ge 400
  (700)\GeV$.  Selecting high diboson CM energies allows one to test a
  possible strongly-interacting regime of the electroweak gauge
  sector.

\end{itemize}

\section{Strategy of the calculation}
\label{sec:strategy}

In this section, we describe the main ingredients of our calculation.
We summarize the adopted approximations and discuss their
domain of applicability.

We consider the process \refeq{eq:process}, with two on-shell
$\PW$~bosons and two neutrinos in the final state. For this process,
exact lowest-order matrix elements are employed in our Monte Carlo,
simultaneously accounting for signal and irreducible background. We
moreover use the complete four-particle phase space and exact
kinematics.

Computing $\Oa$ electroweak corrections in leading-pole approximation,
as in \citeres{beenakker1999,Denner:2000bj,Jadach:1998tz} and
references therein, has revealed successful for analysing $\PW\PW$
physics at LEP2.  A similar philosophy can be adopted for the incoming
bosons in VBS at energies that are large compared to the gauge-boson
masses, since this process is dominated by small invariant masses of
these incoming bosons, which are thus relatively close to their mass
shell, even though these invariant masses are actually negative.

We thus compute the $\Oa$ electroweak corrections to the process 
\refeq{eq:process} in equivalent-vector-boson approximation ($\EVBA$). 
As discussed in the introduction, we do not include real photonic
corrections. For the virtual corrections we work in the logarithmic
approximation and we restrict our calculation to the infrared-finite
part coming from above the electroweak scale. These corrections
correspond to the case where the photon has effectively the mass $\MW$
and are precisely defined in \citere{Denner:2000jv}.  This approach is
sensible since at high energies, the electroweak corrections are
dominated by double and single logarithms of the ratio of the energy
to the electroweak scale.  Hence, keeping only the terms proportional
to $\alpha\log^2(\hat{s}/\MW^2)$ and $\alpha\log(\hat{s}/\MW^2)$,
where $\hat{s}$ is the CM energy of the VBS subprocess, provides the
bulk of the radiative corrections.

\subsection{Equivalent vector-boson approximation}
\label{sec:evba}

In $\EVBA$, the process $\Pep\Pem\to\Pne\Pane\PW^+\PW^-$ is entirely
described by the subset of Feynman diagrams generically represented by
the first graph in \reffi{fi:graphs}. The approximation considers in
fact only those contributions to the final state that come from the
scattering of the two vector bosons emitted by the incoming particles.
The goodness of the approximation thus relies on the assumption that
such contributions are indeed the dominating ones in the considered
kinematical domain. This hypothesis depends on the process at hand,
and can only be checked against an exact computation.  In this
section, we discuss the reliability of the $\EVBA$ in
describing the process $\Pep\Pem\to\Pne\Pane\PW^+\PW^-$ 
and its validity domain.

In implementing the $\EVBA$, we follow the approach of
\citere{kuss1997}.  This method preserves the exact kinematics of the
process, a very useful property for imposing realistic cuts.

We assign the following set of momenta $p_i$ and helicities
$\lambda_i=0,\pm 1$ to the particles involved in the process we are
considering:
\beq
\Pep (p_1,+)\,\Pem (p_2,-)\to\Pne (p_3,-)\,\Pane 
(p_4,+)\,\PW^+(p_5,\lambda_5)\,\PW^-(p_6,\lambda_6).
\label{eq:momenta}
\eeq
In EVBA only left-handed lepton chiralities are relevant,
since the electrons and neutrinos couple always to W bosons.%
\footnote{ Instead, the exact matrix elements that we employ for our
  tree-level predictions receive (background) contributions also from
  right-handed leptons.  } Using the unitary gauge and writing the
propagators of the two incoming $\PW$~bosons, emitted by the initial
$\Pe^\pm$ and exchanged in $t$ channel (see \reffi{fi:graphs}), as a
sum over the vector-boson polarizations
\beq
{1\over{p^2-\MW^2}}\left (-g^{\mu\nu}+{{p^\mu
      p^\nu}\over{\MW^2}}\right ) =
\sum_{\lambda =-1,0,1}{{\epsilon^\mu_\lambda (p)\epsilon^{*\nu}_\lambda
    (p)}\over{p^2-\MW^2}},
\eeq
the exact amplitude corresponding to the first graph in
\reffi{fi:graphs} assumes the form
\beqar\label{eq:amp}
\lefteqn{
\M^{\Pep\Pem\to\Pne\Pane\PW^+\PW^-}(p_1,p_2,p_3,p_4,p_5,p_6;\lambda_5,\lambda_6)=
{1\over{q_+^2-\MW^2}}\,{1\over{q_-^2-\MW^2}}
}\quad&&
\nl&&\times \sum_{\lambda_+,\lambda_- =-1,0,1}
\M^{\Pep\to\Pane\PW^+}(p_1,p_4,q_+;\lambda_+)\;
\M^{\Pem\to\Pne\PW^-}(p_2,p_3,q_-;\lambda_-)
\nl&&\times
\M^{\PW^+\PW^-\to\PW^+\PW^-}(q_+,q_-,p_5,p_6;\lambda_+,\lambda_-,\lambda_5,\lambda_6).
\eeqar

In $\EVBA$ the off-shell amplitude for $\PW\PW$ scattering,
$\M^{\PW^+\PW^-\to\PW^+\PW^-}$, is replaced by a suitably defined
on-shell amplitude. In this way gauge invariance of this amplitude is
ensured, and artifacts from using an incomplete off-shell amplitude
are avoided. Modifications of the on-shell amplitude are necessary in
order to describe the dependence of the off-shell amplitude on the
off-shell masses $q_\pm^2$ to a satisfactory accuracy. We here follow
the approach of \citere{kuss1997} which describes the extrapolation to
off-shell masses by simple proportionality factors for each incoming
vector boson. These are chosen to be equal to 1 for transverse bosons.
For each longitudinal $\PWp$ or $\PWm$ boson, the on-shell amplitude
is multiplied by a factor $\MW/\sqrt{-q_{\pm}^2}$ in order to describe
the singular behaviour $\epsilon^\mu_{\lambda =0}(q_\pm)\sim
1/\sqrt{-q_\pm^2}$ of the off-shell longitudinal polarization vectors.
One can thus write
\beqar\label{eq:ampevba}
\lefteqn{
\M^{\Pep\Pem\to\Pne\Pane\PW^+\PW^-}_{\EVBA}(p_1,p_2,p_3,p_4,p_5,p_6;\lambda_5,\lambda_6)=
{1\over{q_+^2-\MW^2}}\,{1\over{q_-^2-\MW^2}}
}\quad
\nl&&
\times\sum_{\lambda_+,\lambda_- =-1,0,1}
\M^{\Pep\to\Pane\PW^+}(p_1,p_4,q_+;\lambda_+)\;
\M^{\Pem\to\Pne\PW^-}(p_2,p_3,q_-;\lambda_-) 
\nl&&{}\times
\M^{\PW^+\PW^-\to\PW^+\PW^-}(q_+^{\mathrm{on}},q_-^{\mathrm{on}},p_5,p_6;\lambda_+,\lambda_-,\lambda_5,\lambda_6)
\nl&&{}\times
\left[{\MW\over{\sqrt{-q_+^2}}}\,\delta_{\lambda_+,0}+\delta_{\lambda_+,\pm}\right ]
\left[{\MW\over{\sqrt{-q_-^2}}}\,\delta_{\lambda_-,0}+\delta_{\lambda_-,\pm}\right ]
,
\eeqar
where $q_{\pm}^{\mathrm{on}}$ are the on-shell projected momenta of
the two incoming $\PW$~bosons.  The definition of the on-shell
projection is given in \refapp{app:projection}.  Note that the W-boson
momenta in the matrix elements $\M^{\Pep\to\Pane\PW^+}$ and
$\M^{\Pem\to\Pne\PW^-}$ are not projected on shell.

In order to proceed, in \citere{kuss1997} the squared amplitude was
considered and the contributions of the matrix elements
$\M^{\Pep\to\Pane\PW^+}$ and $\M^{\Pem\to\Pne\PW^-}$ were transformed
into vector-boson luminosities. Instead, we work at the matrix element
level and compute the three amplitudes on the right-hand side of
\refeq{eq:ampevba} with the help of {\tt PHACT}
\cite{Ballestrero:1999md}, a routine based on the helicity-amplitude
method of \citere{HAmethod}.

\begin{table}\centering
$$\arraycolsep 5pt
\begin{array}{|c|c|c|c|}
\hline
\multicolumn{4}{|c|}{\sigma_{\Born} (\Pep\Pem\to\Pne\Pane\PW^+\PW^-)}\\
\hline
~\sqrt{s}~[\TeV]~ & 
~~\sigma_{\exact}~[\mathrm{fb}]~~ & 
~~\sigma_{\EVBA}~[\mathrm{fb}]~~ & 
~~\Delta_{\EVBA}~~[\%]~~\\
\hline
\hline
1 & 0.595 & 0.479 & 19.5 \\
\hline
3 & 3.507 & 3.471 & 1.0 \\
\hline
\end{array}$$
\caption {Exact lowest-order cross section (second column) as well as total 
cross section in $\EVBA$ (third column) and their difference in per cent of 
the exact result (fourth column). Kinematical cuts as in \refse{sec:setup} are 
applied.}
\label{ta:sigma_evba}
\end{table}
In \refta{ta:sigma_evba}, we show the comparison between the $\EVBA$
and the exact lowest-order result for the total cross section.  We
select the kinematical domain where we expect new-physics effects
related to strongly interacting vector bosons to be enhanced.  Such a
region, characterized by high diboson invariant masses and large
scattering angles of the two produced $\PW$~bosons, is selected via
appropriate cuts described in detail in \refse{sec:setup}.  At
$\sqrt{s}=1\TeV$ and $\sqrt{s}=3\TeV$ the accuracy of the EVBA for the
total cross section,
$\Delta_{\EVBA}=(\sigma_{\exact}-\sigma_{\EVBA})/\sigma_{\exact}$,
amounts to about 20\% and 1\%, respectively.

\begin{figure}
  \unitlength 1cm
  \begin{center}
  \begin{picture}(16.,15.)
  \put(-2.5,-1){\epsfig{file=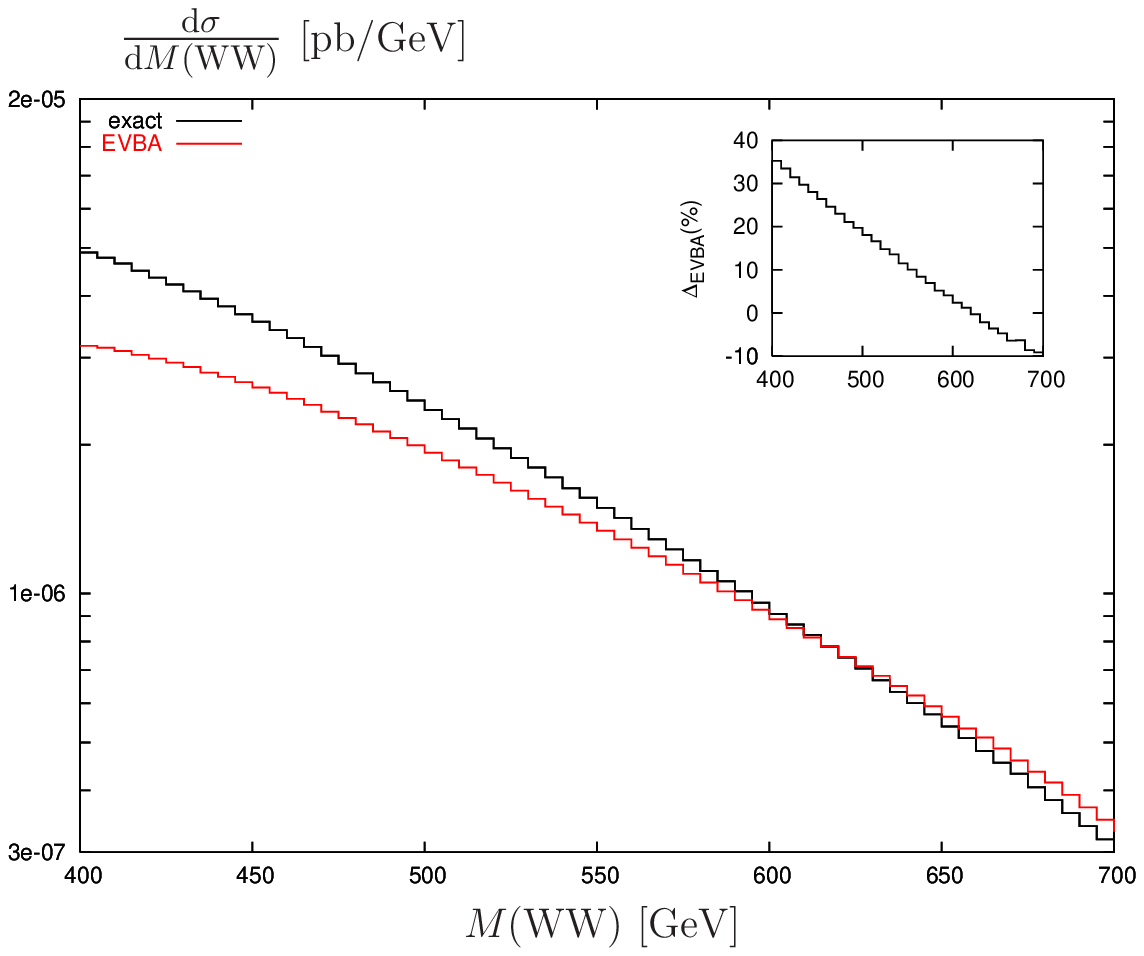,width=14cm}}
  \put(5.5,-1){\epsfig{file=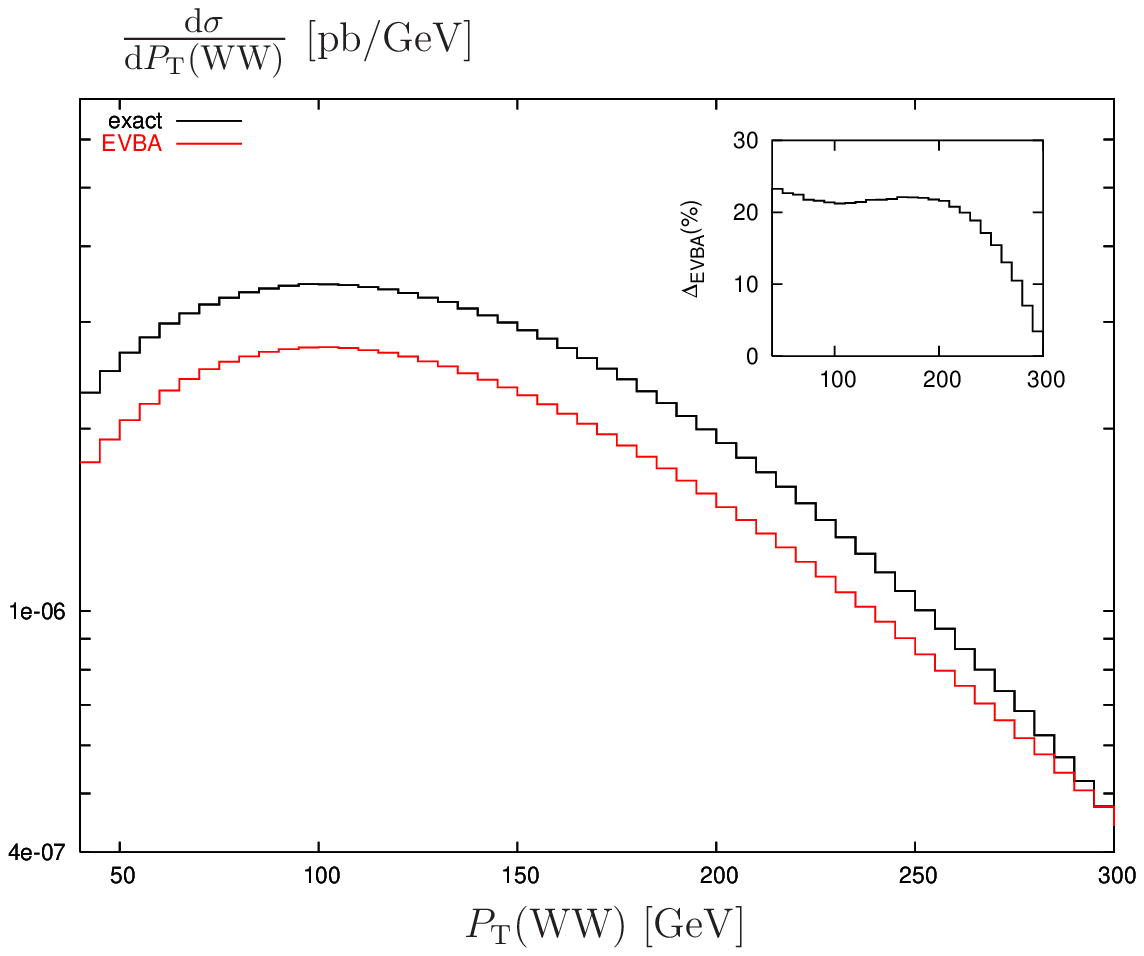,width=14cm}}
  \put(-2.5,-8){\epsfig{file=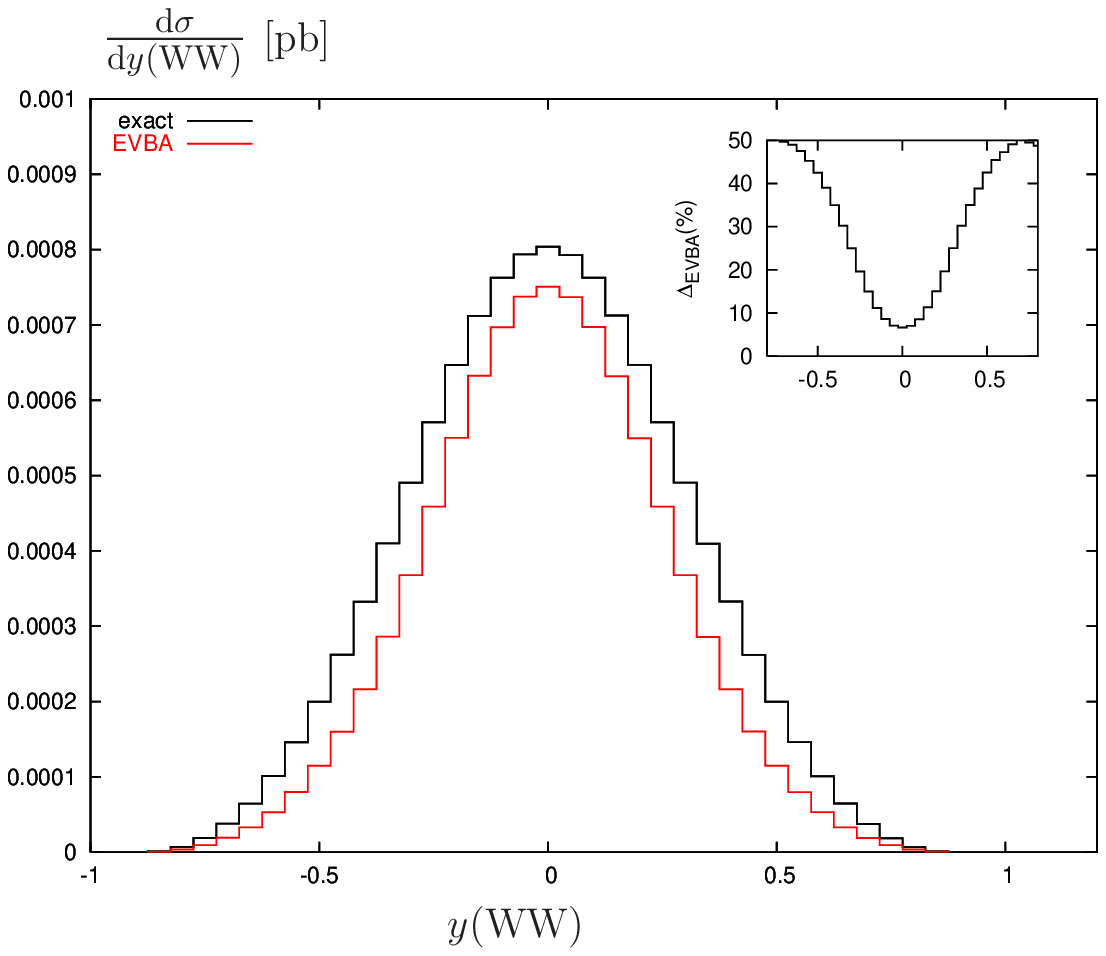,width=14cm}}
  \put(5.5,-8){\epsfig{file=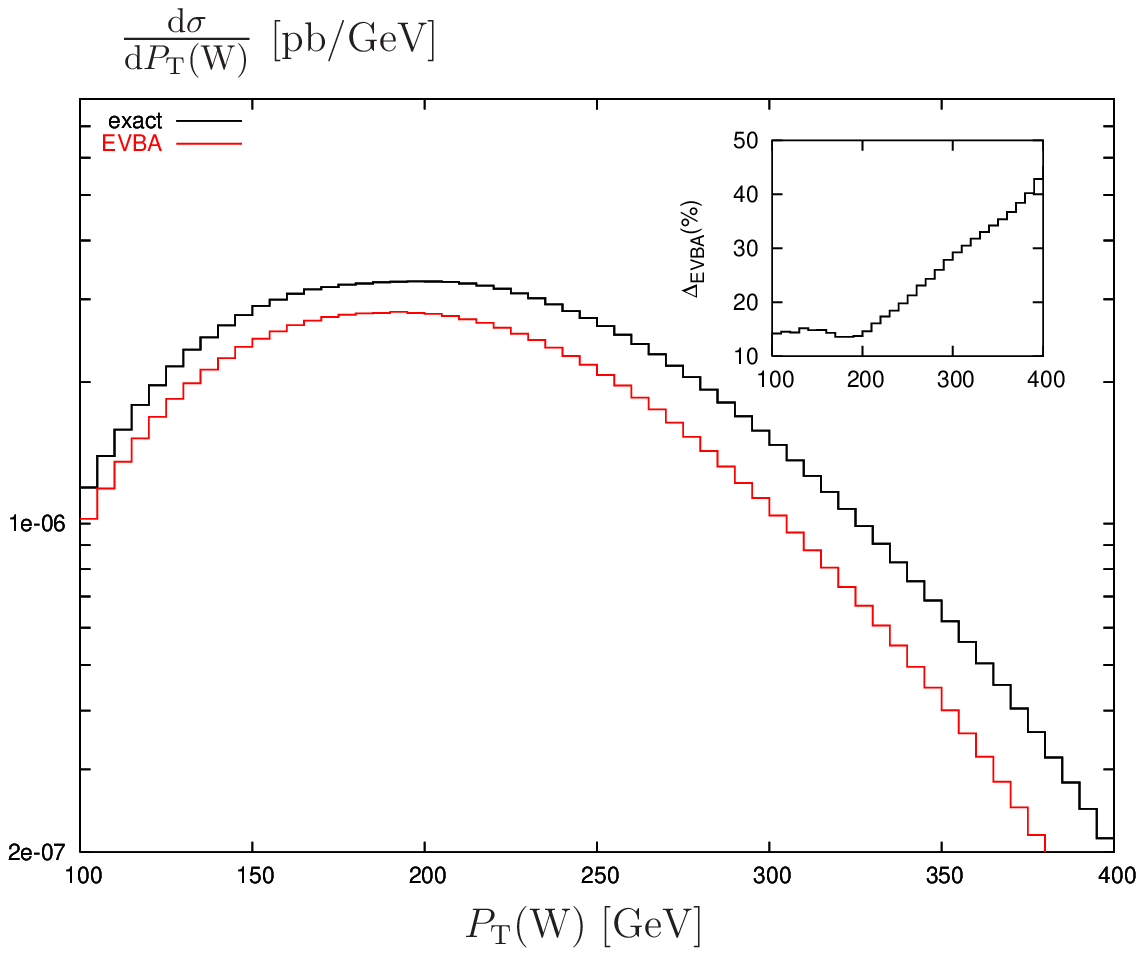,width=14cm}}
  \end{picture}
  \end{center}
\vspace{-2.cm}
\caption{Lowest-order distributions for $\sqrt{s}=1\TeV$ from the
  exact matrix elements and in $\EVBA$: invariant mass of the diboson
  pair (upper left), transverse momentum of the diboson pair (upper
  right), rapidity of the diboson pair (lower left), transverse
  momentum of the produced $\PW$~boson (lower right).  The inset plots
  show the relative difference $\Delta_{\EVBA}$ in per cent.
  Standard cuts are applied. 
}
\label{fi:evba_ilc}
\end{figure}

\begin{figure}
  \unitlength 1cm
  \begin{center}
  \begin{picture}(16.,15.)
  \put(-2.5,-1){\epsfig{file=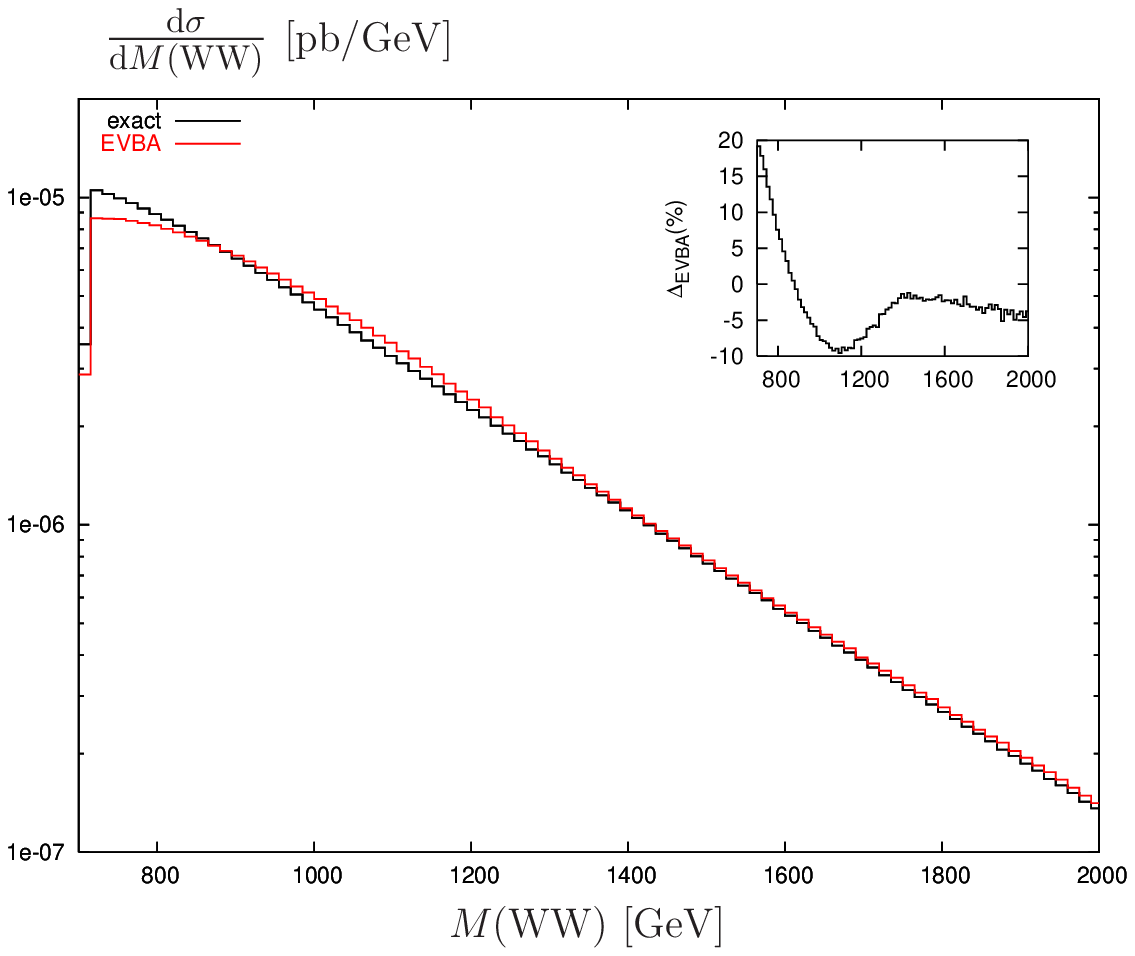,width=14cm}}
  \put(5.5,-1){\epsfig{file=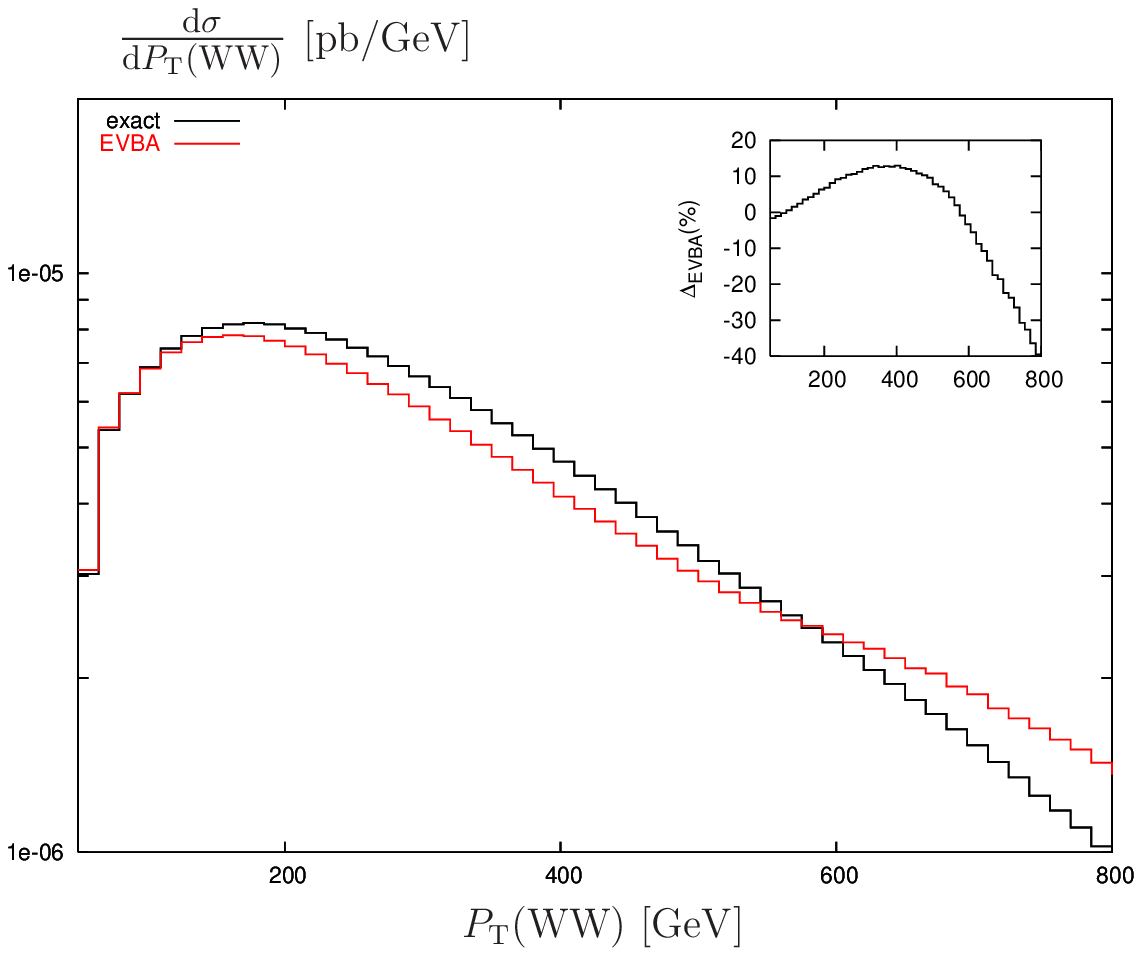,width=14cm}}
  \put(-2.5,-8){\epsfig{file=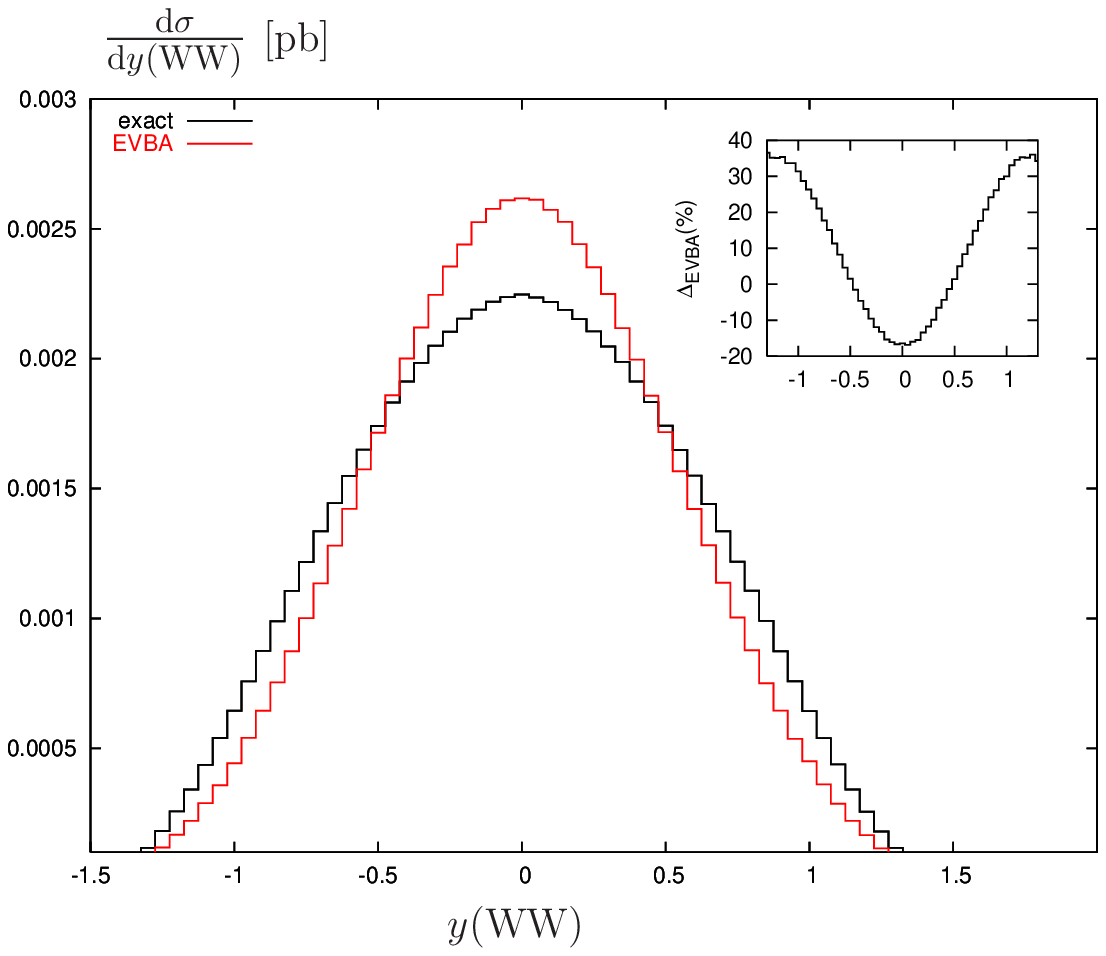,width=14cm}}
  \put(5.5,-8){\epsfig{file=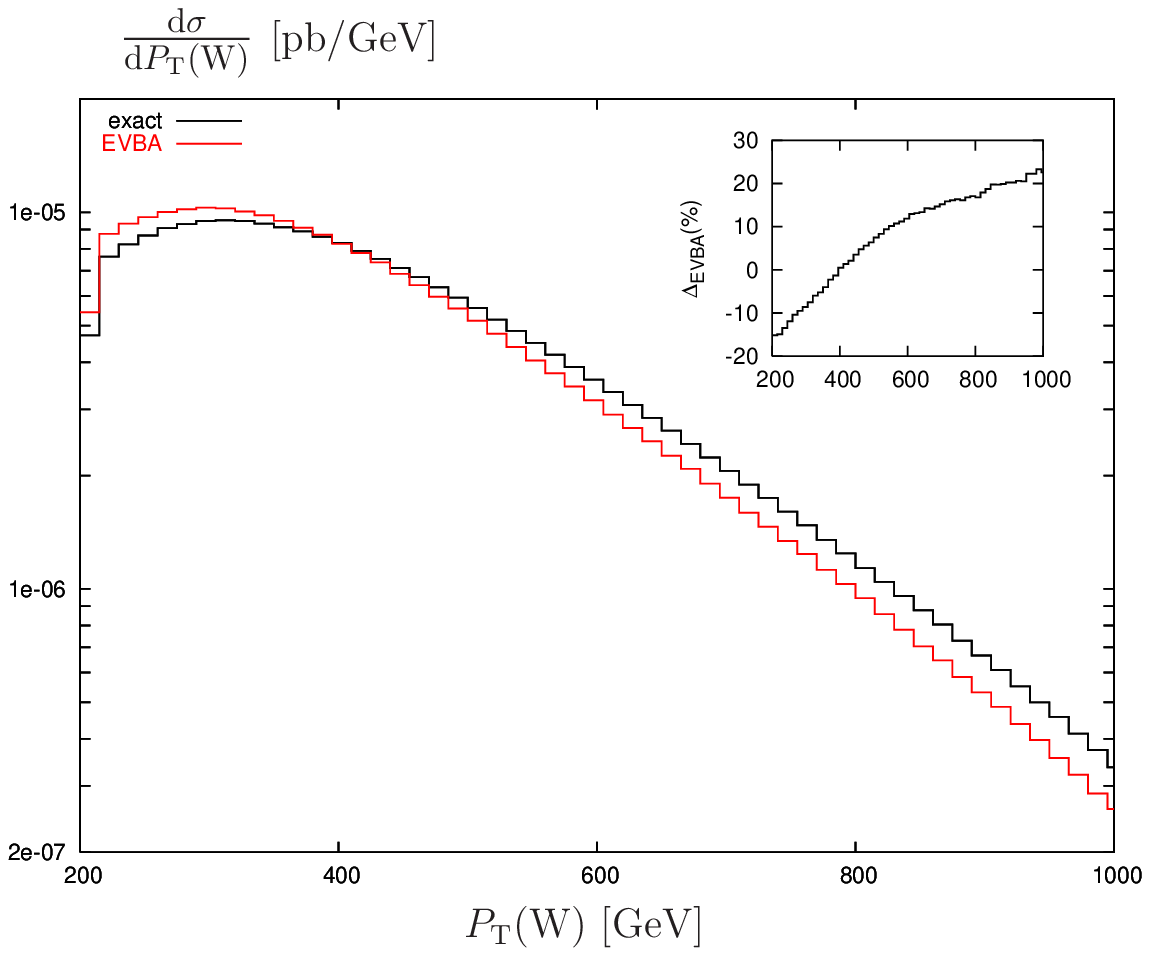,width=14cm}}
  \end{picture}
  \end{center}
\vspace{-2.cm}
\caption{
  Lowest-order distributions for $\sqrt{s}=3\TeV$ from the exact
  matrix elements and in $\EVBA$.  Same conventions as in
  \reffi{fi:evba_ilc}.}
\label{fi:evba_clic}
\end{figure}

The integrated cross section gives 
only a partial information on the goodness of
the $\EVBA$. In order to display more extensively the reliability of this 
approximation in the selected kinematical domain, in \reffis{fi:evba_ilc} and 
\ref{fi:evba_clic} we analyse distributions in both energy-like and 
angular-like variables.
In particular, we consider four observables of interest:
\begin{itemize}
\item {diboson invariant mass $M(\PW\PW)$},
\item {diboson transverse momentum $\PT (\PW\PW)$,} 
\item {diboson rapidity
    $y(\PW\PW)=0.5\ln\left[\frac{E(\PW\PW)+P_{\rL}(\PW\PW)}{E(\PW\PW)-P_{\rL}(\PW\PW)}\right]$,}
\item $\PW$-boson transverse momentum $\PT (\PW)=\PT (\PWp)$.
\end{itemize}
The distributions in $\PT (\PWm)$ and $\PT (\PWp)$ are identical.
We plot the
lowest-order results of the {EVBA} and of the {exact} calculation for
the two collider energies $\sqrt{s}=1\TeV$ and $\sqrt{s}=3\TeV$ in
\reffi{fi:evba_ilc} and \reffi{fi:evba_clic}, respectively.
The inset plots show the difference in per cent between the two
results, \ie $\Delta_{\EVBA}
=(\Pd\sigma_{\exact}-\Pd\sigma_{\EVBA})/\Pd\sigma_{\exact}$. 
For ILC and CLIC, in most of the regions that are not statistically
irrelevant, the difference is below $25\%$ and $20\%$, respectively.
The increase for small $M(\PW\PW)$ is not problematic, since the
radiative corrections are small in this region (see
\refse{sec:results}).
As a remark, let us add that the agreement between $\EVBA$ and exact
result reached in this analysis should not be taken as the best one
can do.  The quality of the approximation is sensibly cut-dependent.
In this paper, following \citere{boos-1997} we have adopted generic
cuts and, in principle, the $\EVBA$ could be improved imposing more
stringent constraints.  The choice of the set-up depends, however, on
the particular analysis to be performed.

In the following, we use the $\EVBA$ only for computing the $\Oa$
electroweak corrections. The lowest-order results are calculated
exactly.  The $\Oa$ inaccuracy associated with the EVBA can be
estimated to be of the order of the product of the $\Oa$ corrections
(computed in EVBA) times the inaccuracy of the EVBA at tree level.
Since the radiative effects typically amount to 10--30\%, the
inaccuracy of the $\EVBA$ translates into a few-percent error for the
full result.  As shown in \refse{sec:results}, apart for the case of
very high $\PT (\PW\PW)$, the $\Oa$ uncertainty associated with the
$\EVBA$ is always smaller than the expected statistical error at the
ILC and CLIC.

\subsection{Virtual electroweak corrections} 
\label{sec:ewcor}

In analogy with the leading-pole approximation for virtual
corrections, the $\EVBA$ allows two types of radiative contributions,
factorizable and non-factorizable ones. The former are those that can
be associated to either the emission of one of the two incoming
$\PW$~bosons from the beam particles or the VBS subprocess. The latter
are those connecting these subprocesses.

The non-factorizable corrections in leading-pole approximation consist
only of photonic contributions. These corrections have been evaluated
for $\PW$-boson pair production in $\Pep\Pem$ annihilation in
\citeres{nonfac1,nonfac2}. There it was found that all infrared and
mass-singular logarithms cancel between virtual and real corrections
and that the remaining effects are small.  In $\EVBA$
non-factorizable corrections do not only result from photon exchange
between the subprocesses of $\PW$-boson-production and $\PW\PW$
scattering but also from analogous exchanges of massive gauge bosons.
These contributions deserve further investigations. In this paper, we
do not consider the non-factorizable corrections but restrict
ourselves to the calculation of the factorizable contributions.

\begin{figure}
  \unitlength 1cm
  \begin{center}
  \begin{picture}(16.,15.)
  \put(-.5,-3){\epsfig{file=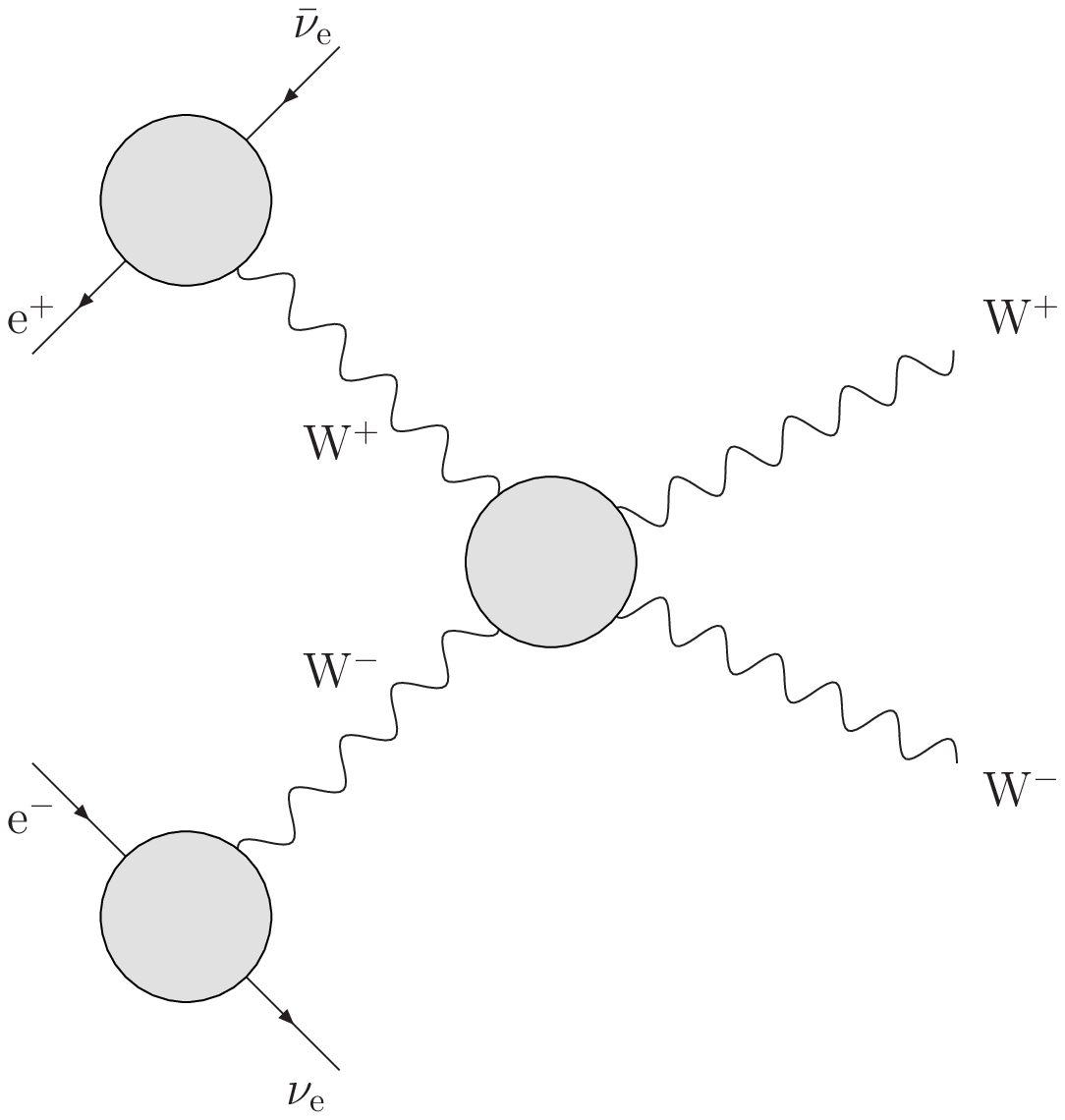,width=14cm}}
  \end{picture}
  \end{center}
\vspace{-6.8cm}
\caption{Structure of the virtual factorizable corrections in $\EVBA$ with
one-loop contributions in the blobs.}
\label{fi:evba_graph}
\end{figure}

The virtual factorizable corrections are represented by the schematic
diagram of \reffi{fi:evba_graph}, in which the big blobs contain all
one-loop corrections to the incoming $\PW$-boson-production and
on-shell $\PW\PW$-scattering subprocesses.
The corresponding matrix element can be written as
\beqar\label{eq:corr}
\de\M_{\virt,\EVBA,\fact}^{\Pep\Pem\to\Pne\Pane\PW_{\lambda_5}^+\PW_{\lambda_6}^-} &=&
\frac{1}{q_+^2-\MW^2}\;\frac{1}{q_-^2-\MW^2}\nl
&&\times
\sum_{\lambda_+,\lambda_-}\biggl\{
\de\M_{\virt}^{\Pep\to\Pane\PW_{\lambda_+}^+}
\M_{\Born}^{\Pem\to\Pne\PW_{\lambda_-}^-}
\M_{\Born,\mathrm{on}}^{\PW_{\lambda_+}^+\PW_{\lambda_-}^-\to\PW_{\lambda_5}^+\PW_{\lambda_6}^- }\nl
&&{}+\M_{\Born}^{\Pep\to\Pane\PW_{\lambda_+}^+}
\de\M_{\virt}^{\Pem\to\Pne\PW_{\lambda_-}^-}
\M_{\Born,\mathrm{on}}^{\PW_{\lambda_+}^+\PW_{\lambda_-}^-\to\PW_{\lambda_5}^+\PW_{\lambda_6}^- }\nl
&&{}+\M_{\Born}^{\Pep\to\Pane\PW_{\lambda_+}^+}
\M_{\Born}^{\Pem\to\Pne\PW_{\lambda_-}^-}
\de\M_{\virt,\mathrm{on}}^{\PW_{\lambda_+}^+\PW_{\lambda_-}^-\to\PW_{\lambda_5}^+\PW_{\lambda_6}^- }\biggr\}
\nl&&{}\times
\left[{\MW\over{\sqrt{-q_+^2}}}\,\delta_{\lambda_+,0}+\delta_{\lambda_+,\pm}\right ]
\left[{\MW\over{\sqrt{-q_-^2}}}\,\delta_{\lambda_-,0}+\delta_{\lambda_-,\pm}\right]
,
\eeqar
where $\delta\M_{\virt}^{\Pep\to\Pane\PW_{\lambda_+}^+}$,
$\delta\M_\virt^{\Pem\to\Pne\PW_{\lambda_-}^-}$, and
$\delta\M_{\virt,\mathrm{on}}^{\PW_{\lambda_+}^+\PW_{\lambda_-}^-\to\PW_{\lambda_5}^+\PW_{\lambda_6}^-
}$ denote the virtual corrections to the matrix elements for the
$\PW$-boson-production and on-shell $\PW\PW$-scattering subprocesses.
The index `on' indicates that on-shell vector-boson momenta are used
to calculate these matrix elements.

We calculate the factorizable $\Oa$ virtual corrections in logarithmic
high-energy approximation including single and double enhanced
logarithms, \ie contributions proportional to
$\alpha\log^2(\hat{s}/\MW^2)$ and $\alpha\log(\hat{s}/\MW^2)$, where
$\hat{s}$ is the $\CM$ energy of the scattering subprocess. The
logarithmic approximation yields the dominant corrections as long as
$\CM$ energies and scattering angles are large.  Pure
angular-dependent logarithms of the form
$\alpha\log^2(\hat{s}/\hat{r})$ and $\alpha\log(\hat{s}/\hat{r})$,
with $\hat{r}$ equal to the Mandelstam variables $\hat{t}$ and
$\hat{u}$ of the $\PW\PW$-scattering subprocess, are not included.
However, angular-dependent terms of the form
$\alpha\log(\hat{s}/\hat{r})\log(\hat{s}/\MW^2)$ are taken into
account.  The validity of the results relies therefore on the
assumption that all invariants are large compared with $\MW^2$ and
approximately of the same size
\beq\label{HEA}
\hat{s}\sim |\hat{t}|\sim |\hat{u}|\gg \MW^2.
\eeq
This implies that the produced gauge bosons should be energetic and
emitted at sufficiently large angles with respect to the beam. This is
precisely the kinematical region where effects due to a possible
strongly interacting regime of the gauge sector are maximally
enhanced. In this region, the accuracy of the logarithmic high-energy
approximation is expected to be of the order of a few per cent. We can
thus reasonably adopt this approximation at $\Pe^+\Pe^-$ colliders
with energy in the 1--$3\TeV$ range, where the experimental error is
at the few-per-cent level.  Since the emission subprocesses of the two
incoming $\PW$~bosons involve no large energy variable (they peak at
$|q^2_\pm |\sim \MW^2$), the corresponding virtual corrections vanish
in the logarithmic approximation. As a consequence, we do not consider
the first two contributions on the right-hand side of \refeq{eq:corr}
in the following.  Moreover, for the $\PW^+\PW^-\to\PW^+\PW^-$
subprocess we take into account only the corrections to the dominating
channels involving four transverse ($\PPT\PPT\PPT\PPT$) or two
transverse and two longitudinal ($\PPL\PPL\PPT\PPT$,
$\PPL\PPT\PPL\PPT$, $\PPT\PPL\PPT\PPL$) gauge bosons.  The
contributions of the channels with an odd number of longitudinally
polarized $\PW$~bosons are suppressed by $\MW/\sqrt{\hat{s}}$, and
those of the channels LTTL and TLLT by $\MW^2/\hat{s}$.  Moreover, the
configurations with two final-state longitudinal \PW~bosons are
numerically small within the SM with a light Higgs boson.  As shown in
\refta{ta:sigmapol}, for $\MH=120\GeV$, the cross section for the
production of two longitudinal \PW~bosons is suppressed by a factor 50
or 100 compared to the full result for $\sqrt{s}=1\TeV$ or $3\TeV$,
respectively.
\begin{table}\centering
$$\arraycolsep 5pt
\begin{array}{|c|c|c|c|c|c|}
\hline
\multicolumn{6}{|c|}{\sigma_{\Born}(\Pep\Pem\to\Pne\Pane\PW^+_\lambda\PW^-_{\lambda^\prime})}\\
\hline
~\sqrt{s}~[\TeV]~ & 
~~\sigma^{\PPT\PPT}~[\mathrm{fb}]~~ & 
~~\sigma^{\PPT\PPL}~[\mathrm{fb}]~~ &
~~\sigma^{\PPL\PPT}~[\mathrm{fb}]~~ &  
~~\sigma^{\PPL\PPL}~[\mathrm{fb}]~~ &  
~~\sigma^{\tot}~[\mathrm{fb}]~~ \\
\hline
\hline
1 & 0.500 & 0.0410 &  0.0410 &  0.0134 & 0.595\\
\hline
3 & 3.111 & 0.1786 & 0.1786  & 0.0390 & 3.507\\
\hline
\end{array}$$
\caption {Born cross section for the process
$\Pep\Pem\to\Pne\Pane\PW^+_\lambda\PW^-_{\lambda^\prime}$ and
various polarizations ($\lambda ,\lambda^\prime = \PPT ,\PPL$) of the produced 
$\PW$~bosons. Kinematical cuts as specified in 
\refse{sec:setup} are applied.}
\label{ta:sigmapol}
\end{table}

The analytical expressions for the $\Oa$ virtual corrections to
$\PW\PW$ scattering in the high-energy limit are given in
\refapp{app:corr}.  The formulas are rather compact and easy to
implement.  In our default set-up, we have used the version with exact
SU(2)-transformed lowest-order matrix elements.  This means that we
use the complete expression as in \refeq{chargedsscgeneral}.  More
precisely, we use \refeq{SCTTLLexact}, \refeq{SCTLTLexact}, and
\refeq{SCTTTTexact}.  We have verified that the numerical results
based on \refeq{SCTTLLhe}, \refeq{SCTLTLhe}, and \refeq{SCTTTThe},
which are obtained by making use of the high-energy approximation for
the SU(2)-transformed Born amplitudes given in \refapp{se:Born}, are
in very good agreement. For all results shown in the following, the
difference between the two methods is in fact at the per-mille level.
This comparison confirms the reliability of the high-energy
approximation for the Born matrix elements, under which the correction
factor can be factorized and expressed in a very simple form, leading
to considerable decrease in $\PCPU$ time.
 
\begin{figure}
  \unitlength 1cm
  \begin{center}
  \begin{picture}(16.,15.)
  \put(-1.5,-9){\epsfig{file=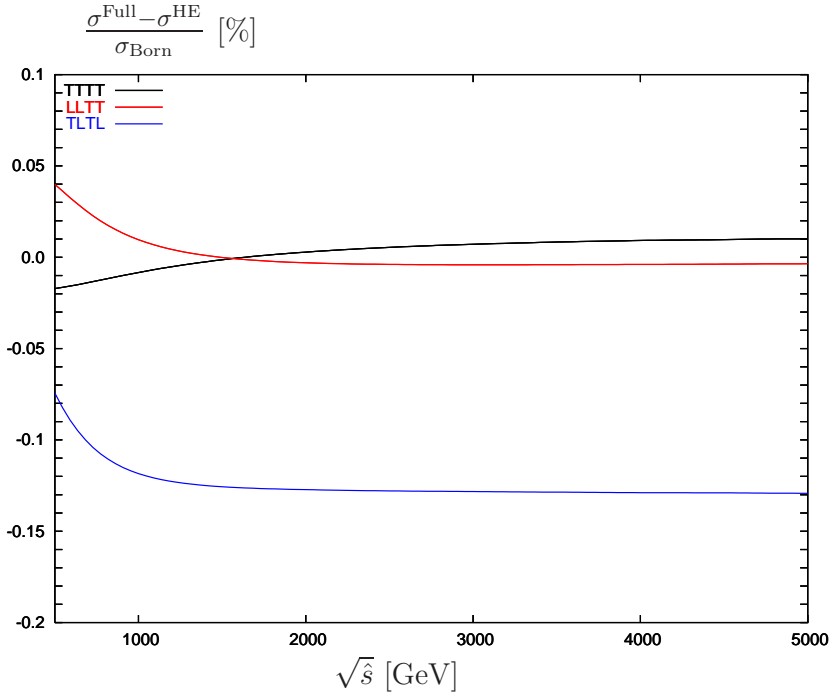,width=20cm}}
  \end{picture}
  \end{center}
\vspace{-5.2cm}
\caption{$\PW\PW$-scattering subprocess $\PW^+_{\lambda_+}\PW^-_{\lambda_-}
  \rightarrow\PW^+_{\lambda_5}\PW^-_{\lambda_6}$. Difference between
  complete $\Oa$ cross section ($\si^{\mathrm{Full}}$) and $\Oa$ cross
  section in logarithmic high-energy approximation
  ($\si^{\mathrm{HE}}$), normalized to the lowest-order result. In the
  legend, $\PPT$ and $\PPL$ refer to the transverse and longitudinal
  polarizations of the two incoming and two outgoing $\PW$~bosons from
  left to right.  }
\label{fi:hahn}
\end{figure}

In order to estimate the accuracy of the logarithmic high-energy
approximation in computing the radiative effects, we have moreover
compared our results with the complete $\Oa$ corrections to the
on-shell
$\PW^+_{\lambda_+}\PW^-_{\lambda_-}\rightarrow\PW^+_{\lambda_5}\PW^-_{\lambda_6}$
process \cite{hahn1997}. For this comparison, we have included the
electromagnetic terms given in \refapp{se:emlogs}, evaluated for a
soft-photon cutoff $\Delta E=0.5\sqrt{\hat s}$.
In \reffi{fi:hahn}, we show the difference between the complete $\Oa$ cross 
section and the $\Oa$ result in logarithmic high-energy approximation, 
normalized to the Born cross section, as a function of the $\CM$ energy.
The three curves represent the three polarized cross-section
differences for $\lambda_+\lambda_-\lambda_5\lambda_6=
\PPT\PPT\PPT\PPT,\PPL\PPL\PPT\PPT,\PPT\PPL\PPT\PPL$, respectively.
They all display a plateau at high energy, showing that the difference
between approximate and exact result is due to terms that do not grow
with energy in that region.  This implies that the high-energy
behaviour of the radiative effects is well reproduced by the adopted
approximation which takes into account only energy-enhanced
logarithmic terms.

The terms, omitted in the logarithmic approximation might be, however,
not negligible. Figure \ref{fi:hahn} shows indeed that the agreement
between exact and approximate $\Oa$ cross section is within a few per
cent for the first two polarization configurations, while it goes up
to about 13$\%$ for the third one. This discrepancy is unexpectedly
large. It could be understood by performing a complete high-energy
approximation including non-logarithmic terms. However, for the
unpolarized cross section this discrepancy is still tolerable.  The
lowest-order cross section is in fact dominated by the
$\PPT\PPT\PPT\PPT$ configuration, and receives only a 10$\%$
contribution from $\PPT\PPL\PPT\PPL$, as shown in \refta{ta:sigmapol}.
We can thus safely assume our $\Oa$
inaccuracy to be of the order of a
few per cent.

\section{Numerical results}
\label{sec:results}

In this section, we illustrate the effect of the logarithmic
electroweak corrections on the production of a $\PW^+\PW^-$ pair plus
missing energy at future $\Pe^+\Pe^-$ colliders.
We consider the process \refeq{eq:process} in the numerical setup given
in \refse{sec:setup}. We analyse the behaviour of the VBS included in 
\refeq{eq:process} over the kinematical region characterized by large diboson
invariant masses and large scattering angles of the produced \PW~bosons.
This is the domain where possible new-physics effects entering the VBS would 
be maximally enhanced.

We focus on the SM predictions for a light Higgs boson.  As pointed
out by Bagger et al.\ \cite{sewsb-analyses1}, in this case one would
observe the production of mostly transversally polarized $\PW$~bosons
in the high $\PW\PW$ invariant-mass region. The longitudinal spin
configuration is in fact strongly suppressed in the presence of a
light Higgs boson.  For the specific case at hand, this is shown in
\refta{ta:sigmapol} for $\MH=120\GeV$ and two possible setups.

\begin{table}\centering
$$\arraycolsep 5pt
\begin{array}{|c|c|c|c|c|c|}
\hline
\multicolumn{6}{|c|}{\sigma (\Pep\Pem\to\Pne\Pane\PW^+\PW^-)}\\
\hline
~\sqrt{s}~[\TeV]~ & 
~~\sigma_{\Born}~[\mathrm{fb}]~~ & 
~~\sigma ~[\mathrm{fb}]~~ &
~~\Delta_{\EW} ~[\%]~~ & 
~~\Delta ~[\%]~~ & ~1/\sqrt{L\sigma_{\Born}}~[\%]~\\
\hline
\hline
1 & 0.595 & 0.556 & ~{-6.7} & 1.3 & 4.1   \\
\hline
3 & 3.507 & 2.897 & -17.4 & 0.2 & 1.7  \\
\hline
\end{array}$$
\caption {Total lowest-order cross section (second column) as well as total 
$\Oa$ cross section (third column) and electroweak corrections in per cent of 
the lowest-order result (fourth column), including their uncertainty (fifth
column). The last entry shows the statistical 
error for an integrated luminosity $L=1\aba^{-1}$. 
Kinematical cuts as in \refse{sec:setup} are applied.}
\label{ta:sigmaoa}
\end{table}

In \refta{ta:sigmaoa}, we show the impact of the corrections on the
total cross section for $\Pep\Pem\to\Pne\Pane\PW^+\PW^-$ and compare
it with the expected statistical error. The second column contains the
lowest-order result. The third and fourth entries respectively display
the $\Oa$-corrected cross section and the contribution of the one-loop
corrections relative to the Born result, $\Delta_{\EW}
=(\sigma-\sigma_{\Born})/\sigma_{\Born}$. 
The fifth column provides an estimate of the uncertainty of the
one-loop contributions due to the EVBA, $\Delta
=\Delta_{\EW}\times\Delta_{\EVBA}$, which is obtained combining the
one-loop corrections with the uncertainty of the EVBA at Born level.
For the considered process, the electroweak radiative effects are
negative and of the order of $-5\%$ to $-20\%$. This has to be
compared with the last column of \refta{ta:sigmaoa}, where we show an
estimate of the statistical error based on an integrated luminosity
$L=1\aba^{-1}$.  As can be seen, the electroweak corrections are quite
important.  Already comparable with the statistical uncertainty at the
ILC, they further increase at higher energies giving rise to a
$10\sigma$~effect at CLIC.
 
\begin{figure}
  \unitlength 1cm
  \begin{center}
  \begin{picture}(16.,15.)
  \put(-2.5,-1){\epsfig{file=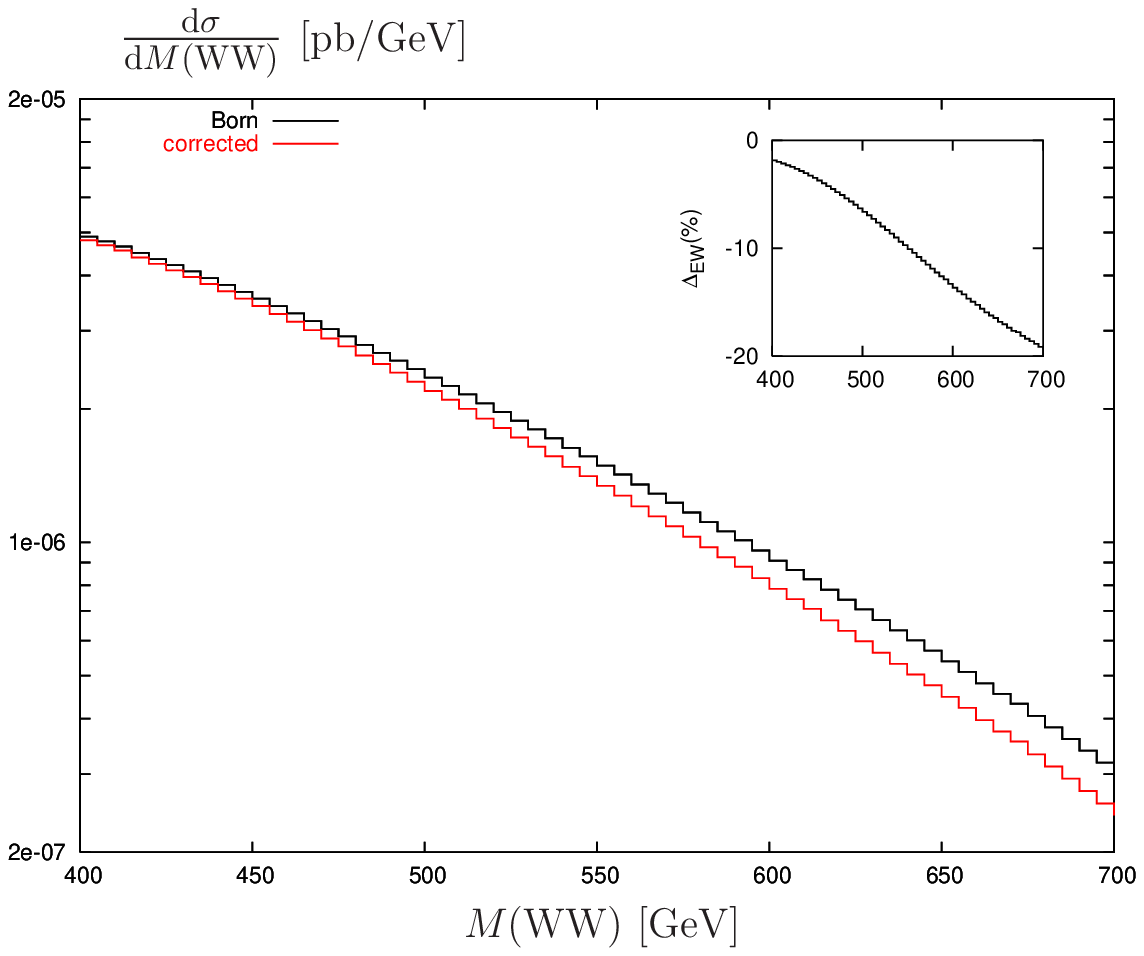,width=14cm}}
  \put(5.5,-1){\epsfig{file=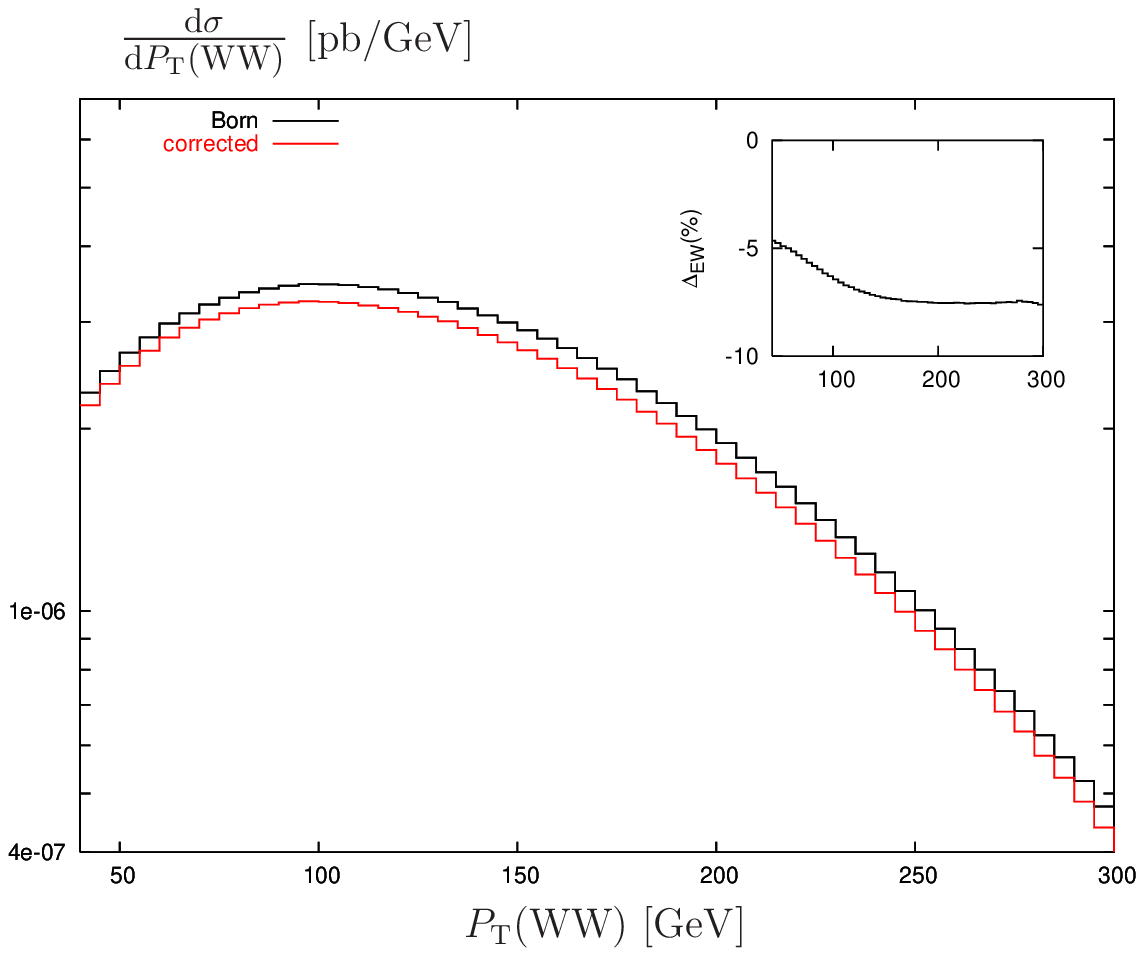,width=14cm}}
  \put(-2.5,-8){\epsfig{file=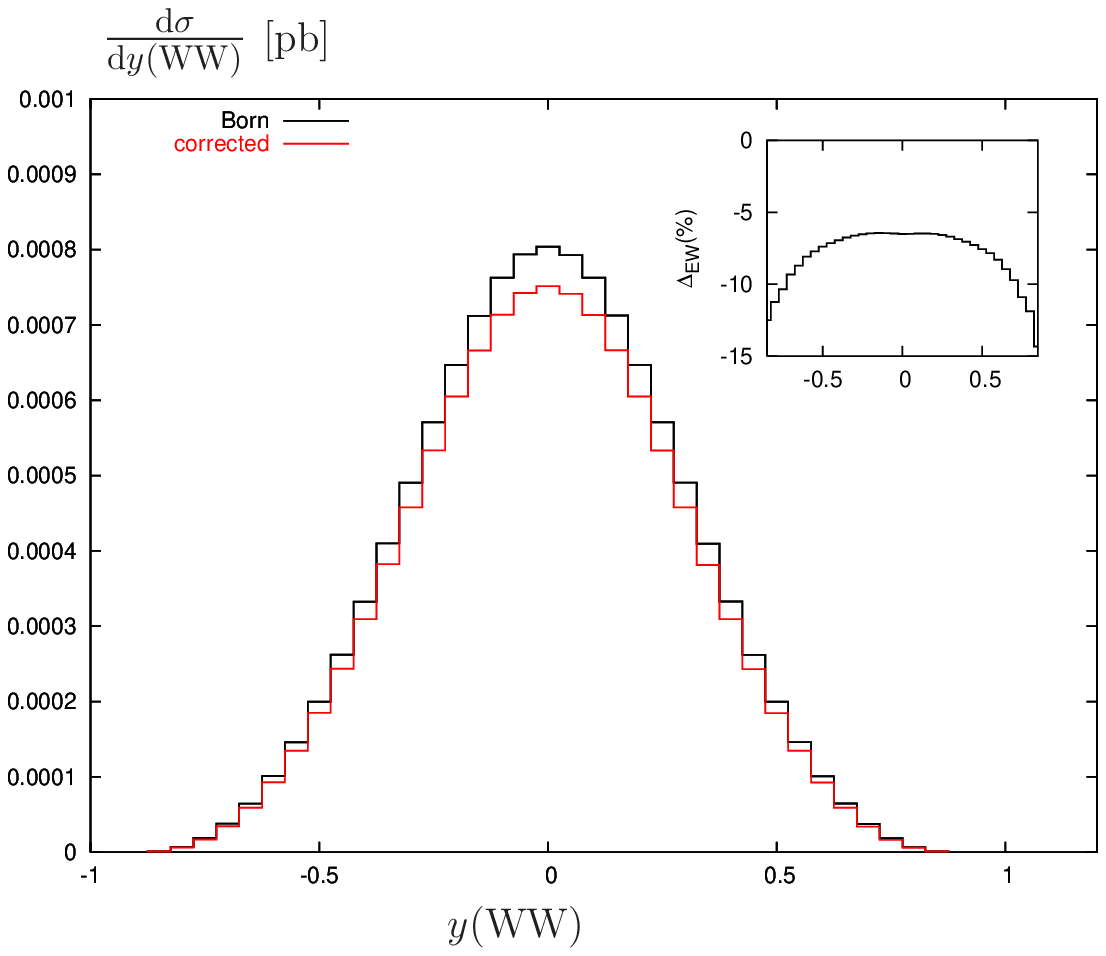,width=14cm}}
  \put(5.5,-8){\epsfig{file=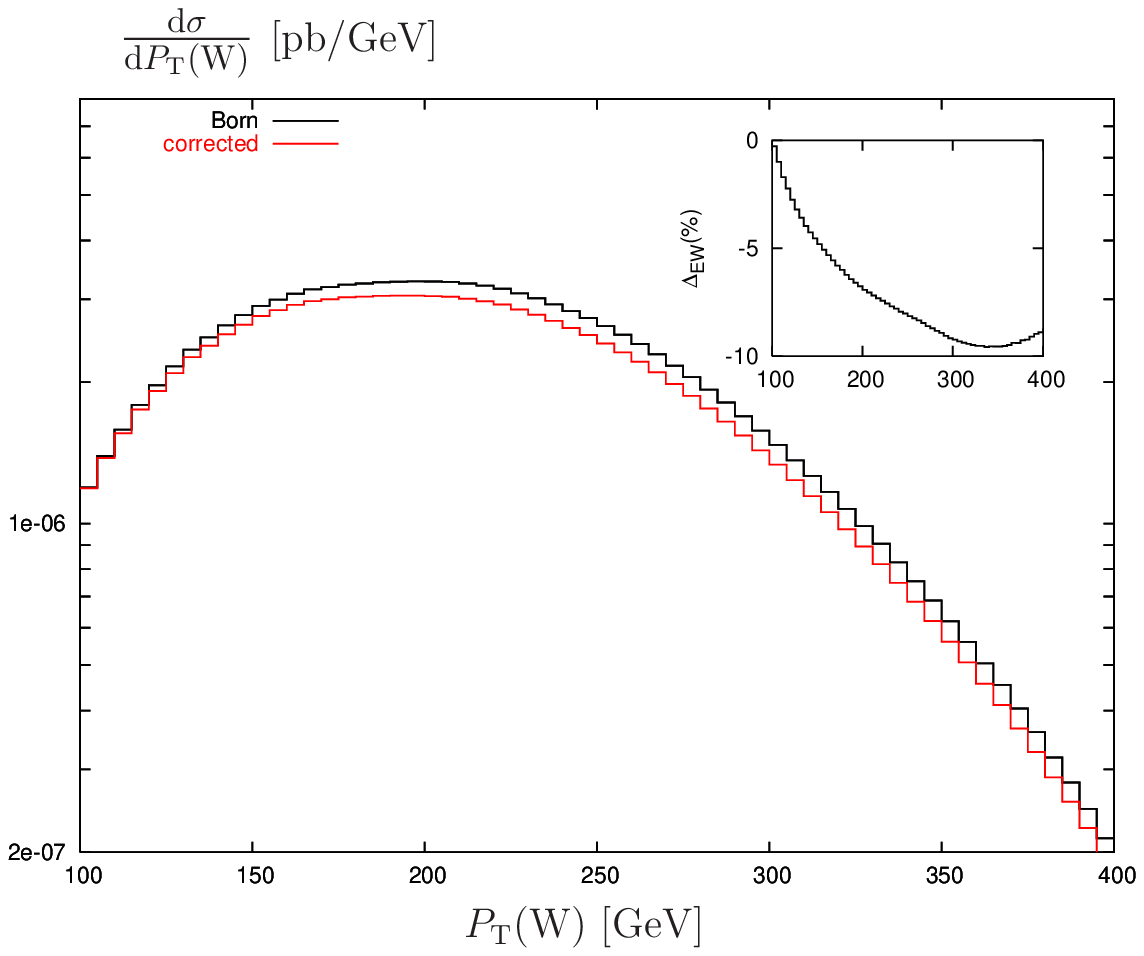,width=14cm}}
  \end{picture}
  \end{center}
\vspace{-2.cm}
\caption{Distributions for $\sqrt{s}=1\TeV$ in lowest order and
  including logarithmic corrections: invariant mass of the diboson
  pair (upper left), transverse momentum of the diboson pair (upper
  right), rapidity of the diboson pair (lower left), transverse
  momentum of the produced $\PW$~boson (lower right).  The inset plots
  show the relative $\Oa$ corrections $\Delta_{\EW}$ in per cent
  normalized to the lowest-order results.  Standard cuts are applied.
}
\label{fi:ewcor_ilc}
\end{figure}
\begin{figure}
  \unitlength 1cm
  \begin{center}
  \begin{picture}(16.,15.)
  \put(-2.5,-1){\epsfig{file=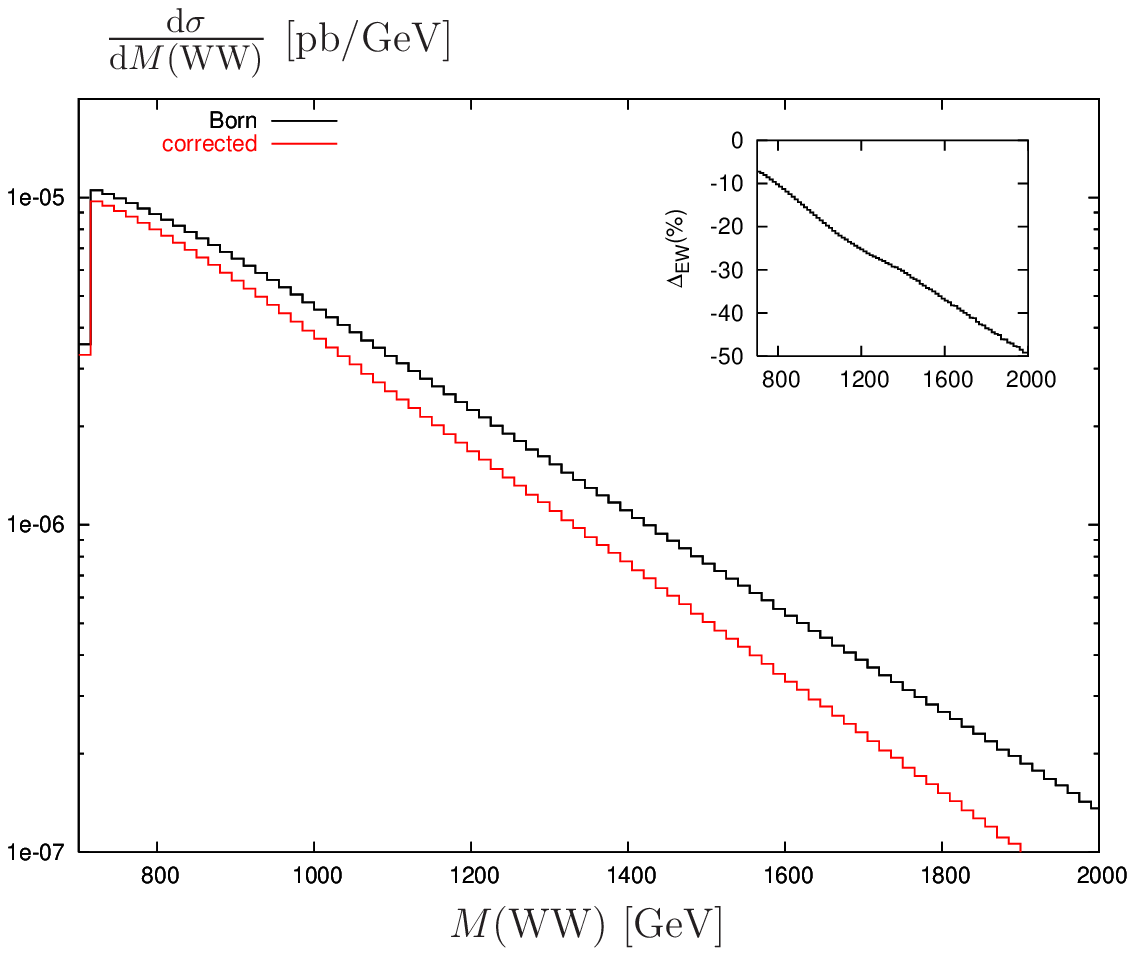,width=14cm}}
  \put(5.5,-1){\epsfig{file=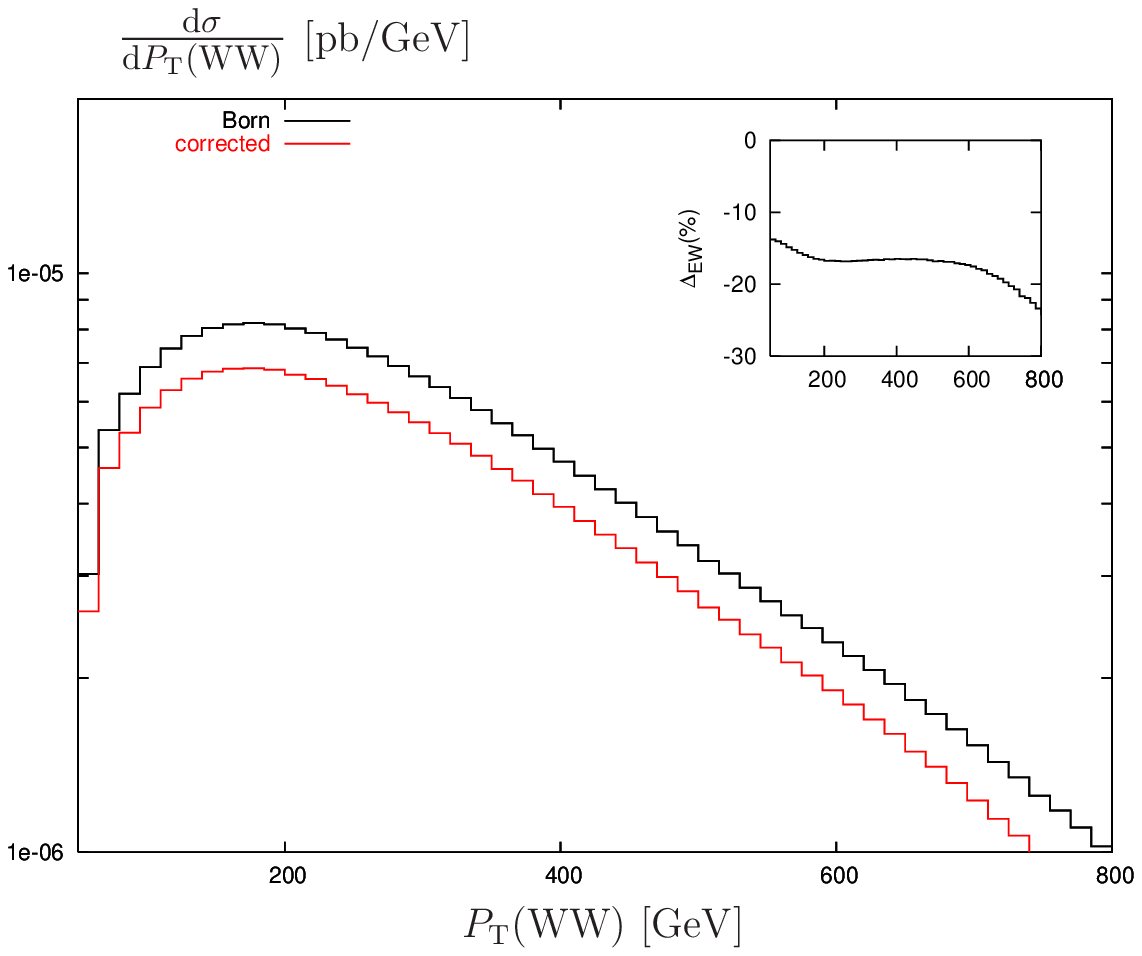,width=14cm}}
  \put(-2.5,-8){\epsfig{file=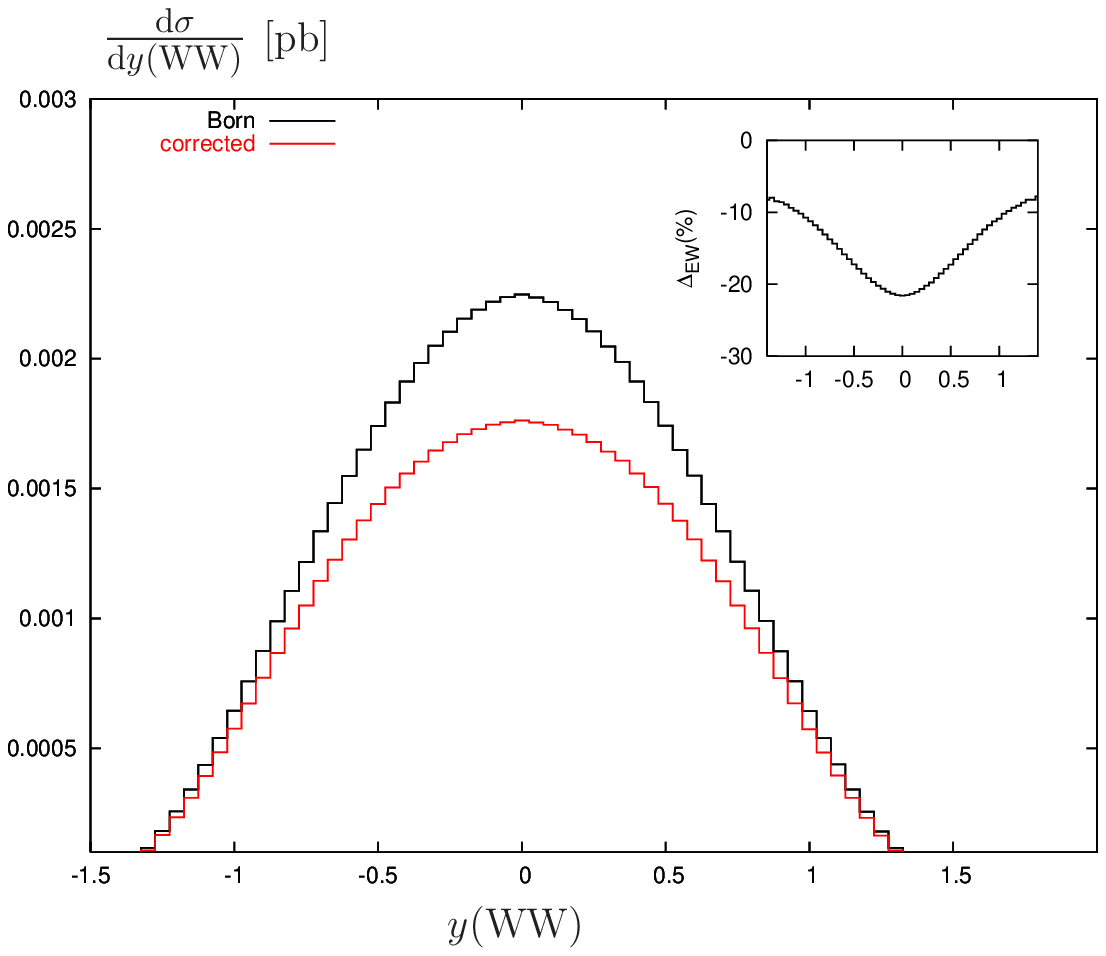,width=14cm}}
  \put(5.5,-8){\epsfig{file=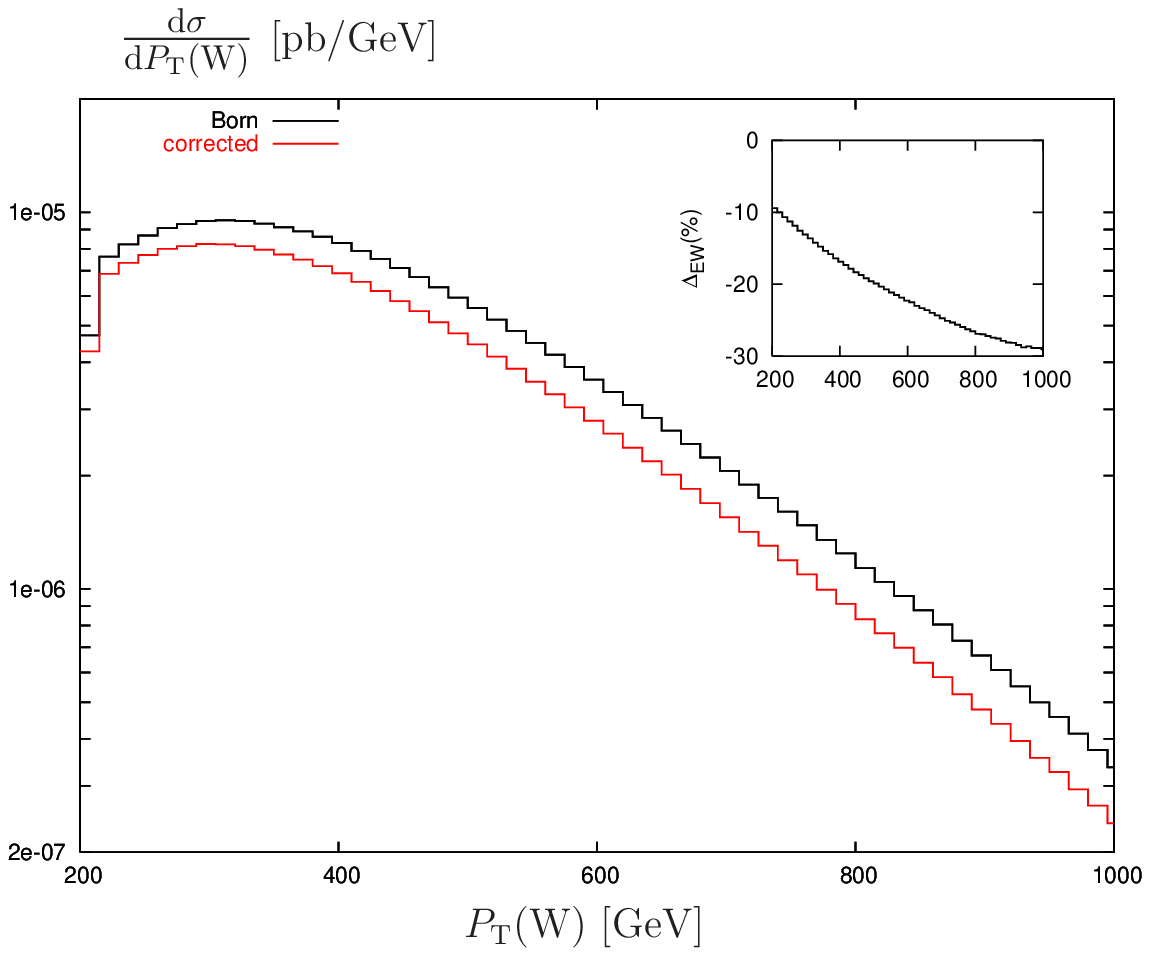,width=14cm}}
  \end{picture}
  \end{center}
\vspace{-2.cm}
\caption{Distributions for $3\TeV$ in lowest order and
  including logarithmic corrections. Same conventions as in
  \reffi{fi:ewcor_ilc}.}
\label{fi:ewcor_clic}
\end{figure}

The influence of the $\Oa$ corrections is highly dependent on the cuts
imposed and the selected kinematical domain. Also, distributions can
be differently affected by the radiative corrections.  We illustrate
this point for the sample variables defined and analysed in
\refse{sec:evba}. The $\Oa$ effects on these four observables at
$\sqrt{s} =1\TeV$ and $3\TeV$ are displayed in \reffi{fi:ewcor_ilc}
and \reffi{fi:ewcor_clic}, respectively.  The upper curves represent
the lowest-order differential cross section, the lower ones
the corresponding corrected result.%

We first consider the case of $\sqrt{s} =1\TeV$.  In the left-upper
plot of \reffi{fi:ewcor_ilc}, we show the distribution in the diboson
invariant mass. This variable, representing the $\CM$ energy of the
VBS subprocess, gives direct access to the energy scale at which new
physics could appear. In absence of a light Higgs boson, the SM VBS
amplitudes would violate perturbative unitarity at high $\CM$
energies.  In order to recover it, new physics should manifest itself
at those scales.  Hence, at future colliders it will be useful to
analyse the diboson production (plus missing energy) at the highest
possible $M(\PW\PW )$ values.  In this region, the $\Oa$ corrections
are enhanced. They can go up to $-18\%$, as shown by the corresponding
inset plot. The increase of the radiative effects with $M(\PW\PW )$ is
a typical effect of large logarithms of Sudakov type. The electroweak
corrections should therefore be included to match the experimental
accuracy.

A second variable of interest, $\PT (\PW\PW)$, combines energy and
angle information. This observable is expected to be sensitive to a
strongly interacting VBS signal at low and intermediate values, say
for $\PT(\PW\PW)$ between 50 and $300\GeV$ \cite{boos-1997}. In this
region, the radiative effects are of order $-5\%$ to $-7\%$, thus
smaller than in the case discussed above. Their size is, however,
comparable with the statistical accuracy in the considered range. The
same conclusion holds for the pure angular-like variable, $y
(\PW\PW)$, shown in the left-lower plot, which gets corrections in
the range between $-7\%$ and $-10\%$. The transverse momentum $\PT
(\PW)$ on the right-lower plot displays instead a behaviour analogous
to $M(\PW\PW )$.  Note that for all considered distributions the
corrections reduce the tree-level SM predictions.

\begin{table}[t]
\begin{center}
\begin{tabular}{||c| |c| c| c| c| |c| |c| c| c| c||}
\hline
\hline
$\mathrm{bin}$ & \multicolumn{4}{|c||}{$M(\PW\PW )$} & $\mathrm{bin}$ & 
\multicolumn{4}{|c||}{$\PT (\PW\PW)$}\\
\hline
\hline
{\footnotesize [GeV]}& {\footnotesize $N_{\mathrm{evt}}$} & {\footnotesize $\Delta_{\mathrm{stat}}[\%]$} & {\footnotesize $\Delta_{\mathrm{EW}} [\%]$} & 
{\footnotesize $\Delta [\%]$} & {\footnotesize [GeV]} & 
{\footnotesize $N_{\mathrm{evt}}$} & {\footnotesize $\Delta_{\mathrm{stat}}[\%]$} & {\footnotesize $\Delta_{\mathrm{EW}} [\%]$} & 
{\footnotesize $\Delta [\%]$}\\
\hline
{\footnotesize 700--900} & {\footnotesize 1719} & {\footnotesize 2.4} & {\footnotesize -10.4} & {\footnotesize 0.7} &
{\footnotesize 50--250} & {\footnotesize 1467} & {\footnotesize 2.6} & {\footnotesize -15.9} & {\footnotesize 0.7}\\
\hline
{\footnotesize 900--1100} & {\footnotesize 916} & {\footnotesize 3.3} & {\footnotesize -18.2} & {\footnotesize 1.2} &
{\footnotesize 250--450} & {\footnotesize 1119} & {\footnotesize 3.0} & {\footnotesize -16.6} & {\footnotesize 2.0}\\
\hline
{\footnotesize 1100--1300} & {\footnotesize 446} & {\footnotesize 4.7} & {\footnotesize -25.1} & {\footnotesize 1.8} &
{\footnotesize 450--650} & {\footnotesize 550} & {\footnotesize 4.3} & {\footnotesize -17.0} & {\footnotesize 0.5}\\
\hline
{\footnotesize 1300--1500} & {\footnotesize 216} & {\footnotesize 6.8} & {\footnotesize -30.4} & {\footnotesize 0.7} &
{\footnotesize 650--850} & {\footnotesize 249} & {\footnotesize 6.3} & {\footnotesize -21.4} & {\footnotesize 6.5}\\
\hline
{\footnotesize 1500--1700} & {\footnotesize 106} & {\footnotesize 9.7} & {\footnotesize -36.8} & {\footnotesize 0.8} &
{\footnotesize 850--1050} & {\footnotesize 92} & {\footnotesize 10.4} & {\footnotesize -32.6} & {\footnotesize 24.9}\\
\hline
{\footnotesize 1700--1900} & {\footnotesize 53} & {\footnotesize 13.7} & {\footnotesize -43.3} & {\footnotesize 1.5} &
{\footnotesize           } & {\footnotesize   } & {\footnotesize     } & {\footnotesize      } & {\footnotesize     }\\
\hline
{\footnotesize 1900--2100} & {\footnotesize 26} & {\footnotesize 19.4} & {\footnotesize -49.5} & {\footnotesize 2.2} &
{\footnotesize           } & {\footnotesize  } & {\footnotesize  } & {\footnotesize  } & {\footnotesize  }\\
\hline
\hline
$\mathrm{bin}$ & \multicolumn{4}{|c||}{$\PT(\PW )$} & $\mathrm{bin}$ & 
\multicolumn{4}{|c||}{$y(\PW\PW)$}\\
\hline
\hline
{\footnotesize [GeV ]}& {\footnotesize $N_{\mathrm{evt}}$} & {\footnotesize $\!\Delta_{\mathrm{stat}}[\%]\!$} & {\footnotesize $\Delta_{\mathrm{EW}} [\%]$} & {\footnotesize $\Delta [\%]$} &   & {\footnotesize $N_{\mathrm{evt}}$} & 
{\footnotesize $\!\Delta_{\mathrm{stat}}[\%]\!$} & {\footnotesize $\Delta_{\mathrm{EW}} [\%]$} & {\footnotesize $\Delta [\%]$}\\
\hline
{\footnotesize 200--400} & {\footnotesize 1765} & {\footnotesize 2.4} & {\footnotesize -13.4} & {\footnotesize 0.9} &
{\footnotesize 0--0.25} & {\footnotesize 549} & {\footnotesize 4.3} & {\footnotesize -21.1} & {\footnotesize 3.2}\\
\hline
{\footnotesize 400--600} & {\footnotesize 1140} & {\footnotesize 3.0} & {\footnotesize -19.4} & {\footnotesize 1.2} &
{\footnotesize 0.25--0.50} & {\footnotesize 486} & {\footnotesize 4.5} & {\footnotesize -18.9} & {\footnotesize 1.1}\\
\hline
{\footnotesize 600--800} & {\footnotesize 424} & {\footnotesize 4.9} & {\footnotesize -24.3} & {\footnotesize 3.6} &
{\footnotesize 0.50--0.75} & {\footnotesize 374} & {\footnotesize 5.2} & {\footnotesize -15.6} & {\footnotesize 1.4}\\
\hline
{\footnotesize 800--1000} & {\footnotesize 134} & {\footnotesize 8.6} & {\footnotesize -27.8} & {\footnotesize 5.5} &
{\footnotesize 0.75--1.00} & {\footnotesize 235} & {\footnotesize 6.5} & {\footnotesize -12.4} & {\footnotesize 3.0}\\
\hline
{\footnotesize 1000--1200} & {\footnotesize 35} & {\footnotesize 16.8} & {\footnotesize -28.6} & {\footnotesize 8.1} &
{\footnotesize 1.00--1.25} & {\footnotesize 101} & {\footnotesize 9.9} & {\footnotesize -9.9} & {\footnotesize 3.3}\\
\hline
\end{tabular}
\caption{
Impact of electroweak corrections on  binned distributions at
  $\sqrt{s}=3\TeV$.  The first and sixth columns show the bin. For each
  variable, the four entries from left to right give the total number
  of events for a luminosity   $L =1\aba^{-1}$, the corresponding statistical accuracy, 
the size of the $\Oa$ electroweak corrections relative 
  to the Born result, and the estimated one-loop uncertainty due to the EVBA.}
\label{ta:dis3}
\end{center}
\end{table}
In \reffi{fi:ewcor_clic}, we show the same set of distributions as
above for $\sqrt{s} =3\TeV$ and the corresponding cuts specified in
\refse{sec:setup}.  As expected the impact of the radiative
corrections increases with the collider energy since this allows for
higher $\CM$ energies, which translate into higher diboson invariant
masses and transverse momenta.  This behaviour is well depicted in all
four plots.  For the distributions in $\PT (\PW\PW)$ and $y(\PW\PW )$
the corrections are in the range between $-10\%$ and $-20\%$, apart
from the region of very high $\PT (\PW\PW)$ where statistics is small.
For the $M(\PW\PW )$ distribution the electroweak corrections grow
from $-10\%$ to $-50\%$ with increasing invariant mass and for the
$\PT (\PW)$ distribution the effect is similar.  For the distribution
in $y(\PW\PW )$, the corrections are large at the maximum of the
distribution at small $y(\PW\PW )$.  For the other distributions, the
cross sections get small where the corrections get large. Still, the
$\Oa$ corrections are statistically relevant.
This is illustrated in \refta{ta:dis3}, where the size of the
electroweak corrections $\Delta_{\mathrm{EW}}$ and the $\Oa$
uncertainty $\Delta =\Delta_{\EW}\times\Delta_{\EVBA}$, which results
from the EVBA, are compared with the statistical accuracy
$\Delta_{\mathrm{stat}}$ based on the projected luminosity $L
=1\aba^{-1}$.
\begin{table}[t]
\begin{center}
\begin{tabular}{||c| |c| c| c| c| |c| |c| c| c| c||}
\hline
\hline
$\mathrm{bin}$ & \multicolumn{4}{|c||}{$M(\PW\PW )$} & $\mathrm{bin}$ & 
\multicolumn{4}{|c||}{$\PT (\PW\PW)$}\\
\hline
\hline
{\footnotesize [GeV]}& {\footnotesize $N_{\mathrm{evt}}$} & {\footnotesize $\Delta_{\mathrm{stat}}[\%]$} & {\footnotesize $\Delta_{\mathrm{EW}} [\%]$} & 
{\footnotesize $\Delta [\%]$} & {\footnotesize [GeV]}& {\footnotesize $N_{\mathrm{evt}}$} & {\footnotesize $\Delta_{\mathrm{stat}}[\%]$} & {\footnotesize $\Delta_{\mathrm{EW}} [\%]$} & {\footnotesize $\Delta [\%]$}\\
\hline
{\footnotesize 400--500} & {\footnotesize 363} & {\footnotesize 5.2} & {\footnotesize -3.5} & {\footnotesize 1.0} &
{\footnotesize 40--140} & {\footnotesize 316} & {\footnotesize 5.6} & {\footnotesize -6.1} & {\footnotesize 1.3}\\
\hline
{\footnotesize 500--600} & {\footnotesize 157} & {\footnotesize 8.0} & {\footnotesize -9.4} & {\footnotesize 1.1} &
{\footnotesize 140--240} & {\footnotesize 215} & {\footnotesize 6.8} & {\footnotesize -7.4} & {\footnotesize 1.6}\\
\hline
{\footnotesize 600--700} & {\footnotesize 57} & {\footnotesize 13.2} & {\footnotesize -16.0} & {\footnotesize 0.5} &
{\footnotesize 240--340} & {\footnotesize 59} & {\footnotesize 13.0} & {\footnotesize -7.6} & {\footnotesize 0.5}\\
\hline
{\footnotesize 700--800} & {\footnotesize 16} & {\footnotesize 24.8} & {\footnotesize -19.9} & {\footnotesize 2.0} &
{\footnotesize         } & {\footnotesize  } & {\footnotesize  } & {\footnotesize  } & {\footnotesize  }\\
\hline
\hline
$\mathrm{bin}$ & \multicolumn{4}{|c||}{$\PT(\PW )$} & $\mathrm{bin}$ & 
\multicolumn{4}{|c||}{$y(\PW\PW)$}\\
\hline
\hline
{\footnotesize [GeV ]}& {\footnotesize $N_{\mathrm{evt}}$} & {\footnotesize $\!\Delta_{\mathrm{stat}}[\%]\!$} & {\footnotesize  
$\Delta_{\mathrm{EW}} [\%]$} & {\footnotesize $\Delta [\%]$} &   & {\footnotesize $N_{\mathrm{evt}}$} 
& {\footnotesize $\!\Delta_{\mathrm{stat}}[\%]\!$} & {\footnotesize $\Delta_{\mathrm{EW}} [\%]$} &  {\footnotesize $\Delta [\%]$}\\
\hline
{\footnotesize 100--200} & {\footnotesize 259} & {\footnotesize 6.2} & {\footnotesize -4.8} & {\footnotesize 0.7} &
{\footnotesize 0--0.25} & {\footnotesize 180} & {\footnotesize 7.4} & {\footnotesize -6.5} & {\footnotesize 0.7}\\
\hline
{\footnotesize 200--300} & {\footnotesize 259} & {\footnotesize 6.2} & {\footnotesize -7.9} & {\footnotesize 1.6} &
{\footnotesize 0.25--0.50} & {\footnotesize 94} & {\footnotesize 10.3} & {\footnotesize -6.9} & {\footnotesize 2.1}\\
\hline
{\footnotesize 300--400} & {\footnotesize 72} & {\footnotesize 11.8} & {\footnotesize -9.4} & {\footnotesize 3.1} &
{\footnotesize 0.50--0.75} & {\footnotesize 23} & {\footnotesize 20.8} & {\footnotesize -8.3} & {\footnotesize 3.8}\\
\hline
\end{tabular}
\caption{Impact of electroweak corrections on  binned distributions at
  $\sqrt{s}=1\TeV$. Same conventions as in \refta{ta:dis3}.}
\label{ta:dis1}
\end{center}
\end{table}
Dividing the range of energy-like variables in 200 GeV intervals, we
find that the influence of the $\Oa$ corrections is not washed out by
the binning.  In the more statistically relevant bins, the radiative
corrections ranging from $-10\%$ to $-30\%$ give rise to a
3--6$\sigma$ effect. In the tails of the distributions, the $\Oa$
contributions can increase up to $-50\%$, still being bigger than the
estimated experimental accuracy.  We also observe that, apart from the
region of very high $\PT (\PW\PW)$, the uncertainty associated with
the EVBA never exceeds the statistical error.

In \refta{ta:dis1} we give analogous results for $\sqrt{s} =1\TeV$. In
this case, the radiative effects are comparable with the statistical
accuracy. Their significance is strictly dependent on the binning.

In our analysis, we have not included any option on the polarization
of the initial beams. In the process
$\Pe^+\Pe^-\rightarrow\Pne\Pane\PW^+\PW^-$, the VBS signal is purely
given by the $\PW$-boson scattering, as illustrated in the first
Feynman diagram of \reffi{fi:graphs}. Only the right--left combination
$\Pe^+_\rR\Pe^-_\rL$ thus contributes to the signal.  The irreducible
background (see for instance the last diagram in \reffi{fi:graphs})
can receive instead contributions also from other helicity
configurations.  The possibility of selecting a given initial
polarization has thus two advantages. In first place, it helps in
suppressing the background.  As a second benefit, it increases the
statistics.  Assuming a polarization efficiency of 80$\%$ and 60$\%$
for electron and positron, respectively, we would have in fact an
increase of about a factor two in the number of events at fixed
luminosity. In this set-up, the relevance of the radiative effects
would be further enhanced.

\section{Conclusions}
\label{sec:conclusions}

If the Higgs boson should be heavy or absent, the scattering of
longitudinal electroweak gauge bosons can provide information on the
mechanism of electroweak symmetry breaking. An irreducible background
to signals of new physics in vector-boson scattering is provided by
the Standard Model contribution for a light Higgs boson. In order to
be able to disentangle possible small new-physics effects a precise
knowledge of the latter is required.

We have studied electroweak radiative corrections to
$\PWp\PWm$~scattering at high-energy $\Pep\Pem$ colliders.  We have
used the equivalent vector-boson approximation and calculated the
factorizable one-loop corrections in the high-energy logarithmic
approximation. Corrections to the splitting of the W~bosons from the
incoming particles as well as non-factorizable corrections have not
been taken into account.  We have presented explicit analytical
results for the logarithmic corrections to $\PWp\PWm\to\PWp\PWm$.
These have been implemented into a Monte Carlo program for the process
$\Pe^+\Pe^-\rightarrow\Pne\Pane\PW^+\PW^-$ together with the complete
lowest-order matrix elements.

We have defined a set of cuts suitable for the analysis of the
scattering of strongly interacting $\PW$~bosons and investigated our
approximations within this setup.  In kinematical regions that are
statistically relevant and receive non-negligible corrections, we find
that the effective vector-boson approximation agrees with the complete
lowest-order prediction within about 20\% to 25\%.  The logarithmic
approximation reproduces the complete one-loop corrections for the
unpolarized cross section of $\PWp\PWm\to\PWp\PWm$ at the level of a
few per cent. These approximations cause an uncertainty of our
predictions at the level of a few per cent.  This uncertainty is
always comparable to the statistical error.  In principle it can be
further reduced by introducing appropriate cuts that improve the
accuracy of the EVBA.  But, in general, for a given set of cuts, a
better precision can be achieved only by means of an exact
calculation.

The size of the electroweak corrections depends strongly on the cuts
and the considered observable.  We have studied their effect on the
total cross section and four physically interesting distributions.
Within our set-up, the corrections to the total cross section amount
to $-7\%$ and $-17\%$ for $\sqrt{s}=1\TeV$ and $\sqrt{s}=3\TeV$,
respectively. For the distributions in the transverse momentum and the
rapidity of the W-boson pair they are of similar size. In the
distributions in the invariant mass of the W-boson pair and the
transverse momentum of a W~boson, the corrections are negative, of the
order of 10\% and increase in magnitude with increasing energy. They
can reach up to $-20\%$ and $-50\%$ for $\sqrt{s}=1\TeV$ and
$\sqrt{s}=3\TeV$, respectively. In summary, the electroweak
corrections reduce the Standard Model predictions by a sizeable amount
that is comparable or larger than the expected statistical error.
Therefore, they should be taken into account when searching for
effects of a strongly interacting scalar sector.
In fact, being negative, the corrections increase the sensitivity to
this kind of effects, which typically appear as an enhancement of the
WW-scattering cross section.
   
\section*{Acknowledgements}
This work was supported by the Italian Ministero dell'Istruzione,
dell'Universit\`a e della Ricerca (MIUR) under contract Decreto MIUR
26-01-2001 N.13 ``Incentivazione alla mobilit\`a di studiosi stranieri
ed italiani residenti all'estero''.

\begin{appendix}
\section{On-shell projection}
\label{app:projection}

In this section we give the explicit form of the on-shell projection
of the two incoming W bosons which induce the VBS.  When performing
the projection, care should be taken that the on-shell projected
W-boson momenta lie in the physical phase-space region. This can be
ensured by fixing the angles of the two incoming bosons while
performing their on-shell limit. We thus fix the direction of the
incoming $\PW^+$ in the CM frame of the W-boson pair. In this way, the
on-shell projected momenta can be written as
\beq
{\hat{q}}^{\rm{on}}_{\pm}={\sqrt{\hat{s}}\over 2}\left(1,\pm
\beta{{\hat{\bf{q}}}_{+}\over{|{\hat{\bf{q}}}_{+}|}}
\right)
,
\eeq 
where $\sqrt{\hat{s}}$ is the VBS CM energy,
$\beta=\sqrt{1-4\MW^2/\hat{s}}$, and ${\hat{\bf q}}_{+}$ is the
three-momentum of the incoming (off-shell ) $\PWp$ boson in the CM
frame of the W-boson pair.  The on-shell momenta are then boosted to
the laboratory frame.

\section{Logarithmic electroweak corrections}
\label{app:corr}

In this appendix, using the general results of
\citeres{Denner:2000jv,Pozzorini:2001rs}, we derive analytical
formulas for the logarithmic electroweak corrections to the subprocess
\beq\label{genprocess}
\PW^{-}_{\la_1}(k_1)
\PW^{+}_{\la_2}(k_2)
 \to 
 \PW^{-}_{-\la_3}(-k_3)
\PW^{+}_{-\la_4}(-k_4).
\eeq
The momenta $k_3, k_4$ and the helicities $\la_{3},\la_{4}$ of the
final states are defined as incoming.  The $2\to 2$ process
\refeq{genprocess} is thus equivalent to the $4\to 0$ process
\beq\label{genprocessb}
\PW^{-}_{\la_1}(k_1)
\PW^{+}_{\la_2}(k_2)
 \PW^{+}_{\la_3}(k_3)
\PW^{-}_{\la_4}(k_4)
 \to  0.
\eeq
This convention facilitates the application of the formalism of
\citeres{Denner:2000jv,Pozzorini:2001rs}, which is based on $n\to 0 $
reactions.
We consider the limit of high energies and large scattering angles,
where all invariants are much larger than the electroweak scale,
\beq\label{sudaklim}
|r_{ij}|=|(k_i+k_j)^2|\simeq |2k_ik_j|
\gg \MW^2\qquad\mbox{for}\quad i\neq j.
\eeq
In order to specify our conventions for the gauge-boson helicities,
which we denote by $\la_{i}=0,\pm 1$, we choose the CM frame.  There,
the gauge-boson momenta can be parametrized as
\beqar
k_1^\mu&=&E(1,0,0,1),\qquad
-k_3^\mu=E(1,\sin{\vartheta},0,\cos{\vartheta}),\nl
k_2^\mu&=&E(1,0,0,-1),\qquad
-k_4^\mu=E(1,-\sin{\vartheta},0,-\cos{\vartheta}),
\eeqar
and for the Mandelstam variables we have  
$r_{12}=r_{34}=4 E^2$,
$r_{13}=r_{24}= -\mas(1-\cos{\vartheta})/2$,
$r_{23}=r_{14}=-\mas(1+\cos{\vartheta})/2$.
Note that mass terms are systematically neglected in the high-energy
limit.  The polarization vectors for transverse gauge bosons
($\la_i=\tau_i=\pm 1$) read
\beqar\label{polvec}
\varepsilon^\mu(k_1,\tau_1)&=&\frac{1}{\sqrt{2}}(0,1,\tau_1 \ri,0),\qquad
\varepsilon^{\mu*}(-k_3,-\tau_3)=\frac{1}{\sqrt{2}}(0,\cos{\vartheta},\tau_3 \ri,-\sin{\vartheta}),\nl
\varepsilon^\mu(k_2,\tau_2)&=&\frac{1}{\sqrt{2}}(0,-1,\tau_2 \ri,0),\qquad
\varepsilon^{\mu*}(-k_4,-\tau_4)=\frac{1}{\sqrt{2}}(0,-\cos{\vartheta},
\tau_4 \ri,\sin{\vartheta}).\quad
\eeqar
Here and in the following, the symbol $\tau_i=\pm 1$ is used to denote
the helicity of transversely polarized gauge bosons.  Longitudinal
gauge bosons $(\la_i=0)$ have to be related to corresponding would-be
Goldstone bosons using the Goldstone Boson Equivalence Theorem (GBET)
as discussed in \refapp{se:GBET}.  There, we also introduce effective
couplings for longitudinal gauge bosons, which permit to apply the
general results of \citere{Denner:2000jv} directly to physical matrix
elements.

The matrix element for the process \refeq{genprocess} or,
equivalently, \refeq{genprocessb} is denoted as
\beq\label{melgenprocess}
\csmel{-}{\la_1}{+}{\la_2}{+}{\la_3}{-}{\la_4}
\equiv\csmel{-}{\la_1}{+}{\la_2}{+}{\la_3}{-}{\la_4}
\,
(\mas,\mat,\mau).
\eeq
In the high-energy limit, we restrict ourselves to the matrix elements
that are not mass-suppressed by factors of order $\MW/\sqrt{\mas}$.
By means of the GBET, it can easily be seen that only the matrix
elements involving helicity combinations with an even number of
longitudinally (L) and transversely (T) polarized gauge bosons are not
suppressed. The reason is that all vertices involving an odd number of
Goldstone bosons are suppressed by coupling factors proportional to
masses.  In addition, the matrix elements with helicities $\rT\rL\to
\rL\rT$ and $\rL\rT\to \rT\rL$ are suppressed, \ie
\beqar
\csmel{-}{\tau_1}{+}{0}{+}{0}{-}{\tau_4}=
\csmel{-}{0}{+}{\tau_2}{+}{\tau_3}{-}{0}=0,
\eeqar
up to terms of order $\MW^2/r_{12}$.  Therefore, in the following we
restrict ourselves to the non-suppressed combinations
\beqar\label{nonsupphelicities}
\rL\rL\to \rL\rL &:\quad&\procLL
,\nl
\rT\rT\to \rL\rL &:\quad&\procmixTT
,\nl
\rL\rL\to \rT\rT &:\quad&\procmixLL
,\nl
\rT\rL\to \rT\rL &:\quad&\procmixTL
,\nl
\rL\rT\to \rL\rT &:\quad&\procmixLT
,\nl
\rT\rT\to \rT\rT &:\quad&\procTT.
\eeqar
The amplitudes for $\rL\rL\to \rT\rT$, $\rT\rL\to \rT\rL$, and
$\rL\rT\to \rL\rT$ can be obtained from the amplitude $\rT\rT\to
\rL\rL$ using the relations
\beqar
\M^{\cprocmixLL}&=&
\M^{
\iwbos{-}{\tau_4}{1}
\iwbos{+}{\tau_3}{2}
\iwbos{+}{0}{3}
\iwbos{-}{0}{4}
}
,\nl
\M^{\cprocmixTL}&=&
\left.\M^{
\iwbos{-}{\tau_1}{1}
\iwbos{+}{\tau_3}{2}
\iwbos{+}{0}{3}
\iwbos{-}{0}{4}
}\right|_{
\mas\leftrightarrow \mat},\nl
\M^{\cprocmixLT}&=&
\left.\M^{
\iwbos{-}{\tau_4}{1}
\iwbos{+}{\tau_2}{2}
\iwbos{+}{0}{3}
\iwbos{-}{0}{4}
}\right|_{
\mas\leftrightarrow \mat
},
\eeqar
which follow from crossing symmetry. 

\subsection{Structure of the one-loop logarithmic corrections}
\label{app:one-loopcorr}

In the following sections, our results for the one-loop logarithmic
corrections
are given either in explicit form, or as correction factors 
\beq \label{relco}
\de_{\proc{-}{\la_1}{+}{\la_2}{-}{-\la_3}{+}{-\la_4}}
=
\frac{
\de\M_1^{\csproc{-}{\la_1}{+}{\la_2}{+}{\la_3}{-}{\la_4}}
}{
\M_0^{\csproc{-}{\la_1}{+}{\la_2}{+}{\la_3}{-}{\la_4}}
},
\eeq
relative to the Born matrix elements.  Following
\citere{Denner:2000jv}, we split the logarithmic corrections
\refeq{relco} according to their origin as
\beq
\de=\de^{\mathrm{LSC}}+\de^{\mathrm{SSC}}+\de^{\mathrm{C}}+\de^{\mathrm{PR}}.
\eeq
The double logarithms originating from soft-collinear gauge bosons are
split into leading contributions $\de^{\mathrm{LSC}}$ of the type
$\alpha\log^2{(|\mas|/M^2)}$ and subleading contributions
$\de^{\mathrm{SSC}}$ of the type
$\alpha\log{(|\mas|/M^2)}\log{(|r_{ij}/\mas|)}$, with
$r_{ij}=\mat,\mau$. These latter depend on ratios of Mandelstam
variables, $|r_{13}/\mas|= (1-\cos{\vartheta})/2$,
$|r_{23}/\mas|=(1+\cos{\vartheta})/2$, and thus on the scattering
angle $\vartheta$ between initial and final states.  Purely
angular-dependent logarithms of the type
$\alpha\log^2{(|r_{ij}/\mas|)}$ and $\alpha\log{(|r_{ij}/\mas|)}$ are
neglected in our approximation.  The part $\de^{\mathrm{C}}$ contains
the single-logarithmic contributions from collinear (or soft)
particles to loop diagrams and wave-function renormalization
constants.  Finally, $\de^{\mathrm{PR}}$ consists of the single
logarithms that originate from parameter renormalization.

All these logarithmic contributions depend on various mass scales
$M$=\MW, \MZ, $M_\gamma$, \MH, \Mt. This mass dependence is separated
from the energy dependence by writing
\beq\label{splitting}
\log{\left(\frac{|r_{ij}|}{M^2}\right)}
=\log{\left(\frac{|r_{ij}|}{\MW^2}\right)}
-
\log{\left(\frac{M^2}{\MW^2}\right)},
\eeq
\ie we spilt all logarithms into a contribution with scale $\MW$ 
and  a remaining part that depends on the ratio $M^2/\MW^2$.
The logarithms of $\Mt/\MW$ and $\MH/\MW$ can be found in
\citere{Pozzorini:2001rs}.  Here we give all logarithms of $\Mt/\MW$
as well as the universal logarithms of $\MH/\MW$, \ie those that
originate from parameter renormalization and collinear singularities.
Also the subleading logarithms of the type $\alpha\log{(|\mas|/\MW^2)}
\log{(\MZ^2/\MW^2)} $ are included.

In \citere{Denner:2000jv}, the virtual electromagnetic corrections
have been regularized by an infinitesimal photon mass $M_\gamma=\la$
and split as in \refeq{splitting} into a contribution corresponding to
a heavy photon ($\la=\MW$) and a remaining part which originates from
the mass gap $\la\ll\MW$ in the gauge sector. The heavy-photon
contribution has been combined with the weak corrections resulting
into the so-called symmetric-electroweak part of the corrections.  The
remaining mass-gap contribution has been isolated into the
infrared-divergent logarithms $L^\elm(s,\la^2,m_k^2)$, $\lWla$,
$l^\elm(m_k^2)$ which appear in Eqs.\ (3.7), (3.8), (3.10), (3.12),
(4.6), (4.7), (4.10) and (4.33) of \citere{Denner:2000jv}.  These
logarithms contain only contributions from virtual photons.

In \refapp{se:emlogs} we provide simple substitutions that permit to
generalize the results of \citere{Denner:2000jv} to semi-inclusive $2
\to 2$ processes, by including the soft-photon bremsstrahlung
corrections.  The resulting logarithms, defined in \refeq{LSCemlogs1},
\refeq{angularbrems1}, and \refeq{NLLbrems1}, are infrared finite and
depend on the soft-photon cutoff $\Delta E$.

\subsubsection*{Notation}
The coefficients of the various logarithms are expressed in terms of
the eigenvalues $I^{V^a}_\varphi$, or of the matrix components
$I^{V^a}_{\varphi\varphi'}$, of the generators\footnote{A detailed
  list of the gauge-group generators and of related quantities that
  are used in the following can be found in App.~B of
  \citere{Denner:2000jv} and App.~B of \citere{Pozzorini:2001rs}.}
\beq
I^A=-Q=-\frac{Y}{2}-T^3,\qquad 
I^Z=-\frac{\sw}{\cw}\frac{Y}{2}+\frac{\cw}{\sw}T^3,\qquad
I^{W^\pm}=
\frac{T^1\pm\ri T^2}{\sqrt{2}\sw},
\eeq
where $\cw^2=1-\sw^2=\MW^2/\MZ^2$.
Another group-theoretical object that often appears in our results is the
electroweak Casimir operator 
\beq
\cew=
\sum_{V^a=A,Z,W^\pm}
I^{V^a} I^{\bar V^{a}}=
\frac{1}{\cw^2}
\left(\frac{Y}{2}\right)^2
+\frac{1}{\sw^2}
T(T+1),
\eeq
where $T$ represents the total isospin.

\subsection{Leading soft--collinear corrections}

Below we list the angular-independent leading soft--collinear (LSC)
corrections for various polarizations of the gauge bosons. These
results are obtained from Eqs.~(3.6) and (3.7) of
\citere{Denner:2000jv} and depend on the eigenvalues of the
electroweak Casimir operator
\beq
\cew_\Phi=\frac{1+2\cw^2}{4\sw^2\cw^2},\qquad
\cew_W=\frac{2}{\sw^2},
\eeq
as well as on the squared Z-boson couplings
\beqar
(I^Z_{W^\pm})^2&=&\frac{\cw^2}{\sw^2},\qquad
(I^Z_{\phi^\pm})^2= \frac{(\cw^2-\sw^2)^2}{4\sw^2\cw^2}.
\eeqar
The explicit expressions for the electromagnetic logarithms
$L^\EM(\MW^2)$, which contain contributions from both virtual and real
soft photons, are given in \refeq{LSCemlogs1}.

\subsubsection*{Purely longitudinal polarizations}
\beqar\label{SCWWlong}
\de^{\SC}_{\procLL}
&=&
-\frac{\alpha}{2\pi}
\left[
\cew_{\Phi}\losw{2}
-2\left(I^Z_{\phi^\pm}\right)^2
\loZW\losw{}
\right]
\nl&&{}
-2 L^\EM(\MW^2)
.
\eeqar
\subsubsection*{Mixed polarizations}
\beqar\label{SCWWmixed}
&&\hspace{-12mm}
\de^{\SC}_{\procmixTT}
=
\de^{\SC}_{\procmixLL}=
\de^{\SC}_{\procmixTL}=
\de^{\SC}_{\procmixLT}=
\nl&=&
-\frac{\alpha}{4\pi}
\left[
\left(\cew_{W}+\cew_{\Phi}\right)\losw{2}
\right.\nl&&\left.{}
-2\left[\left(I^Z_{W^\pm}\right)^2+\left(I^Z_{\phi^\pm}\right)^2\right]
\loZW\losw{}
\right]
-2 L^\EM(\MW^2)
.
\eeqar
\subsubsection*{Purely transverse polarizations}
\beqar\label{SCtrans}
\de^{\SC}_{\procTT}
&=&
-\frac{\alpha}{2\pi}
\left[
\cew_{W}\losw{2}
-2\left(I^Z_{W^\pm}\right)^2
\loZW\losw{}
\right]
\nl&&{}
-2 L^\EM(\MW^2)
.
\eeqar

\subsection{Subleading soft--collinear corrections}
\newcommand{\J}{J}

The angular-dependent subleading soft--collinear ($\SS$) corrections
to the $2\to 2$ processes \refeq{nonsupphelicities} are obtained by
applying the formula (3.12) of \citere{Denner:2000jv} [see also
\refeq{SSCformula}] to the corresponding $4\to 0$ processes which
result from reversing the outgoing particles by a crossing
transformation, \ie the charges of the outgoing states have to be
reversed as in \refeq{melgenprocess}.  The corrections to matrix
elements involving longitudinal gauge bosons are obtained via the GBET
applying Eq.~(3.12) of \citere{Denner:2000jv} to corresponding matrix
elements that involve would-be Goldstone bosons.  Alternatively, as
discussed in \refapp{se:GBET}, one can directly apply Eq.~(3.12) of
\citere{Denner:2000jv} to matrix elements for longitudinal gauge
bosons using the effective couplings \refeq{longcouplings}.

The SSC corrections originating from soft neutral gauge bosons $N=A,Z$
result in
\beqar\label{neutralsscgeneral}
\sum_{N=A,Z}\de^{N,\SS}_{\genproc}
&=&\frac{\alpha}{2\pi}\losw{}\sum_{N=A,Z}
\left[\left(
I^N_{W^-_{\la_1}}
I^N_{W^+_{\la_3}}+
I^N_{W^+_{\la_2}}
I^N_{W^-_{\la_4}}\right)\lots
\right.
\nl
&&\hspace{-4cm}\left.+
\left(
I^N_{W^-_{\la_1}}
I^N_{W^-_{\la_4}}+
I^N_{W^+_{\la_2}}
I^N_{W^+_{\la_3}}\right)\lous
\right]
-4\lotu
l^\EM_\SS
,
\eeqar
where the couplings $I^N_{W^\pm_0}$ for longitudinal gauge bosons are
given in \refeq{effbeutcouplings1}, whereas for transverse gauge
bosons $I^A_{W^\pm_{\tau}}=\mp 1$ and $I^Z_{W^\pm_{\tau}}=\pm
\cw/\sw$.  The electromagnetic logarithms $l^\EM_\SS$ are defined in
\refeq{angularbrems1}.

Soft virtual W bosons can be exchanged only in the $\mat$ channel, and yield
\beqar\label{chargedsscgeneral}
\lefteqn{\sum_{V=W^\pm}\de^{V,\SS} \M^{\genproccrossed}=
  \frac{\alpha}{2\pi}\losw{}\lots }\quad&&\nl &&{}\times
\left[\sum_{N_{\la_1}}\sum_{N'_{\la_3}}
  \J^-_{N_{\la_1}}\J^+_{N'_{\la_3}} \M_0^{{N_{\la_1}}W_{\la_2}^+
    N'_{\la_3}W^-_{\la_4}} +\sum_{N_{\la_2}}\sum_{N'_{\la_4}}
  \J^+_{N_{\la_2}}\J^-_{N'_{\la_4}}
  \M_0^{{W^-_{\la_1}}N_{\la_2}W^+_{\la_3}N'_{\la_4}} \right],\qquad
\eeqar
where the sums over $N^{}_{\la}$ and $N^{'}_{\la}$ depend on the
polarization $\la$.  In the case $\la=0$, they run over $N_0=H,Z_0$,
whereas for $\la=\tau=\pm 1$ they run over $N_\tau=A_\tau,Z_\tau$.
The corresponding couplings $\J^\pm_{H},\J^\pm_{Z_0}$ and
$\J^\pm_{A_\tau},\J^\pm_{Z_\tau}$ are given in
\refeq{chargedeffcouplings} and \refeq{chargedtransvcouplings},
respectively.  The $\SUtwo$-transformed Born matrix elements that
appear on the right-hand side of \refeq{chargedsscgeneral} are
evaluated in the high-energy limit in \refapp{se:Born}.

In the following we list the explicit results corresponding to all
helicity combinations \refeq{nonsupphelicities}.  Note that, owing to
crossing symmetry, both contributions \refeq{neutralsscgeneral} and
\refeq{chargedsscgeneral} are invariant with respect to simultaneous
exchange of the polarizations $\la_1\leftrightarrow
\la_4,\la_2\leftrightarrow \la_3$, \ie
\beqar\label{trivialssccrossing}
\sum_{N=A,Z}\de^{N,\SS}_{\PW^-_{\la_4}
\PW^+_{\la_3}
\to
\PW^-_{-\la_2}
\PW^+_{-\la_1}
}&=&
\sum_{N=A,Z}\de^{N,\SS}_{
\PW^-_{\la_1}
\PW^+_{\la_2}
\to
\PW^-_{-\la_3}
\PW^+_{-\la_4}
},\nl
\sum_{V=W^\pm}\de^{V,\SS} \M^{W^-_{\la_4}
W^+_{\la_3}
W^+_{\la_2}
W^-_{\la_1}
}&=&
\sum_{V=W^\pm}\de^{V,\SS} \M^{W^-_{\la_1}
W^+_{\la_2}
W^+_{\la_3}
W^-_{\la_4}
}.
\eeqar
Therefore, the corrections for $\rL\rL\to \rT\rT$ and $\rL\rT\to
\rL\rT$ can easily be obtained from those for $\rT\rT\to \rL\rL$ and
$\rT\rL\to \rT\rL$, respectively.

\subsubsection*{Purely longitudinal polarizations}
The contribution \refeq{neutralsscgeneral}
of soft neutral gauge bosons gives
\beqar 
\sum_{N=A,Z}\de^{N,\SS}_{\procLL}
=-\frac{\alpha}{4\pi\sw^2\cw^2}\losw{}\lotu
 -4\lotu l^\EM_\SS 
.
\eeqar
The contribution \refeq{chargedsscgeneral}
of soft W 
bosons yields
\beqar\label{SCLLLLexact}
\lefteqn{
\sum_{V=W^\pm}\de^{V,\SS} \M^{\cprocLL}
=
}
\quad&&\nl
&=&
\frac{\alpha}{2\pi}\losw{}\lots 
\sum_{S,S'=H,Z_0}
\J^-_S\J^+_{S'} \left[\M_0^{SW_0^+ S'W_0^-}
+\M_0^{W_0^-S' W_0^+S}
\right]\nl
&=&
\frac{\alpha}{8\pi\sw^2}\losw{}\lots 
\left\{
\left[\left(\M_0^{HW_0^+ HW_0^-}
+\M_0^{W_0^-H W_0^+H}\right)-(H\to Z_0)
\right]
\right.\nl&&\left.
{}-
\left[\left(\M_0^{HW_0^+ Z_0W_0^-}
+\M_0^{W_0^-Z_0 W_0^+H}\right)
-(H\leftrightarrow Z_0)
\right]
\right\}.
\eeqar
Using our expressions \refeq{LLLLBornampli2}, \refeq{LLLLBornampli3}
for the $\SUtwo$-transformed Born matrix elements in the {high-energy
  limit}, and dividing by the Born matrix element \refeq{4phibornmel}
we obtain
\beqar\label{SCLLLLhe}
\lefteqn{
\sum_{V=W^\pm}\de^{V,\SS}_{\procLL}
=-\frac{\alpha}{2\pi\sw^2}\losw{}\lots\left[
\frac{\lah}{2}+\frac{e^2}{2\sw^2}\As
\right.
}\quad&&
\nl&&\left.{}
+ e^2\frac{\sw^2-\cw^2}{4\sw^2\cw^2} \At
\right]
\left[\lah+\frac{e^2}{4\sw^2\cw^2}\left(\As+\At\right)\right]^{-1},
\eeqar
where $\lah$ is the scalar self coupling defined in \refeq{scalarcouplingdef}.

\subsubsection*{Mixed  polarizations:  $\rT\rT \to \rL\rL$ and $\rL\rL \to \rT\rT$
}
For the $\rT\rT \to \rL\rL$ configuration,
the contribution \refeq{neutralsscgeneral} of soft  neutral gauge bosons gives
\beqar 
\sum_{N=A,Z}\de^{N,\SS}_{\procmixTT}
=-\frac{\alpha}{2\pi\sw^2} \losw{}\lotu -4\lotu l^\EM_\SS 
.
\eeqar
Soft virtual W bosons \refeq{chargedsscgeneral} yield
\beqar\label{SCTTLLexact}
\lefteqn{
\sum_{V=W^\pm}\de^{V,\SS} \M^{\cprocmixTT}
=
}\quad&&\nl
&=&
\frac{\alpha}{2\pi}\losw{}\lots 
\sum_{N =A,Z}\sum_{S=H,Z_0}
\left[\J^-_{N}\J^+_{S} \M_0^{{N_{\tau_1}}W_{{\tau_2}}^+ SW_0^-}
+\J^+_{N}\J^-_{S}\M_0^{W_{{\tau_1}}^-{N_{\tau_2}} W_0^+S}
\right]\nl
&=&
\frac{\alpha}{4\pi\sw}\losw{}\lots \left\{
\left[
-\M_0^{A_{\tau_1}W_{\tau_2}^+ HW_0^-}
+\M_0^{W_{\tau_1}^-A_{\tau_2} W_0^+H}
\right.\right.\nl&&\left.\left.{}
+
\M_0^{A_{\tau_1}W_{\tau_2}^+ Z_0W_0^-}
+\M_0^{W_{\tau_1}^-A_{\tau_2} W_0^+Z_0}
\right]
-\frac{\cw}{\sw}\left[A \rightarrow Z\right]
\right\}.
\eeqar
Using the Born amplitudes \refeq{TTLLBornampli1}, \refeq{TTLLBornampli2} 
in the high-energy limit we obtain the relative correction
\beqar\label{SCTTLLhe}
\sum_{V=W^\pm}\de^{V,\SS}_{\procmixTT}
&=&
\frac{\alpha}{2\pi \sw^2}
\left(\frac{\mat}{\mau}-1\right)
\losw{}\lots 
.
\eeqar
These results can directly be extended to the  $\rL\rL \to \rT\rT$
configuration using \refeq{trivialssccrossing}.

\subsubsection*{Mixed polarizations: $\rT\rL \to \rT\rL$ and $\rL\rT
  \to \rL\rT$} 
For the $\rT\rL \to \rT\rL$ configuration, the contribution
\refeq{neutralsscgeneral} of soft neutral gauge bosons gives
\beqar 
\lefteqn{
\sum_{N=A,Z}\de^{N,\SS}_{\procmixTL}
=
}\quad&&\nl
&=&
-\frac{\alpha}{2\pi\sw^2} \losw{}\left[\lotu+\frac{1}{4\cw^2}\lots \right]
-4
\lotu l^\EM_\SS .\qquad
\eeqar
Soft virtual W bosons \refeq{chargedsscgeneral} yield
\beqar\label{SCTLTLexact}
\lefteqn{
\sum_{V=W^\pm}\de^{V,\SS} \M^{\cprocmixTL}
=\frac{\alpha}{2\pi}\losw{}\lots 
}\quad&&\nl
&&{}\times
\Biggl[
\sum_{N=A,Z} \sum_{N'=A,Z}
\J^-_{N}\J^+_{N'} \M_0^{{N_{\tau_1}}W_0^+ N'_{\tau_3} W_0^-}
+\sum_{S=H,Z_0}\sum_{S'=H,Z_0}
\J^+_{S}\J^-_{S'}
\M_0^{W_{{\tau_1}}^-S W_{\tau_3}^+S'}
\Biggr]\nl
&=&
\frac{\alpha}{2\pi}\losw{}\lots 
\Biggl\{
 -\M_0^{{A_{\tau_1}}W_0^+ A_{\tau_3} W_0^-}
 -\frac{\cw^2}{\sw^2}\M_0^{{Z_{\tau_1}}W_0^+ Z_{\tau_3} W_0^-}
\nl&&{}
+\frac{\cw}{\sw}\Biggl[
\M_0^{{A_{\tau_1}}W_0^+ Z_{\tau_3} W_0^-}
+(A\leftrightarrow Z)\Biggr]
+
\frac{1}{4\sw^2}\Biggl[
\left(
\M_0^{W_{{\tau_1}}^-H W_{\tau_3}^+H}
+\M_0^{W_{{\tau_1}}^-H W_{\tau_3}^+Z_0}
\right)
\nl&&{}
-(H\leftrightarrow Z_0)\Biggr]
\Biggr\}.
\eeqar
Using the Born amplitudes \refeq{TTLLBornampli1cross},
\refeq{TTLLBornampli2cross} in the { high-energy limit} we obtain the
relative correction
\beqar\label{SCTLTLhe}
\sum_{V=W^\pm}\de^{V,\SS}_{\procmixTL}
&=&
-\frac{\alpha}{2\pi \sw^2}
\left(\frac{\mas}{\mau}+\frac{1}{2}\right)
\losw{}\lots 
.
\eeqar
These results can directly be extended to the $\rL\rT \to \rL\rT$
configuration using \refeq{trivialssccrossing}.

\subsubsection*{Purely transverse polarizations}
The contribution \refeq{neutralsscgeneral} of soft neutral gauge
bosons gives
\beq 
\sum_{N=A,Z}\de^{N,\SS}_{\procTT}
=-\frac{\alpha}{\pi\sw^2} \losw{}\lotu -4\lotu l^\EM_\SS .
\eeq
Soft virtual W bosons \refeq{chargedsscgeneral} yield
\beqar\label{SCTTTTexact}
\lefteqn{\sum_{V=W^\pm}\de^{V,\SS} \M^{\cprocTT}=}\quad&&\nl
&=&
\frac{\alpha}{2\pi}\losw{}\lots 
\sum_{N,N'=A,Z}
\J^-_N \J^+_{N'} \left[
\M_0^{N_{\tau_1}W_{\tau_2}^+ N'_{\tau_3}W_{\tau_4}^-}
+\M_0^{W_{\tau_1}^-N'_{\tau_2} W^+_{\tau_3}N_{\tau_4}}
\right]\nl
&=&
\frac{\alpha}{2\pi}\losw{}\lots \left\{
-\left(\M_0^{A_{\tau_1}W^+_{\tau_2} A_{\tau_3}W^-_{\tau_4}}
+\M_0^{W^-_{\tau_1}A_{\tau_2} W^+_{\tau_3}A_{\tau_4}}\right)
\right.\nl&&\left.{}
+\frac{\cw}{\sw}\left[\left(
\M_0^{A_{\tau_1}W^+_{\tau_2} Z_{\tau_3}W^-_{\tau_4}}
+\M_0^{W^-_{\tau_1}A_{\tau_2} W^+_{\tau_3}Z_{\tau_4}}\right)
+(A\leftrightarrow Z)\right]
\right.\nl&&\left.{}
-\frac{\cw^2}{\sw^2}\left(
\M_0^{Z_{\tau_1}W^+_{\tau_2} Z_{\tau_3}W^-_{\tau_4}}
+\M_0^{W^-_{\tau_1}Z_{\tau_2} W^+_{\tau_3}Z_{\tau_4}}\right)
\right\}.
\eeqar
In the high-energy limit, using \refeq{TTTTBornampli3} 
we obtain the relative correction factor
\beqar\label{SCTTTThe}
\sum_{V=W^\pm}\de^{V,\SS}_{\procTT}
&=&
\frac{\alpha}{\pi\sw^2}
\frac{r_{13}}{r_{23}}
\losw{}\lots 
.
\eeqar

\subsection{Single logarithms from collinear singularities}

The single-logarithms originating from collinear singularities
associated to external transverse or longitudinal \PW~bosons can be
obtained from Eqs.~(4.10) and (4.33) in \citere{Denner:2000jv}.  Here
we have included also the logarithms of $\Mt/\MW$ and $\MH/\MW$, which
can be found in \citere{Pozzorini:2001rs}.  The resulting corrections
to all processes \refeq{nonsupphelicities} can be expressed by the
following general formula, which only depends on the numbers
$n_\rT,n_\rL$ of transversely and longitudinally polarized W~bosons,
respectively.
\beqar\label{decc}
\de^\cc_{\proc{-}{\la_1}{+}{\la_2}{-}{-\la_3}{+}{-\la_4}}
&=&\sum_{i=1}^4 \de^\cc_{W_{\la_i}}=
n_\rT \de^\cc_{W_{\rT}}+ n_\rL \de^\cc_{W_0},
\eeqar
where
\beqar\label{decc2}
\de^\cc_{W_{\rT}}
&=&\frac{\alpha}{4\pi}
\left\{
\frac{1}{2}\bew_W\losw{}
+\frac{1}{24\sw^2}\loHw{}
+\frac{1}{2\sw^2}\lotw{}
\right\}
+l^\EM(\MW^2),
\nl
\de^\cc_{W_0}
&=&\frac{\alpha}{4\pi}
\left\{
2\cew_\Phi\losw{}
-\frac{3}{4\sw^2}\frac{\Mt^2}{\MW^2}
\lost{}
+\frac{1}{8\sw^2}\loHw{}
\right\}
+l^\EM(\MW^2),
\nln
\eeqar
the one-loop coefficient of the SU(2) $\beta$-function reads
$\bew_W=19/(6\sw^2)$, and the electromagnetic contributions
$l^\EM(\MW^2)$ are defined in \refeq{NLLbrems1}.

\subsection{Single logarithms  from parameter renormalization}

\subsubsection*{Purely longitudinal polarizations}

The renormalization of the parameters $g=e^2/(4\sw^2\cw^2)$ and $\lah$
\refeq{scalarcouplingdef} in the Born amplitude \refeq{4phibornmel}
gives rise to the relative correction
\beqar\label{deprelong}
\de^\pre_{\procLL}
&=&
\frac{\de g}{g}+\left(\frac{\de \lah}{\lah}-\frac{\de g}{g}\right)\frac{\lah}{\lah+g A},
\eeqar
where 
\beq
A=
\As+\At
=
2\left(\frac{\mau^2}{\mas\mat}-1\right)
\eeq
in the high-energy limit, and the 't~Hooft scale of dimensional
regularization has to be set to $\mu^2=\mas$ in the counterterms
\cite{Denner:2000jv,Denner:2001gw}. In the on-shell scheme (including
the tadpole contributions in the renormalization of $\lah$) these read
\cite{Pozzorini:2001rs}
\newcommand{\lprtop}{l_{\pre,\Pt}}
\beqar\label{dealpha1}
\left.\frac{\de g}{g}\right|_{\mu^2=\mas}
&=&
\frac{\alpha}{4\pi}
\left\{
-\bew_{ZZ}\losw{}
+
\frac{\cw^2-\sw^2}{\sw^2\cw^2}
\left[
\frac{5}{6}\loHw{}
\right.\right.\nl&&\left.\left.{}
-
\frac{9+6\sw^2-32\sw^4}{18\sw^2}
\lotw{}
\right]
\right\}
+\Delta\alpha(\MW^2),
\nl
\left. \frac{\de \lah}{\lah}\right|_{\mu^2=\mas}
&=&
\Delta\alpha(\MW^2)+
\frac{\alpha}{4\pi}
\left\{
\left\{
\frac{3}{2\sw^2}\biggl[\frac{\MW^2}{\MH^2}\left(2+\frac{1}{\cw^4}\right)
-\left(2+\frac{1}{\cw^2}\right) + \frac{\MH^2}{\MW^2}\biggr]
\right.\right.\nl&&\left.\left.{}
+\frac{\NCt}{\sw^2}\frac{\Mt^2}{\MW^2}\left(1-2\frac{\Mt^2}{\MH^2}\right)\right\}\losw{}
\right.\nl&&\left.{}
+
\frac{3}{2\sw^2}\biggl[-\frac{3}{2}\frac{\MW^2}{\MH^2}
\left(2+\frac{1}{\cw^4}\right)
+\frac{1+2\cw^2}{4\cw^2}
+\frac{10}{9}
-\frac{\MH^2}{\MW^2}
\biggr]\loHw{}
\right.\nl&&\left.\hspace{5mm} {} 
-\left[
\frac{\NCt}{\sw^2}\frac{\Mt^2}{\MW^2}\left(1-2\frac{\Mt^2}{\MH^2}\right)
+\frac{9-12\sw^2-32\sw^4}{18\sw^4}\right] \lotw{}
\right\}
,\qquad
\eeqar
where
$\bew_{ZZ}=\sw^2\bew_{B}+\cw^2\bew_{W}=(19-38\sw^2-22\sw^4)/(6\sw^2\cw^2)$,
with the $\Uone$ and $\SUtwo$ one-loop $\beta$-function coefficients
$\bew_{B}$ and $\bew_{W}$ defined in \citere{Denner:2000jv}, and
$\Delta\alpha(\MW^2)$ represents the running of the electromagnetic
coupling constant from the scale $0$ to $\MW$.  Within the $G_\mu$
scheme, these effects of the running are already included in the
definition of $\alpha_{G_\mu}$. Thus, the $\Delta \alpha (\MW^2)$-terms
appearing in \refeq{dealpha1}--\refeq{depretransv} were not included
in our implementation.

\subsubsection*{Mixed polarizations}
In this case, the Born matrix element is proportional to the squared
SU(2) coupling $g_2^2=e^2/\sw^2$. As a result, the parameter
renormalization yields
\beqar\label{depremixed}
\de^{\pre}_{\procmixTT}&=&
\de^{\pre}_{\procmixLL}=
\de^{\pre}_{\procmixTL}=
\de^{\pre}_{\procmixLT}=
\nl
&=&\left. 2 \frac{\de
g_2}{g_2}\right|_{\mu^2=\mas}
=\frac{\alpha}{4\pi}\Biggl[-\bew_W\losw{}
+\frac{5}{6\sw^2}
\loHw{}
\nl&&{}-\frac{9+6\sw^2-32\sw^4}{18\sw^4} \lotw{}
\Biggr]
+ \Delta\alpha(\MW^2),
\eeqar
with $\bew_W=19/(6\sw^2)$.  Note that the contribution 
$\bew_W
\log{(|\mas|/\MW^2)}$ from parameter renormalization cancels a
corresponding contribution $\bew_W \log{(|\mas|/\MW^2)}$ in
\refeq{decc}, which is associated to the two transverse gauge bosons
($n_{\rT}=2$).  This cancellation is analogous to the one observed in
Eq.~(A.11) in \citere{Denner:2000jv}.

\subsubsection*{Purely transverse polarizations}
Also in  this case, the Born matrix element is proportional to the squared SU(2) coupling $g_2^2$ and we have
\beqar\label{depretransv}
\de^{\pre}_{\procTT}
&=&
\frac{\alpha}{4\pi}\left[-\bew_W\losw{}
+\frac{5}{6\sw^2}\loHw{}\right.
\nl&&\left.{}
-\frac{9+6\sw^2-32\sw^4}{18\sw^4} \lotw{}
\right]
 + \Delta\alpha(\MW^2)
.
\eeqar
In this case the contributions $\bew_W \log{(|\mas|/\MW^2)}$ from
parameter renormalization cancel only two of the four ($n_{\rT}=4$)
contributions $\bew_W \log{(|\mas|/\MW^2)}$ in \refeq{decc}.

\section{Longitudinal gauge bosons}\label{se:GBET}
In order to apply the general formulas of \citere{Denner:2000jv} to
matrix elements involving longitudinal gauge bosons
$V^a_0=W^\pm_0,Z_0$, these have to be transformed first into
corresponding matrix elements involving would-be Goldstone bosons
$\Phi_a=\phi^\pm,\chi$ by means of the Goldstone-Boson Equivalence
Theorem (GBET) in its naive lowest-order form\footnote{The relevant
  quantum corrections to the GBET, which contribute to the collinear
  single-logarithms are already taken into account into the
  corresponding corrections factors (4.33) in
  \citere{Denner:2000jv}.},
\beqar\label{logbet}
\M^{\varphi_{i_1}\dots V^a_0 \dots \varphi_{i_n}}
&=&\ri^{1-Q_{V^a}}
\M^{\varphi_{i_1}\dots \Phi_a \dots \varphi_{i_n}},
\eeqar
where
$\varphi_{i_1},\dots\varphi_{i_n}$ represent  arbitrary particles with
arbitrary polarizations and $Q_{V^a}$ is the gauge-boson charge, \ie
$Q_{W^\pm}=\pm 1$ and $Q_{Z}=0$.

Particular care must be taken  of the angular-dependent subleading
soft-collinear 
corrections 
(see Eqs.~(3.9)--(3.12) of
 \citere{Denner:2000jv}) 
\beqar\label{SSCformula}
\lefteqn{
\de^{\SS} \M^{\varphi_{i_1} \dots \varphi_{i_n}}
=}\quad&&\nl
&=&
\frac{\alpha}{2\pi}
\sum_{k=1}^n
\sum_{l<k}
\sum_{V^a=A,Z,W^\pm}
\log{\left(\frac{|\mas|}{M_a^2}\right)} 
\log{\left(\frac{|r_{kl}|}{|\mas|}\right)} 
I^{V^a}_{\varphi_{i'_l}\varphi_{i_l}} 
I^{\bar{V}^a}_{\varphi_{i'_k}\varphi_{i_k}} 
\M_0^{\varphi_{i_1} \dots\varphi_{i'_l} \dots\varphi_{i'_k} \dots \varphi_{i_n}},\quad
\eeqar
which involve non-abelian couplings $I^{V^a}_{\varphi_{i'}\varphi_{i}}
$ 
that are in general non-diagonal and lead therefore to
$\SUtwo$-transformed Born matrix elements 
with $\varphi_{i'}\neq\varphi_{i}$
on the right-hand side of
\refeq{SSCformula}.
For processes involving longitudinal gauge bosons and Higgs bosons,
this formula has to be applied to corresponding processes involving
would-be Goldstone bosons and Higgs bosons.  Then the unphysical
matrix elements on the left- and right-hand sides of
\refeq{SSCformula} have to be transformed into physical matrix
elements by means of the GBET. In doing this, the factors
$\ri^{1-Q_{V^a}}$ originating from the GBET \refeq{logbet} must be
carefully taken into account, since the would-be Goldstone bosons and
Higgs bosons appearing on the left- and right-hand side of
\refeq{SSCformula} (and thus the corresponding GBET factors) can be
different.

In order to avoid these complications related to the explicit use of
the GBET, in the following we introduce effective couplings for
longitudinal gauge bosons and Higgs bosons, which permit to apply
\refeq{SSCformula} directly to physical matrix elements.  To this end
the factors $\ri^{1-Q_{V^a}}$ from the GBET are combined with the
gauge couplings for would-be Goldstone bosons and Higgs bosons,
resulting into the 
effective couplings
\beqar\label{longcouplings}
I^{V^c}_{V^b_0 V^a_0}
=
(-\ri)^{1-Q_{V^b}}
I^{V^c}_{\Phi_b\Phi_a}
\ri^{1-Q_{V^a}} 
,
\qquad
I^{V^c}_{H V^a_0}
=
I^{V^c}_{H \Phi_a}
\ri^{1-Q_{V^a}} 
,
\qquad
I^{V^c}_{V^b_0 H}
=
(-\ri)^{1-Q_{V^b}} I^{V^c}_{\Phi_b H}, \nln
\eeqar 
where $V^c=A,Z,W^\pm$, and $\Phi_{a}=\phi^\pm,\chi$ are the would-be
Goldstone bosons corresponding to $V^{a}_0=W^\pm_0,Z_0$.  We observe
that the relations \refeq{logbet} and \refeq{longcouplings} can be
regarded as the result of a reparametrization of the scalar sector
through the the unitary transformation
\beqar\label{unitarytransf}
\Phi_a=\ri^{1-Q_{V^a}} V_0^a.
\eeqar
If one performs this transformation directly at the level of the
Lagrangian one can express the Feynman rules in terms of the
longitudinal gauge-boson fields.  This simplifies the calculation of
matrix elements for longitudinal gauge bosons in the high-energy limit
(see \refapp{se:Born}).  The Feynman rules for the scalar and gauge
interactions of the fields $S_i=W_0^\pm,Z_0,H$ and their propagators
in the 't~Hooft--Feynman gauge read
\beqar
\label{feynmanrules}
\vcenter{\hbox{
\begin{picture}(110,100)(-50,-50)
\Text(-45,5)[lb]{$V^a_{\mu}$}
\Text(35,30)[cb]{$S_{i_1} (p_1)$}
\Text(35,-30)[ct]{$S_{i_2}(p_2)$}
\Vertex(0,0){2}
\DashLine(0,0)(35,25){5}
\DashLine(0,0)(35,-25){5}
\Photon(0,0)(-45,0){2}{3}
\end{picture}}}
&&
=\ri e I^{V^{a_1}}_{{S_{i_1}^+} S_{i_2}}
(p_2-p_1)_{\mu_1},
\\
\label{feynmanrules2}
\vcenter{\hbox{
\begin{picture}(110,100)(-50,-50)
\Text(-35,30)[cb]{$S_{i_1}$}
\Text(-35,-30)[ct]{$S_{i_2}$}
\Text(35,30)[cb]{$V^{a_1}_{\mu_1}$}
\Text(35,-30)[ct]{$V^{a_2}_{\mu_2}$}
\Vertex(0,0){2}
\Photon(0,0)(35,25){2}{3}
\Photon(0,0)(35,-25){-2}{3}
\DashLine(0,0)(-35,25){5}
\DashLine(0,0)(-35,-25){5}
\end{picture}}}
&&
=\ri e^2 g_{\mu_1\mu_2} \left\{I^{V^{a_1}}, I^{V^{a_2}}\right\}_{{ S_{i_1}^+} S_{i_2}},
\\
\label{feynmanrules3}
\vcenter{\hbox{
\begin{picture}(110,100)(-50,-50)
\Text(-35,30)[cb]{$S_{i_1}$}
\Text(-35,-30)[ct]{$S_{i_3}$}
\Text(35,30)[cb]{$S_{i_2}$}
\Text(35,-30)[ct]{$S_{i_4}$}
\Vertex(0,0){2}
\DashLine(0,0)(35,25){5}
\DashLine(0,0)(35,-25){5}
\DashLine(0,0)(-35,25){5}
\DashLine(0,0)(-35,-25){5}
\end{picture}}}
&&
=-\ri \frac{\la_{\PH}} {2} \left[
\de_{{ S_{i_1}^+} S_{i_2}}
\de_{{ S_{i_3}^+} S_{i_4}}
+(2\leftrightarrow 3)
+(2\leftrightarrow 4)
\right],
\\
\label{feynmanrules4}
\vcenter{\hbox{
\begin{picture}(100,40)(-40,-20)
\DashLine(35,0)(-35,0){5}
\Vertex(35,0){2}
\Vertex(-35,0){2}
\Text(-34,5)[cb]{$S_i(p)$}
\Text(34,5)[cb]{$S_j(-p)$}
\end{picture}}}
&&
=
\frac{\ri\, \de_{S_i^+S_j}}{p^2-M^2_{S_i}}.
\eeqar
Here, all fields are incoming, the curly brackets represent an
anticommutator and products of couplings have to be understood as
\beqar
\left(I^{V^{a_1}} I^{V^{a_2}}\right)_{{ S_{i}} S_{j}}
=\sum_{ S_{k}=W_0^\pm,Z_0,H}
I^{V^{a_1}}_{{ S_{i}} S_{k}} I^{V^{a_2}}_{{ S_{k}} S_{j}}.
\eeqar 
The Feynman rules \refeq{feynmanrules}--\refeq{feynmanrules4} are
closely analogous to those in \citere{Pozzorini:2001rs}.  However
here, as a result of the unitary transformation \refeq{unitarytransf},
the hermitian conjugation of the fields $V_0^a=Z_0,W_0^\pm$ generates
a minus sign,
\beqar
{V^a_0}^+
= -\bar{V}^a_0,
\eeqar
where $\bar{V}^a_0=(-\ri)^{(1-Q_{\bar{V}^a})} \Phi^+_a$ or,
equivalently, $\bar W^\pm_0=W^\mp_0$ and $\bar Z_0=Z_0$.  As a
consequence, the coupling matrices and the Kronecker symbols that
involve hermitian conjugate fields in
\refeq{feynmanrules}--\refeq{feynmanrules4} have to be understood as
\beqar
I^{V^a}_{S_i^+ S_j}
=
-I^{V^a}_{{\bar S}_i S_j}
,\qquad
\de_{S_i^+ S_j}
=
-\de_{{\bar S}_i S_j}
\eeqar 
for $S_i=W^\pm_0,Z_0$.  Instead, $H^+=\bar{H}=H$ for the Higgs field,
and
\beqar
I^{V^a}_{H^+ S_j}
=
I^{V^a}_{H S_j}
,\qquad
\de_{H^+ S_j}
=
\de_{HS_j}.
\eeqar 
Below we list the explicit expression for all non-vanishing effective
couplings \refeq{longcouplings}, which can be easily derived from
Eqs.~(B.21)--(B-23) in \citere{Denner:2000jv}.  For neutral gauge
bosons ($N=A,Z$) we have
\beqar\label{effbeutcouplings1}
I^N_{W^\si_0 W^{\si'}_0}=\de_{\si\si'}I^N_{W_0^\si}
,\qquad
\mbox{with}\qquad
I^A_{W_0^\si}=-\si
,\qquad
I^Z_{W^\si}=\si\frac{1-2\sw^2}{2\sw\cw},
\eeqar
and
\beqar
I^N_{Z_0 H}=
I^N_{H Z_0}=
\de_{NZ}
\frac{1}{2\sw\cw}.
\eeqar
For charged gauge bosons coupling to $S=Z_0,H$ we obtain,
\beqar\label{chargedeffcouplings}
I^{W^{\si'}}_{W^\si_0 S}=
I^{W^{-\si'}}_{S W^\si_0}
=
-\de_{\si\si'}J^\si_{S}
,\qquad
J^\si_{Z_0}=\si\frac{1}{2\sw}
,\qquad
J^\si_{H}=-\frac{1}{2\sw}.
\eeqar
Note that the couplings \refeq{chargedeffcouplings} are analogous to
the couplings for transverse gauge bosons defined in (B.26) of
\citere{Denner:2000jv}, which can also be written as
\beqar\label{chargedtransvcouplings}
I^{W^{\si'}}_{W^\si_\rT N_\rT}=
I^{W^{-\si'}}_{N_\rT W^\si_\rT}
=-\de_{\si\si'}J^\si_{N_\rT}
,\qquad
J^\si_{A_\rT}=-\si
,\qquad 
J^\si_{Z_\rT}=\si\frac{\cw}{\sw}.
\eeqar

\section{Born matrix elements in the high-energy limit}
\label{se:Born}

As input for the evaluation of the angular-dependent subleading
corrections \refeq{chargedsscgeneral}, which originate from exchange
of soft-collinear W bosons, we need the $\SUtwo$-transformed Born
matrix elements that appear on the right-hand side of
\refeq{chargedsscgeneral}.  The needed amplitudes are evaluated in the
following in high-energy approximation, \ie omitting mass-suppressed
terms.  For longitudinal gauge bosons or Higgs bosons, which are
denoted as
\beqar
S_{i}=W_0^\pm,Z_0,H,\qquad \bar{S}_{i}=W_0^\mp,Z_0,H,
\eeqar
we use the couplings introduced in \refapp{se:GBET}.  Transverse gauge
bosons are denoted as
\beqar
V^{a}_\tau=A_\tau,Z_\tau,W^\pm_\tau
,\qquad
\bar{V}^{a}_\tau=A_\tau,Z_\tau,W^\mp_\tau,
\eeqar
with $\tau=\pm 1$.

\subsubsection*{Purely longitudinal polarizations: $\rL\rL\to \rL\rL$
}

For a generic process
\beq
S_{i_1}(k_1)
S_{i_2}(k_2)
\to
\bar{S}_{i_3}(-k_3)
\bar{S}_{i_4}(-k_4),
\eeq
in the high-energy limit we obtain the amplitude
\beqar
\M_0^{
S_{i_1}
S_{i_2}
S_{i_3}
S_{i_4}
}
&=& 
 \Biggl(e^2 \,
\As
\sum_{V^a=A,Z,W^\pm}I^{V^{a}}_{S_{i_1}^+S_{i_2}} 
I^{\bar V^a}_{S_{i_3}^+S_{i_4}}
-\frac{\lah}{2}\,
\de_{S_{i_1}^+S_{i_2}} 
\de_{S_{i_3}^+S_{i_4}}
\Biggr)
\nl&&{}
+(2\leftrightarrow 3)
+(1\leftrightarrow 3)
,
\eeqar
where 
\beq\label{scalarcouplingdef}
\lah=\frac{e^2}{2\sw^2}\frac{\MH^2}{\MW^2},
\eeq
is the scalar self-coupling.

The Born matrix element for W-boson scattering reads
\beq\label{4phibornmel} 
\M_0^{W_0^-W_0^+W_0^+W_0^-}
=-\left[\lah+\frac{e^2}{4\sw^2\cw^2}\left(\As+\At\right)\right].
\eeq
The $\SUtwo$-transformed Born matrix elements involving 
two equal neutral states,  $S=H$ or $Z_0$, give
\beqar\label{LLLLBornampli2}
\M_0^{S W_0^+ S W_0^-}
=
\M_0^{W_0^-S W_0^+S}
=\epsilon_S \left[\frac{\lah}{2}+\frac{e^2}{4\sw^2}\left(\As+\Au\right)\right],
\eeqar
with $\epsilon_H=1$ and $\epsilon_{Z_0}=-1$.
For the case of different neutral states, 
$S\neq S'$ with $S,S'=H$ or $Z_0$, 
we have
\beqar\label{LLLLBornampli3}
\M_0^{S W_0^+  S' W_0^-}
&=&
\M_0^{W_0^-S' W_0^+ S}
=
\nl
&=&-\epsilon_S
e^2\left[
\frac{1}{4\sw^2}\left(\As-\Au\right)
+\frac{\sw^2-\cw^2}{4\sw^2\cw^2}\At\right]
.\quad
\eeqar

\subsubsection*{Mixed polarizations: $\rT\rT\to \rL\rL$ and $\rL\rL\to \rT\rT$}
For a generic  $\rT\rT\to \rL\rL$ process,
\beq
V^{a_1}_{\tau_1}(k_1)
V^{a_2}_{\tau_2}(k_2)
\to
\bar{S}_{i_3}(-k_3)
\bar{S}_{i_4}(-k_4),
\eeq
in the high-energy limit we obtain the amplitude
\beq
\M_0^{
V^{a_1}_{\tau_1}
V^{a_2}_{\tau_2}
S_{i_3}
S_{i_4}
}
=
2 e^2 (1-\de_{\tau_1\tau_2})\left[
\frac{r_{23}}{r_{12}}
\left(I^{V^{a_1}} I^{V^{a_2}}\right)_{S_{i_3}^+S_{i_4}}
+\frac{r_{13}}{r_{12}}
\left(I^{V^{a_2}} I^{V^{a_1}}\right)_{S_{i_3}^+S_{i_4}}
\right].\quad
\eeq
Inserting the explicit values of the couplings we obtain 
\beqar\label{TTLLBornampli1}
\M_0^{
W^{-}_{\tau_1}
W^{+}_{\tau_2}
W^{+}_{0}
W^{-}_{0}
}
&=&  
-
\frac{e^2}{\sw^2}
(1-\de_{\tau_1\tau_2})
\frac{r_{23}}{r_{12}},
\eeqar
and the $\SUtwo$-transformed amplitudes 
\beqar\label{TTLLBornampli2}
\M_0^{
N_{\tau_1}
W^{+}_{\tau_2}
H
W_{0}^{-}
}
&=&  
-\M_0^{
W^{-}_{\tau_1}
N_{\tau_2}
W_{0}^{+}
H
}=
- \M_0^{
N_{\tau_1}
W^{+}_{\tau_2}
Z_{0}
W_{0}^{-}
}
=  
- \M_0^{
W^{-}_{\tau_1}
N_{\tau_2}
W_{0}^{+}
Z_{0}
}\nl
&=&
\frac{ e^2}{\sw}
(1-\de_{\tau_1\tau_2})
\frac{r_{23}}{r_{12}}D_N,
\eeqar
with $N_\tau=A_\tau,Z_\tau$, and 
\beq\label{Dfun}
D_A=\frac{r_{13}}{r_{23}},\qquad
D_Z=\frac{1}{2\sw\cw}-\frac{r_{13}}{r_{23}}\frac{1-2\sw^2}{2\sw\cw}.
\eeq

Corresponding amplitudes for $\rL\rL\to \rT\rT$
processes  are directly obtained using
$
\M_0^{
S_{i_1}
S_{i_2}
V^{a_3}_{\tau_3}
V^{a_4}_{\tau_4}
}
=
\M_0^{
V^{a_4}_{\tau_4}
V^{a_3}_{\tau_3}
S_{i_2}
S_{i_1}
}
$.

\subsubsection*{Mixed polarizations: $\rT\rL\to \rT\rL$ and $\rL\rT\to \rL\rT$}
For a generic  $\rT\rL\to \rT\rL$ process
\beq
V^{a_1}_{\tau_1}(k_1)
{S}_{i_2}(k_2)
\to
\bar{V}^{a_3}_{-\tau_3}(-k_3)
\bar{S}_{i_4}(-k_4),
\eeq
in the high-energy limit we obtain the amplitude
\beq
\M_0^{
V^{a_1}_{\tau_1}
S_{i_2}
V^{a_3}_{\tau_3}
S_{i_4}
}
=
2 e^2 (1-\de_{\tau_1\tau_3})\left[
\frac{r_{23}}{r_{13}}
\left(I^{V^{a_1}} I^{V^{a_3}}\right)_{S_{i_2}^+S_{i_4}}
+\frac{r_{12}}{r_{13}}
\left(I^{V^{a_3}} I^{V^{a_1}}\right)_{S_{i_2}^+S_{i_4}}
\right].\quad
\eeq
Inserting the explicit values of the couplings we obtain the Born
amplitude
\beqar\label{TTLLBornampli1cross}
\M_0^{
W^{-}_{\tau_1}
W^{+}_{0}
W^{+}_{\tau_3}
W^{-}_{0}
}
&=&  
-  \frac{e^2}{\sw^2}
(1-\de_{\tau_1\tau_3})
\frac{r_{23}}{r_{13}},
\eeqar
and the $\SUtwo$-transformed amplitudes 
\beqar\label{TTLLBornampli2cross}
\M_0^{
N_{\tau_1}
W^{+}_{0}
N'_{\tau_3}
W^{-}_{0}
}
&=&  
2{ e^2}
(1-\de_{\tau_1\tau_3})I^N_{W_0^+}
I^{N'}_{W_0^+}
,\nl  
\M_0^{
W^{-}_{\tau_1}
H
W^{+}_{\tau_3}
H
}
&=&
-\M_0^{
W^{-}_{\tau_1}
Z_0
W^{+}_{\tau_3}
Z_0
}
=- \frac{ e^2}{2\sw^2}
(1-\de_{\tau_1\tau_3})
,\nl  
\M_0^{
W^{-}_{\tau_1}
H
W^{+}_{\tau_3}
Z_0
}
&=&
-\M_0^{
W^{-}_{\tau_1}
Z_0
W^{+}_{\tau_3}
H
}
= \frac{ e^2}{2\sw^2}
(1-\de_{\tau_1\tau_3})
\frac{r_{12}-r_{23}}{r_{13}},
\eeqar
with $N_\tau=A_\tau,Z_\tau$,
$I^A_{W_0^+}=-1$
and
$
I^{Z}_{W_0^+}=(1-2\sw^2)/(2\sw\cw)
$.

Corresponding amplitudes for 
{$\rL\rT\to \rL\rT$ }
processes are directly obtained using
$
\M_0^{
S_{i_1}
V^{a_2}_{\tau_2}
S_{i_3}
V^{a_4}_{\tau_4}
}
=
\M_0^{
V^{a_4}_{\tau_4}
S_{i_3}
V^{a_2}_{\tau_2}
S_{i_1}
}
$.

\subsubsection*{Purely transverse polarizations: $\rT\rT\to \rT\rT$}
For a generic process
\beq
V^{a_1}_{\tau_1}(k_1)
V^{a_2}_{\tau_2}(k_2)
\to
\bar{V}^{a_3}_{-\tau_3}(-k_3)
\bar{V}^{a_4}_{-\tau_4}(-k_4),
\eeq
in the high-energy limit we obtain the amplitude
\newcommand{\detwo}{\de^\SUtwo}
\beqar
\M_0^{
V^{a_1}_{\tau_1}
V^{a_2}_{\tau_2}
V^{a_3}_{\tau_3}
V^{a_4}_{\tau_4}
}
&=& 2  \frac{e^2}{\sw^2} 
\left[
\left(
\detwo_{\bar{V}^{a_1}V^{a_2}}
\detwo_{\bar{V}^{a_3}V^{a_4}}
-
\detwo_{\bar{V}^{a_1}V^{a_3}}
\detwo_{\bar{V}^{a_2}V^{a_4}}
\right)
A(\vec{\tau},\{r_{ij}\})
\right.\nl&&\left.{}
+
\left(
\detwo_{\bar{V}^{a_1}V^{a_2}}
\detwo_{\bar{V}^{a_3}V^{a_4}}
-
\detwo_{\bar{V}^{a_1}V^{a_4}}
\detwo_{\bar{V}^{a_3}V^{a_2}}
\right)
B(\vec{\tau},\{r_{ij}\})
\right],
\eeqar
where the matrix $\detwo$ is defined in Appendix B.2 of
\citere{Pozzorini:2001rs}, and has the non-vanishing components
$\detwo_{AA}=\sw^2$, $\detwo_{ZZ}=\cw^2$,
$\detwo_{AZ}=\detwo_{ZA}=-\cw\sw$,
$\detwo_{W^+W^+}=\detwo_{W^-W^-}=1$.  The functions $A,B$ depend on
the invariants ${r_{ij}}$ and the polarizations
$\vec{\tau}=(\tau_1,\tau_2,\tau_3,\tau_4)$.  The only combinations of
polarizations that yield non-vanishing contributions are
\beqar
\vec{\tau}_a=(\si,\si,-\si,-\si),\qquad
\vec{\tau}_b=(\si,-\si,\si,-\si),\qquad
\vec{\tau}_c=(-\si,\si,\si,-\si)
\eeqar
with $\si=\pm$, and we have 
\beq\label{Bdef}
A(\vec{\tau}_a,\{r_{ij}\})=-\frac{\mas}{\mau},\qquad
A(\vec{\tau}_b,\{r_{ij}\})=
-\frac{\mat^2}{\mas\mau}
,\qquad
A(\vec{\tau}_c,\{r_{ij}\})=-\frac{\mau}{\mas},
\eeq
and
\beq\label{Cdef}
B(\vec{\tau}_a,\{r_{ij}\})=-\frac{\mas}{\mat},\qquad
B(\vec{\tau}_b,\{r_{ij}\})=-\frac{\mat}{\mas},\qquad
B(\vec{\tau}_c,\{r_{ij}\})=
-\frac{\mau^2}{\mas\mat}
,
\eeq
whereas if $\vec{\tau}\neq\vec{\tau}_a,\vec{\tau}_b,\vec{\tau}_c$ then
$ A(\vec{\tau},\{r_{ij}\})=B(\vec{\tau},\{r_{ij}\})=0.  $ Inserting
the explicit values of the couplings we obtain
\beqar\label{TTTTBornampli1}
\M_0^{
W^{-}_{\tau_1}
W^{+}_{\tau_2}
W^{+}_{\tau_3}
W^{-}_{\tau_4}
}
&=& 
\frac{2  e^2}{\sw^2}B(\vec{\tau},\{r_{ij}\}),
\eeqar
and the $\SUtwo$-transformed matrix elements
\beqar\label{TTTTBornampli2}
\M_0^{
N_{\tau_1}
W^{+}_{\tau_2}
N'_{\tau_3}
W^{-}_{\tau_4}
}
&=& 
\M_0^{
W^{-}_{\tau_1}
N'_{\tau_2}
W^{+}_{\tau_3}
N_{\tau_4}
}
=-\frac{2  e^2}{\sw^2}\detwo_{N'N}A(\vec{\tau},\{r_{ij}\}),
\eeqar
for $N,N'=A,Z$.  We note that the ratio between the Born amplitudes
\refeq{TTTTBornampli2} and \refeq{TTTTBornampli1},
\beqar\label{TTTTBornampli3}
\frac{\M_0^{
N_{\tau_1}
W^{+}_{\tau_2}
N'_{\tau_3}
W^{-}_{\tau_4}
}}
{\M_0^{
W^{-}_{\tau_1}
W^{+}_{\tau_2}
W^{+}_{\tau_3}
W^{-}_{\tau_4}
}}
&=& 
\frac{\M_0^{
W^{-}_{\tau_1}
N'_{\tau_2}
W^{+}_{\tau_3}
N_{\tau_4}}}
{\M_0^{
W^{-}_{\tau_1}
W^{+}_{\tau_2}
W^{+}_{\tau_3}
W^{-}_{\tau_4}
}}
=-\detwo_{N'N}
\frac{A(\vec{\tau},\{r_{ij}\})}
{B(\vec{\tau},\{r_{ij}\})}
=-\detwo_{N'N}
\frac{r_{13}}{r_{23}}
,
\eeqar
is independent of the polarizations.

\section{Electromagnetic virtual and real contributions}
\label{se:emlogs}

In this appendix we provide simple substitutions that permit to
generalize the results of \citere{Denner:2000jv} to semi-inclusive
$2\to 2$ processes, by including the soft-photon bremsstrahlung
corrections\footnote{ The soft bremsstrahlung corrections to squared
  matrix elements factorize into the squared Born matrix elements
  times correction factors.  These latter have been divided by 2 and
  combined with the virtual correction factors to (non-squared) matrix
  elements given in \citere{Denner:2000jv}.}.  These substitutions
concern the infrared-divergent logarithms $L^\elm(s,\la^2,m_k^2)$,
$\lWla$, $l^\elm(m_k^2)$ that appear in Eqs.~(3.7), (3.8), (3.10),
(3.12), (4.6), (4.7), (4.10) and (4.33) of \citere{Denner:2000jv} and
have to be replaced with the logarithms $L^\EM(m_k^2)$, $l_\SS^\elm$
and $l^\EM(m_k^2)$, defined in the following.

These results are valid for arbitrary $2\to 2$ processes in the CM
frame, with a soft-photon cut-off $\Delta E$.  The contributions from
virtual photons (superscript `em') and real bremsstrahlung
(superscript `brems') as well as their sum (superscript `EM') are
given separately and split into leading-, subleading-soft-collinear
and collinear (or soft) parts.

\subsubsection*{Leading soft-collinear contributions}
The terms 
$L^\elm(s,\la^2,m_k^2)$, which are defined in 
Eq.~(3.8) of  \citere{Denner:2000jv} and contribute to 
Eq.~(3.7) of  \citere{Denner:2000jv} have to be substituted by
$
L^\EM(m_k^2)=
L^\elm(m_k^2)+
L^\brems(m_k^2)
$,
with
\beqar\label{LSCemlogs1}
L^\elm(m_k^2)
&=& \frac{\alpha}{4\pi}
\Biggl\{
2 \log{\left(\frac{|\mas|}{m_k^2}\right)}
\log{\left(\frac{\MW^2}{\la^2}\right)}
-
\log^2{\left(\frac{\MW^2}{m_k^2}\right)}
\Biggr\}
,\nl
L^\brems(m_k^2)
&=& \frac{\alpha}{4\pi}
\Biggl\{
\log{\left(\frac{|\mas|}{m_k^2}\right)}
\left[2\log{\left(\frac{\la^2}{4\Delta E^2}\right)}
+
\log{\left(\frac{|\mas|}{m_k^2}\right)}
\right]
\Biggr\}
,\nl
L^\EM(m_k^2)
&=& \frac{\alpha}{4\pi}
\Biggl\{
-\log^2{\left(\frac{|\mas|}{\MW^2}\right)}
+
2 \log{\left(\frac{|\mas|}{4\Delta E^2}\right)}
\log{\left(\frac{|\mas|}{m_k^2}\right)}
\Biggr\}
.
\eeqar

\subsubsection*{Subleading soft-collinear contributions}
The terms $\lWla$ in Eqs.~(3.10) and (3.12) of \citere{Denner:2000jv}
have to be substituted by $ l_\SS^\EM= l_\SS^\elm+ l_\SS^\brems $,
with
\beq\label{angularbrems1}
l_\SS^\elm=\frac{\alpha}{4\pi}\log{\left(\frac{\MW^2}{\la^2}\right)}
,\qquad
l_\SS^\brems=\frac{\alpha}{4\pi}\log{\left(\frac{\la^2}{4\Delta E^2}\right)}
,\qquad
l_\SS^\EM=\frac{\alpha}{4\pi}\log{\left(\frac{\MW^2}{4\Delta E^2}\right)}
.
\eeq

\subsubsection*{Collinear and soft single logarithms}

The terms $l^\elm(m_k^2)$, which are defined in Eq.~(4.7) of
\citere{Denner:2000jv} and contribute to Eqs.~(4.6), (4.10) and (4.33)
of \citere{Denner:2000jv}, have to be substituted by $ l^\EM(m_k^2)=
l^\elm(m_k^2)+ l^\brems(m_k^2) $, with
\beqar\label{NLLbrems1}
l^\elm(m_k^2)&=&\frac{\alpha}{4\pi}
\left[\log{\left(\frac{\MW^2}{\la^2}\right)}
+\frac{1}{2}\log{\left(\frac{\MW^2}{m_k^2}\right)}
\right],\nl
l^\brems(m_k^2)&=&\frac{\alpha}{4\pi}
\left[\log{\left(\frac{\la^2}{4\Delta E^2}\right)}
+\log{\left(\frac{|\mas|}{m_k^2}\right)}
\right],\nl
l^\EM(m_k^2)&=&\frac{\alpha}{4\pi}
\left[\log{\left(\frac{|\mas|}{4\Delta E^2}\right)}
+\frac{3}{2}\log{\left(\frac{\MW^2}{m_k^2}\right)}
\right]
.
\eeqar

\end{appendix}


\begin{thebibliography}{9}

 \newcommand{\vj}[4]{{#1 }{\bf #2 }\ifnum#3<100 (19#3) \else (#3) \fi #4}
 \newcommand{\epjc}[3]{\vj{Eur. Phys. J. C}{#1}{#2}{#3}}
 \newcommand{\npb}[3]{\vj{Nucl. Phys. B}{#1}{#2}{#3}}
 \newcommand{\plb}[3]{\vj{Phys. Lett. B}{#1}{#2}{#3}}
 \newcommand{\prp}[3]{\vj{Phys. Rept.}{#1}{#2}{#3}}
 \newcommand{\prd}[3]{\vj{Phys. Rev. D}{#1}{#2}{#3}}
 \newcommand{\prl}[3]{\vj{Phys. Rev. Lett.}{#1}{#2}{#3}}
 \providecommand{\etal}{{\it et al.}}
 \newcommand{\hepph}[1]{{hep-ph/#1}}
 \newcommand{\hepex}[1]{{hep-ex/#1}}

\bibitem{SEWSB}
B.~W.~Lee, C.~Quigg and H.~Thacker, Phys.\ Rev.\ D {\bf 16} (1977) 1519;\\
M.~Veltman, Acta\ Phys.\ Polon.\ B {\bf 8} (1977) 475;\\
M.~S.~Chanowitz and M.~K.~Gaillard, Nucl.\ Phys.\ B {\bf 261} (1985) 379. 

\bibitem{Bagger:1993zf}
  J.~Bagger {\it et al.},
  Phys.\ Rev.\ D {\bf 49} (1994) 1246
  [hep-ph/9306256].


\bibitem{Butterworth:2002tt}
  J.~M.~Butterworth, B.~E.~Cox and J.~R.~Forshaw,
  Phys.\ Rev.\ D {\bf 65} (2002) 096014
  [hep-ph/0201098].

\bibitem{sewsb-analyses1}
J.~Bagger {\it et al.},
Phys.\ Rev.\ D {\bf 52} (1995) 3878.

\bibitem{Barger:1995cn}
  V.~D.~Barger, K.~Cheung, T.~Han and R.~J.~N.~Phillips,
  Phys.\ Rev.\ D {\bf 52} (1995) 3815
  [hep-ph/9501379].

\bibitem{sewsb-analyses2}
M.~Golden, T.~Han and G.~Valencia, in ``Electroweak Symmetry Breaking and New 
Physics at the TeV Scale'', ed. T.~L.~Barklow, hep-ph/9511206. 

\bibitem{boos-1997}
E.~Boos, H.~J.~He, W.~Kilian, A.~Pukhov, C.~P.~Yuan, P.~M.~Zerwas, Phys.\ 
Rev.\ D {\bf 57} (1998) 1553; 
Phys.\ Rev.\ D {\bf 61} (2000) 077901.

\bibitem{sewsb-analyses4}
E.~Accomando, A.~Ballestrero, S.~Bolognesi, E.~Maina, C.~Mariotti, JHEP 
(2006) 0603;
E.~Accomando, A.~Ballestrero, A.~Belhouari, E.~Maina, hep-ph/0603167.


\bibitem{ilc}
%
J.~A.~Aguilar-Saavedra {\it et al.},
TESLA Technical Design Report Part III: Physics at an $\mathrm{e^+e^-}$
Linear Collider,
hep-ph/0106315;\\
%
T.~Abe {\it et al.}  [American Linear Collider Working Group Collaboration],
in {\it Proc. of the APS/DPF/DPB Summer Study on the Future of
  Particle Physics (Snowmass 2001) } ed. R.~Davidson and C.~Quigg,
SLAC-R-570, {\it Resource book for Snowmass 2001},
[hep-ex/0106055, hep-ex/0106056, hep-ex/0106057, hep-ex/0106058];\\
%
K.~Abe {\it et al.}  [ACFA Linear Collider Working Group Collaboration],
ACFA Linear Collider Working Group report,
[hep-ph/0109166];\\
%
Report from the International Linear Collider Technical Review 
Committee, G.~A.~Loew (SLAC), SLAC-PUB-10024, Jul 2003;\\
%
R.~D.~Heuer, Nucl.\ Phys.\ Proc.\ Suppl.\ 154 (2006) 131.

\bibitem{Accomando:2004sz}
  E.~Accomando {\it et al.}  [CLIC Physics Working Group],
  CERN-2004-005, hep-ph/0412251.


\bibitem{Kuroda:1991wn}
M.~Kuroda, G.~Moultaka and D.~Schildknecht,
\npb{350}{1991}{25};\\
%
G.~Degrassi and A.~Sirlin,
\prd{46}{1992}{3104};\\
%
W.~Beenakker \etal,
\npb{410}{1993}{245};
%
\plb{317}{1993}{622};\\
%
A.~Denner, S.~Dittmaier and R.~Schuster,
\npb{452}{1995}{80};\\
%
A.~Denner, S.~Dittmaier and T.~Hahn,
\prd{56}{1997}{117};\\
%
A.~Denner and T.~Hahn,
\npb{525}{1998}{27};\\
%
M.~Beccaria \etal,
\prd{58}{1998}{093014}
[\hepph{9805250}];\\
%
P.~Ciafaloni and D.~Comelli,
Phys.\ Lett.\ B {\bf 446} (1999) 278
[hep-ph/9809321].


\bibitem{Denner:2000jv}
A.~Denner and S.~Pozzorini,
\epjc{18}{2001}{461} 
[\hepph{0010201}].

\bibitem{Denner:2001gw}
A.~Denner and S.~Pozzorini,
\epjc{21}{2001}{63}
[\hepph{0104127}].

\bibitem{Pozzorini:2001rs}
S.~Pozzorini, {\em doctoral thesis, Universit\"at Z\"urich, 2001},
\hepph{0201077}.


\bibitem{Fadin:2000bq}
V.~S.~Fadin, L.~N.~Lipatov, A.~D.~Martin and M.~Melles,
\prd{61}{2000}{094002} 
[\hepph{9910338}];\\
%
J.~H.~K\"uhn, A.~A.~Penin and V.~A.~Smirnov,
\epjc{17}{2000}{97}
[\hepph{9912503}];\\
%
%
M.~Ciafaloni, P.~Ciafaloni and D.~Comelli,
\prl{84}{2000}{4810}
[\hepph{0001142}];\\
%
M.~Melles,
\prd{63}{2001}{034003}
[\hepph{0004056}];\\
%
%
%
W.~Beenakker and A.~Werthenbach,
%
\npb{630}{2002}{3}
[\hepph{0112030}];\\
%
A.~Denner, M.~Melles and S.~Pozzorini,
{\sl Nucl.\ Phys.\ B} {\bf 662} (2003) 299
[\hepph{0301241}].
%
%
%

\bibitem{Denner:2006jr}
A.~Denner, B.~Jantzen and S.~Pozzorini,
{\sl Nucl.\ Phys.\ B}  DOI: 10.1016/j.nuclphysb.2006.10.014,
hep-ph/0608326.

%
%
%
%
%


\bibitem{Beccaria:2000fk}
M.~Beccaria \etal,
\prd{61}{2000}{073005}
[\hepph{9906319}];
%
\prd{61}{2000}{011301}
[\hepph{9907389}];\\
%
M.~Beccaria, F.~M.~Renard and C.~Verzegnassi,
\prd{63}{2001}{053013}
[\hepph{0010205}];
%
\prd{64}{2001}{073008}
[\hepph{0103335}];
%
{\sl Nucl.\ Phys.\ B} {\bf 663} (2003) 394
[hep-ph/0304175].

\bibitem{Layssac:2001ur}
J.~Layssac and F.~M.~Renard,
\prd{64}{2001}{053018}
[\hepph{0104205}];\\
%
G.~J.~Gounaris, J.~Layssac and F.~M.~Renard,
Phys.\ Rev.\ D {\bf 67} (2003) 013012
[\hepph{0211327}].

\bibitem{Dittmaier:2001ay}
S.~Dittmaier and M.~Kr\"amer,
Phys.\ Rev.\ D {\bf 65} (2002) 073007
[hep-ph/0109062].

\bibitem{Accomando:2001fn}
 E.~Accomando, A.~Denner and S.~Pozzorini,
{\sl Phys.\ Rev.\ D} {\bf 65}, 073003 (2002)
[hep-ph/0110114];\\
%
W.~Hollik and C.~Meier,
{\sl Phys.\ Lett.\ B} {\bf 590} (2004) 69
[hep-ph/0402281];\\
%
  E.~Accomando, A.~Denner and C.~Meier,
  {\sl Eur.\ Phys.\ J.\ C} {\bf 47} (2006) 125
  [hep-ph/0509234];\\
%
E.~Accomando and A.~Kaiser,
{\sl Phys.\ Rev.\ D} {\bf 73}, 093006 (2006)
[hep-ph/0511088].

\bibitem{Accomando:2004de}
 E.~Accomando, A.~Denner and A.~Kaiser,
{\sl  Nucl.\ Phys.\ B} {\bf 706}, 325 (2005)
[hep-ph/0409247].
%

\bibitem{Maina:2003is}
E.~Maina, S.~Moretti, M.~R.~Nolten and D.~A.~Ross,
{\sl Phys.\ Lett.\ B} {\bf 570} (2003) 205
[hep-ph/0307021];\\
%
E.~Maina, S.~Moretti and D.~A.~Ross,
{\sl Phys.\ Lett.\ B} {\bf 593} (2004) 143
[Erratum-ibid.\ {\sl B} {\bf 614} (2005) 216]
[hep-ph/0403050];\\
%
S.~Moretti, M.~R.~Nolten and D.~A.~Ross,
{\sl Phys.\ Lett.\ B} {\bf 639} (2006) 513
[hep-ph/0603083];
%
hep-ph/0606201.

\bibitem{Baur:2004ig}
U.~Baur and D.~Wackeroth,
Phys.\ Rev.\ D {\bf 70} (2004) 073015
[hep-ph/0405191].


\bibitem{Kuhn:2004em}
J.~H.~K\"uhn, A.~Kulesza, S.~Pozzorini and M.~Schulze,
{\sl Phys.\ Lett.\ B} {\bf 609}, 277 (2005)
[hep-ph/0408308];
%
{\sl Nucl.\ Phys.\ B} {\bf 727}, 368 (2005)
[hep-ph/0507178];
%
{\sl JHEP} {\bf 0603}, 059 (2006)
[hep-ph/0508253].

\bibitem{Kuhn:2006vh}
J.~H.~K\"uhn, A.~Scharf and P.~Uwer,
hep-ph/0610335.

\bibitem{Bernreuther:2006vg}
W.~Bernreuther, M.~F\"ucker and Z.~G.~Si,
hep-ph/0610334.

\bibitem{moenig-2005}
P.~Krstonosic, K.~M\"onig, M.~Beyer, E.~Schmidt, H.~Schr\"oder, hep-ph/0508179.

\bibitem{zerwas-2006}
W.~Kilian and P.~M.~Zerwas, ECONF C0508141:PLEN0003,2005 [hep-ph/0601217].

\bibitem{kuss1997}
I.~Kuss and H.~Spiesberger, Phys.\ Rev.\ D {\bf 53} (1996) 6078 
[hep-ph/9507204].

\bibitem{Yao:2006px}
  W.~M.~Yao {\it et al.}  [Particle Data Group],
  J.\ Phys.\ G {\bf 33} (2006) 1.


\bibitem{beenakker1999}
W.~Beenakker, F.~A.~Berends and A.~P.~Chapovsky, Nucl.\ Phys.\ B {\bf 548} 
(1999) 3 [hep-ph/9811481].

\bibitem{Denner:2000bj}
  A.~Denner, S.~Dittmaier, M.~Roth and D.~Wackeroth,
  Nucl.\ Phys.\ B {\bf 587} (2000) 67
  [hep-ph/0006307];
%
  Comput.\ Phys.\ Commun.\  {\bf 153} (2003) 462
  [hep-ph/0209330].

\bibitem{Jadach:1998tz}
  S.~Jadach, W.~Placzek, M.~Skrzypek, B.~F.~L.~Ward and Z.~Was,
  Phys.\ Rev.\ D {\bf 61} (2000) 113010
  [hep-ph/9907436]; 
%
  Phys.\ Rev.\ D {\bf 65} (2002) 093010
  [hep-ph/0007012].


\bibitem{Ballestrero:1999md}
  A.~Ballestrero,
  hep-ph/9911318.


\bibitem{HAmethod}
A. Ballestrero and E. Maina,
Phys.\ Lett.\ B {\bf 350} (1995) 225 [hep-ph/9403244]. 

\bibitem{nonfac1}
W.~Beenakker, A.~P.~Chapovsky and F.~A.~Berends, Phys.\ Lett.\ B {\bf 411} 
(1997) 203 [hep-ph/9706339]; Nucl.\ Phys.\ B {\bf 508} (1997) 17 
[hep-ph/9707326].
\bibitem{nonfac2}
A.~Denner, S.~Dittmaier and M.~Roth, Nucl.\ Phys.\ B {\bf 519} (1998) 39
[hep-ph/9710521]; Phys.\ Lett.\ B {\bf 429} (1998) 145 [hep-ph/9803306]. 
\bibitem{hahn1997}
A.~Denner and T.~Hahn, Nucl.\ Phys.\ B {\bf 525} (1998) 27 [hep-ph/9711302].  

\end{thebibliography}
\end{document}